\documentclass[11pt]{article}
\pdfoutput=1
\usepackage{jheppub}
\usepackage{graphicx}
\usepackage{color}
\usepackage{verbatim}
\usepackage{subfigure}
\usepackage{acronym}
\usepackage{amsfonts}
\usepackage{slashed}
\usepackage{epsfig}
\usepackage{latexsym,amssymb,amsmath,makeidx,mathrsfs,multirow,lineno,bm,bbm}

\allowdisplaybreaks[4]
\usepackage{hyperref}

\newcommand{\fr}[2]{\mbox{$\frac{\,{#1}\,}{#2}$}}

\def\bge{\begin{equation}}
\def\ede{\end{equation}}
\def\bga{\begin{aligned}}
\def\eda{\end{aligned}}
\newcommand{\beq}{\begin{equation}}
\newcommand{\eeq}{\end{equation}}
\newcommand{\bq}{\begin{equation}}
\newcommand{\eq}{\end{equation}}
\newcommand{\ba}{\begin{array}}
\newcommand{\ea}{\end{array}}
\newcommand{\beqa}{\begin{eqnarray}}
\newcommand{\eeqa}{\end{eqnarray}}
\newcommand{\beqs}{\begin{subequations}}
\newcommand{\eeqs}{\end{subequations}}

\def\nn{\nonumber}

\def\dis{\displaystyle}

\def\({\left(}
\def\){\right)}
\def\[{\left[}
\def\]{\right]}

\def\End{\end{document}}

\def\over{\overline}

\def\leqq{\leqslant}
\def\geqq{\geqslant}

\def\al{\alpha}

\def\ga{\gamma}

\def\tanb{\tan\!\beta}

\def\cosba{\cos(\beta\!-\!\alpha)}

\def\to{\rightarrow}

\def\gaga{\gamma\gamma}

\def\ZZZ{\mathbb{Z}_2^{}}

\def\gaga{\gamma\gamma}

\def\H{\mathbb{H}}
\def\Ha{\mathbb{H}_1^{}}
\def\Hb{\mathbb{H}_2^{}}

\def\ZZ{\mathbb{Z}}
\def\tanb{\tan\beta}
\def\MH{M_H^{}}

\def\Mh{M_h^{}}

\def\MA{M_A^{}}

\def\SZZ{\mathcal{Z}}
\def\gaga{\gamma\gamma}

\def\End{\end{document}}
\renewcommand{\thefootnote}{\fnsymbol{footnote}}
\title{LHC Search of New Higgs Boson via Resonant Di-Higgs Production
       with Decays into $\boldsymbol{4W}$}

\author[a]{Jing Ren,}
\emailAdd{jren@physics.utoronto.ca}
\author[b,c]{~Rui-Qing Xiao,}
\emailAdd{ruiqingxiao@lbl.gov}
\author[d,e]{~Maosen Zhou,}
\emailAdd{maosenzhou@ihep.ac.cn}
\author[d,e]{\hspace{6cm}~Yaquan Fang,\footnote{Corresponding authors.}}
\emailAdd{fangyq@ihep.ac.cn}
\author[b,f,g,h]{~Hong-Jian He,$^{\dagger}$}
\emailAdd{hjhe@tsinghua.edu.cn}
\author[c]{~Weiming Yao$^{\dagger}$}
\emailAdd{wmyao@lbl.gov}
\affiliation[a]{Department of Physics, University of Toronto, Toronto, Ontario, Canada M5S1A7}
\affiliation[b]{Institute of Modern Physics and Center for High Energy Physics, \\
Tsinghua University, Beijing 100084, China}
\affiliation[c]{Lawrence Berkeley National Laboratory, Berkeley, California 94720, USA}
\affiliation[d]{Institute of High Energy Physics, Beijing 100049, China}
\affiliation[e]{University of Chinese Academy of Sciences, Beijing 100049, China}
\affiliation[f]{T.\ D.\ Lee Institute, Shanghai 200240, China}
\affiliation[g]{School of Physics and Astronomy, Shanghai Jiao Tong University, Shanghai 200240, China}
\affiliation[h]{Center for High Energy Physics, Peking University, Beijing 100871, China
}
\abstract{
\\[1mm]
Searching for new Higgs particle beyond the observed light Higgs boson $h^0$(125GeV)
will unambiguously point to new physics beyond the standard model.
We study the resonant production of a CP-even heavy Higgs state $H^0$
in the di-Higgs channel, $gg\to H^0\to h^0h^0\to WW^*WW^*$,\,
at the LHC Run-2 and the high luminosity LHC (HL-LHC).
We analyze two types of the $4W$ decay modes,
one with the same-sign di-leptons
($4W\!\!\to\ell^\pm\nu\ell^\pm\nu 4q$) and the other with tri-leptons
($4W\!\!\to\ell^\pm\nu\ell^\mp\nu\ell^\pm\nu 2q$).
We perform a full simulation for the signals and backgrounds,
and estimate the discovery potential of the heavy Higgs state
at the LHC Run-2 and the HL-LHC,
in the context of generic two-Higgs-doublet models (2HDM).
We determine the viable parameter space of the 2HDM as allowed by
the theoretical constraints and the current experimental limits.
We systematically analyze the allowed parameter space of the 2HDM
which can be effectively probed by the heavy Higgs searches of the LHC,
and further compare this with the viable parameter region under the
current theoretical and experimental bounds.
}

\keywords{Higgs Physics, Beyond Standard Model. %\hfill
\\[3mm]
JHEP~1806\,(2018)\,090~$[$\,arXiv:1706.05980\,$]$
}

\begin{document}
\maketitle

\renewcommand{\thefootnote}{\arabic{footnote}}
\setcounter{page}{2}

%\tableofcontents

%----------------------------------------------------------------------
\vspace*{5mm}
%\newpage

\section{\hspace*{-2mm}Introduction}
\label{sec:1}

Since the LHC discovery of the light Higgs boson $h^0$(125GeV)
in 2012\,\cite{ATLAS2012}\cite{CMS2012}, both ATLAS and CMS collaborations have
much improved the measurements on its mass and couplings
which behave fairly standard-model-like\,\cite{Higgs-sum}.
But, so far its self-interactions have not yet been tested at the LHC.
The cubic Higgs coupling of $\,h^0$ can be directly probed via the di-Higgs production
at hadron colliders\,\cite{h3}, though it would be much harder.
At the LHC(14TeV), the di-Higgs production cross section
in the standard model (SM) is small.
But, most extensions of the SM contain an enlarged Higgs sector
and predict new cubic interaction $H^0h^0h^0$
between a heavier Higgs state $H^0$ and the light Higgs pair $h^0h^0$.
For $\,\MH > 2\Mh$,\, the di-Higgs cross section can be
significantly enhanced via the resonant production $\,pp\to H^0\to h^0h^0\,$,\,
which simultaneously serves as an important discovery channel of new Higgs boson.
Such an extended Higgs sector may include additional new singlets, doublets, or triplets
under the SM gauge group $\text{SU(2)}_{\!L}^{}\otimes \text{U(1)}_Y^{}$,\,
or under an enlarged gauge group with extra SU(2)\,\cite{SU2x}.
Among these, the two-Higgs-doublet model (2HDM)\,\cite{2HDM}
is a minimal extension by adding the second Higgs doublet to the SM.
It is a necessary ingredient of the minimal supersymmetric SM (MSSM)\,\cite{MSSM}
and its next-to-minimal extension (NMSSM)\,\cite{NMSSM}.
It is common to impose a discrete $\ZZZ$ symmetry on the 2HDM for preventing
the tree-level flavor changing neutral currents. There are at least four kinds
of model setup due to the different assignments of fermion Yukawa couplings with
each Higgs doublet, namely, Type-I, Type-II, lepton-specific, and flipped.
For the current study, we will consider the 2HDM Type-I and Type-II for
demonstrations.

The LHC collaborations have searched the resonant heavy Higgs production
$\,pp\!\to\! H^0\!\to\! h^0h^0\,$ for a number of di-Higgs decay channels.
At the LHC Run-1 with 8\,TeV collision energy,
the di-Higgs decay final states $\,h^0h^0\!\to\! bbbb$~\cite{Aad:2015uka}, $bb\gamma\gamma$~\cite{Aad:2014yja}\cite{CMSHhh},
$bb\tau\tau$~\cite{Aad:2015xja}, $\gamma\gamma WW^*$~\cite{Aad:2015xja} were analyzed.
In the $\,bb\gamma\gamma\,$ channel,  ATLAS found an excess of $2.4\sigma$ at
$\,M(bb\gamma\gamma)\sim 300$\,GeV\,\cite{Aad:2014yja}. The Run-2 of LHC(13\,TeV)
has searched the resonant di-Higgs production with
$\,bbbb$ \cite{ATLAS:2016ixk}\cite{CMS:2016pwo},
$bb\gamma\gamma$ \cite{ATLASRun2bbgg}\cite{CMS:2016vpz},
$bb\tau\tau$~\cite{CMS:2017orf}, $bbWW^*$~\cite{CMS:2016rec},
and $\gamma\gamma WW^*$~\cite{ATLAS:2016qmt} final states,
where the $bb\tau\tau$ and $bbWW^*$
analyses are updated with 35.9\,fb$^{-1}$ data from CMS.
The sensitivity to larger $\MH$ range is also studied
for the high luminosity LHC (HL-LHC) with an integrated luminosity of 3000\,fb$^{-1}$, in
the $bbWW^*$ channel\,\cite{LMPHhh} and the $\gamma\gamma WW^*$ channel\,\cite{llc}.
There are many recent phenomenological studies
on resonant di-Higgs production with the above-mentioned di-Higgs decay channels and
for testing different new physics scenarios\,\cite{hh-other}.
There are also studies of the $4W$ channel for
non-resonant di-Higgs production within the SM\,\cite{4W-SM} or
non-resonant production $gg\!\to\!SS\,$
for the SM with extra singlet $S$ \cite{SM+SS4W}, and the resonant production
$gg\!\to\!H(S)\!\!\to\!\!SS,Sh\!\to\!4W\!\!\to\! 4(\ell\nu)$
in the context of 2HDM plus extra singlet $S$ \cite{s}.

In this work, we study the new Higgs boson $H^0$ production
via the di-Higgs channel with $4W$-decays,
$gg\!\to\! H^0\!\!\to\! h^0h^0\!\!\to\! WW^*WW^*$.\,
We analyze two kinds of $4W$ decay products, one with the same-sign di-leptons
($\,WW^*WW^*\!\!\to\!\ell^\pm\nu\ell^\pm\nu 4q$\,) and the other with tri-leptons
($\,WW^*WW^*\!\!\to\!\ell^\pm\nu\ell^\mp\nu\ell^\pm\nu 2q$\,).
The advantage of these channels is that requiring the same-sign di-leptons
or the tri-leptons in the final states can significantly suppress QCD backgrounds.
We perform a full analysis of signals and backgrounds
by generating the events at parton level,
and then use Pythia for hadronization and parton shower,
followed by fast Delphes detector simulations.
We will study the discovery potential of the heavy Higgs state
at the LHC Run-2 and HL-LHC, in the context of the generic 2HDM Type-I and Type-II.
We derive the theoretical constraints and
the current experimental limits on the 2HDM parameter space.
We systematically analyze which part of the 2HDM parameter space can be probed by the
new Higgs boson searches at the LHC Run-2 and the HL-LHC.
We note that for the LHC discovery of $H^0$
via resonant di-Higgs production, it is valuable
to include the $\,h^0h^0 \!\!\to\!\! WW^*WW^*$
channel in addition to other di-Higgs decay modes.
This will allow a combined analysis of all di-Higgs decay channels to enhance the
new Higgs discovery reach of the LHC.

This paper is organized as follows.
Section\,\ref{sec:2} will provide the 2HDM setup
and present the production and decays of the new Higgs boson $H^0$.
We will analyze the relevant theoretical constraints
and the current direct/indirect experimental limits
on the 2HDM parameter space. We set up the benchmarks for the later collider analysis.
In Section\,\ref{sec:3}, we will study the new Higgs boson production,
$\,gg\!\to\! H^0\!\!\to\! h^0h^0\!\!\to\! WW^*WW^*$, via two kinds of $4W$ decay modes.
For each $4W$ decay channel, we perform full simulations for three
Higgs benchmark scenarios.
In Section\,\ref{sec:4}, we will analyze the 2HDM parameter space which can be probed
by the new Higgs boson searches in the $4W$ channel at the LHC Run-2 and the HL-LHC.
We further present the current theoretical and experimental constraints
on the 2HDM parameter space, and combine them with the direct searches of the
new Higgs boson in the $4W$ channel.
Finally, we will conclude in Section\,\ref{sec:5}.

%%%%%%%%%%%%%%%%%%%%%%%%%%%%%%%%%%%%%%%%%%%%%%
%%=============== Section 2 ================%%
%%%%%%%%%%%%%%%%%%%%%%%%%%%%%%%%%%%%%%%%%%%%%%

\vspace*{3mm}
\section{\hspace*{-2mm}Heavy Higgs Boson H$^0$ in 2HDM: Decays and Production}
\label{sec:2}

In this section, we will first define the model setup of the 2HDM
and its parameter space in Section\,\ref{sec:2.1}.
Then, in Section\,\ref{sec:2.2}, we analyze the relevant theoretical constraints
and the current direct/indirect experimental limits on the 2HDM parameter space.
Finally, in Section\,\ref{sec:2.3}, we present the decays and production
of the heavy Higgs boson $H^0$ at the LHC. With these, we set up three
benchmarks for our LHC studies in the subsequent sections.

%%%%%%%%%%%%%%%%%%%%%%%% Sec.2.1 %%%%%%%%%%%%%%%%%%%%%%%%%
\vspace*{2.5mm}
\subsection{\hspace*{-2mm}The Model Setup}
\vspace*{2mm}
\label{sec:2.1}

The 2HDM\,\cite{2HDM} is a minimal extension of the SM Higgs sector.
To avoid the tree-level flavor changing neutral currents,
it is common to impose a discrete $\,\ZZ_{2}^{}$\, symmetry
on the Higgs sector, with the Higgs doublets $\Ha$ and $\Hb$ being
$\,\ZZZ$\, odd and even, respectively.
Thus, the CP conserving Higgs potential under $\,\ZZZ$\,
can be written as,
\beqa
%\begin{array}{ll}
V &=& \dis M_{11}^2 \H_1^\dagger\H_1^{} + M_{22}^2 \H_2^\dagger\H_2^{}
- M_{12}^2 (  \H_1^\dagger\H_2^{} + \H_2^\dagger\H_1^{})
+  \fr{\,\lambda_1^{}}{2}(  \H_1^\dagger\H_1^{})^2
+  \fr{\,\lambda_2^{}}{2}(  \H_2^\dagger\H_2^{})^2
\hspace*{8mm}
\nn\\[1mm]
&& + \dis\lambda_3^{} \H_1^\dagger\H_1^{} \H_2^\dagger\H_2^{}
+ \lambda_4^{} \H_1^\dagger\H_2^{} \H_2^\dagger\H_1^{}
+ \fr{\,\lambda_5^{}}{2} \left[ (  \H_1^\dagger\H_2^{})^2
 + (\H_2^\dagger\H_1^{})^2 \right] \!,
%\end{array}
\label{eq:2HDM_potential}
\label{eq:V}
\eeqa
where all parameters are real\,\cite{2HDM}\cite{lambda}\cite{Bernon:2015qea}.
The potential $V$ respects the $\,\ZZ_{2}^{}$\, symmetry
except the mixing mass-term of $\,M_{12}^2\,$
which provides a soft breaking of $\,\ZZ_{2}^{}$.\,
This potential contains eight free parameters from the start,
including three mass parameters $(M_{11}^2,\,M_{22}^2,\,M_{12}^2)$
and five quartic couplings
$(\lambda_1^{},\,\lambda_2^{},\,\lambda_3^{},\,\lambda_4^{},\,\lambda_5^{})$.

The vacuum is determined by the potential minimum, with the
Higgs vacuum expectation values (VEVs),
$\,\langle \H_j^{} \rangle = (0,\, v_j^{}/\!\sqrt{2})$\,.\,
Both Higgs fields contribute to the electroweak symmetry breaking,
with their VEVs obeying the condition
$\,v=(v_1^2\!+\!v_2^2)^{1/2}_{}\simeq 246\,$GeV.
Defining $\,v_{1}^{} =v\cos\beta$\, and $\,v_{2}^{} =v\sin\beta$\,,\,
we see that the VEV ratio is described by the parameter
$\,\tan\!\beta = {v_2^{}}/{v_1^{}}\,$.
The two Higgs doublets contain eight real components in total,
\begin{eqnarray}
\H_{j}^{}  \,=\, \(\!\! \begin{array}{c} \pi_{j}^{+}
\\[1mm]
\frac{1}{\sqrt{2}\,}(v_{j}^{} \!+\! h_j^{0} \!+\!\tt{i}\pi_j^{0})
\end{array}  \!\!\)\! , ~~~~~~~(j=1,2)\,.
\end{eqnarray}
The Higgs VEVs satisfy the extremal conditions,
\beqs
\label{eq:Vmin}
\begin{eqnarray}
\frac {\partial V} {\partial v_1^{}} &\,=\,&
M_{11}^2 v_1^{}- M_{12}^2 v_2^{} + \fr{1}{2} \lambda_1^{} v_1^3
+ \fr{1}{2} (\lambda_{3}^{}\!+\! \lambda_{4}^{}\!+\!\lambda_{5}^{}) v_1^{} v_2^2 \,=\, 0 \,,
\\[2mm]
\frac {\partial V} {\partial v_2^{}} &\,=\,&
M_{22}^2 v_2^{} -  M_{12}^2 v_1^{} + \fr{1}{2}\lambda_2^{} v_2^3
+ \fr{1}{2}(\lambda_3^{}\!+\! \lambda_4^{}\!+\!\lambda_5^{})v_1^2v_2^{} \,=\, 0 \,.
\end{eqnarray}
\eeqs
These are equivalent to the Higgs tadpole conditions
$\left<h_1^0\right>=\left<h_2^0\right>=0$\,,\, and
determine the two VEVs $(v_1^{},\,v_2^{})$ or $(v,\,\tanb )$
as functions of the eight parameters in the Higgs potential \eqref{eq:V}.

With the three massless would-be Goldstone bosons $\,(\pi_j^\pm,\,\pi^0_j)$\,
eaten by the weak gauge bosons $(W^{\pm}\!,\,Z^0)$,
the physical spectrum consists of five states:
two CP-even neutral scalars $(h_1^0,\,h_2^0)$,
one pseudoscalar $A^0$, and a pair of charged scalars $H^{\pm}$.\,
The CP-even sector involves a generic mass-mixing between $(h_1^0,\,h_2^0)$,
and the mass eigenstates $(h^0,\,H^0)$ are given by the orthogonal rotation
with mixing angle $\,\alpha\,$,
\beqa
\label{eq:evenHiggs}
\left(\!\begin{array}{c}
h \\[1mm] H   \end{array} \!\right)
=
\left(\!\begin{array}{rr}
\cos\al~ & -\sin\al \\[1mm]
\sin\al~ & \cos\al
\end{array} \!\right)\!\!
\left(\!\begin{array}{c} h_2^{} \\[1mm]
h_1^{} \end{array} \!\right)\! .
\eeqa
Given the current LHC data, it is most natural to identify lighter Higgs state
$\,h^0\,$ as the observed Higgs boson of mass 125\,GeV
\cite{ATLAS2012}\cite{CMS2012}.
The heavier state $\,H^0\,$ is a brandnew Higgs boson beyond the SM,
and can have sizable di-Higgs decays $H\to hh$ for the mass-range $\,\MH>2\Mh$.\,
The current LHC measurements show that the Higgs boson $h^0$(125GeV) behaves
rather SM-like. The favored 2HDM parameter space is then pushed to the region
around alignment limit $\,\cosba \sim 0$\,.\,

As mentioned above, the Higgs potential \eqref{eq:V} contains eight parameters in total,
including three mass parameters and five dimensionless self-couplings.
We can reexpress the eight parameters in terms of four Higgs masses
$(\Mh,\MH,M_{H^\pm}^{},M_A^{})$, the combined VEV $\,v\!=\!\sqrt{v_1^2\!+\!v_2^2\,}$,\,
the VEV ratio $\tanb =v_2^{}/v_1^{}$,\,
the mixing angle $\,\alpha\,$,\, and the mixing mass parameter
$M_{12}^2$\,.\,
Imposing the experimental inputs
$v\simeq 246\,$GeV and $\,\Mh \simeq 125\,$GeV, we note that
the Higgs sector is described by six parameters in total: the VEV ratio $\tanb$,
the mixing angle $\alpha$, heavy Higgs masses $(\MH, \MA, M_{H^\pm}^{})$,\,
and the mass-mixing parameter $M_{12}^{}$\,.
Thus, we can express the five dimensionless Higgs couplings $\,\lambda_j^{}\,$ as follows,
\beqs
\label{eq:lambda-all}
\begin{eqnarray}
\label{eq:lambda}
\label{eq:lambda-1}
\lambda_1^{} &\,=\,&
\frac{1}{\,v^2\!\cos^2\!\beta\,}\(\sin^2\!\alpha\,M_h^2+\cos^2\!\alpha\,
 M_H^2-M_{12}^2\tan\!\beta\),
\\[1mm]
\label{eq:lambda-2}
\lambda_2^{} &=& \frac{1}{\,v^2\sin^2\!\beta\,} \left(\cos^2\!\alpha\,M_h^2
+\sin^2\!\alpha\, M_H^2-M_{12}^2\cot\!\beta\right),
\\[1.5mm]
\label{eq:lambda-3}
\lambda_3^{} &=&
\frac{\,\sin 2\alpha\!\left(M_H^2\!-\!M_h^2\right)\,}
     {v^2\sin 2\beta}+\frac{\,2 M_{H^\pm}^2\,}{v^2}-\frac{2M_{12}^2}{\,v^2\sin2\beta\,} \,,
\\[1mm]
\label{eq:lambda-4}
\lambda_4^{} &=& \frac{\,M_A^2\!-\!2 M_{H^\pm}^2\,}{v^2}+\frac{2M_{12}^2}{\,v^2\sin2\beta\,} \,,
\\[1mm]
\label{eq:lambda-5}
\lambda_5^{} &=& \frac{2M_{12}^2}{\,v^2\sin2\beta\,}-\frac{M_A^2}{v^2} \,,
\end{eqnarray}
\eeqs
which are consistent with \cite{lambda}.
Then, the masses $M_{11}^2$ and $M_{22}^2$ can be solved from Eqs.(2.3)
and (2.5), so they are not independent parameters.

For the later numerical analyses in this section and in Sec.\,\ref{sec:4},
we will consider the 2HDM parameter space in the following ranges,
\begin{eqnarray}
\label{eq:2HDM-scan}
&& \tan\beta \in [1, 10],~~
\cos(\beta\!-\!\alpha) \in [-0.5,\, 0.5],~~
\nn\\
&& M_{12}^2 \in [-10^5,\, 4\!\times\!10^5] ,~~
\MH \in [300, 800],~~
\\
&&
M_A^{} \in [M_H^{}\!-\!M_Z^{},\, 1000],~~
 M_{H^\pm}^{} \in [M_{H^\pm}^{\min},\, 1000],
\hspace*{16mm}
\nn
\end{eqnarray}
where $\,M_{H^\pm}^{\min}= \max (M_H^{}-M_W^{},\,\over{M}_{\pm}^{})$,\,
and all the mass parameters are in the unit of GeV.
Here we choose the value of $\over{M}_{\pm}^{}$ to be consistent with the bound
of the weak radiative $B$-meson decays\,\cite{Misiak:2017bgg}.
It was found that the $B$-decay constraint is quite weak for 2HDM-I, since it
only requires
$M_{H^\pm}^{}>450$\,GeV for $\tanb\geqq 1$ and quickly drops below the LEP bound
($\simeq 80$GeV) for $\tanb\gtrsim 2$ \cite{Misiak:2017bgg}.
For 2HDM-II, the analysis of $B$-decay measurement
gives a stronger limit $\,M_{H^\pm}^{}\gtrsim 580$\,GeV \cite{Misiak:2017bgg}.
Thus, taking these into account, we set $\over{M}_{\pm}^{}=500$GeV for 2HDM-I and
$\over{M}_{\pm}^{}=580$GeV for 2HDM-II.

\vspace*{1mm}

\begin{table}[t]
\centering
\vspace*{1.5mm}
\begin{tabular}{c||c|c|c}
\hline\hline
&&&\\[-3mm]
 Couplings~ &  $\xi_{H}^{u}$ (\,$\xi_{h}^{u}$\,) & $\xi_{H}^{d}$ (\,$\xi_{h}^{d}$\,)
 & $\xi_{H}^{\ell}$ (\,$\xi_{h}^{\ell}$\,)
\\[-4mm]
&&&
\\
\hline
&&& \\[-3.5mm]
 2HDM-I~\,
 & $\dis\frac{\sin\!{\alpha}}{\,\sin\!{\beta}\,}$
   $\dis\(\!\frac{\cos\!{\alpha}}{\,\sin\!{\beta}\,}\!\)$
 & $\dis\frac{\sin\!{\alpha}}{\,\sin\!{\beta}\,}$
   $\dis\(\!\frac{\cos\!{\alpha}}{\,\sin\!{\beta}\,}\!\)$
 & $\dis\frac{\sin\!{\alpha}}{\,\sin\!{\beta}\,}$
   $\dis\(\!\frac{\cos\!{\alpha}}{\,\sin\!{\beta}\,}\!\)$
\\[3.9mm]
 2HDM-II
 & $\dis\frac{\sin\!\alpha}{\,\sin\!\beta\,}$
   $\dis\(\!\frac{\cos\!{\alpha}}{\,\sin\!{\beta}\,}\!\)$
 & $\dis\frac{\cos\!{\alpha}}{\,\cos\!{\beta}\,}$
   $\dis\(\!\!-\frac{\sin\!{\alpha}}{\,\cos\!{\beta}\,}\!\)$
 & $\dis\frac{\cos\!{\alpha}}{\,\cos\!{\beta}\,}$
   $\dis\(\!\!-\frac{\sin\!{\alpha}}{\,\cos\!{\beta}\,}\!\)$
\\[3.5mm]
\hline\hline
\end{tabular}
\caption{Yukawa couplings $\,\xi_H^f\,$ ($\,\xi_h^f\,$)
between the heavy Higgs boson $H^0$ (light Higgs boson $h^0$) and
the SM fermions are shown for the 2HDM-I and 2HDM-II,
where a common factor $\,m_f/v\,$
(corresponding to the SM Higgs Yukawa coupling) is factorized out.}
%  \vspace*{3mm}
\label{tab:1}
\end{table}

The 2HDM type-I and type-II are defined according to their different assignments
for the Yukawa sector under $\,\ZZ_{2}^{}$\, symmetry.
In the 2HDM type-I, all the SM fermions are defined as
$\,\ZZZ\,$ even, thus only the Higgs doublet $\H_2^{}$ joins Yukawa
interactions and generates all the fermion masses. For the 2HDM type-II, all
the right-handed down-type fermions are assigned as $\ZZZ$ odd,
while all other fermions are $\ZZZ$ even. Thus, the 2HDM-II has
the Higgs doublets $\,\H_2^{}\,$ and $\,\H_1^{}\,$ couple
to the up-type and down-type fermions, respectively.
Under the $\mathbb{Z}_2^{}$ assignments,
the $H^0$ Yukawa couplings for 2HDM-I and 2HDM-II can be expressed in the form,
$\,G_{Hff}^{}=-\xi_{H}^{f}\fr{m_f^{}}{v}$,\,
$(f=u, d, \ell)$, where the dimensionless coefficients $\,\xi_{H}^{f}\,$
only depends on $\,\alpha\,$ and $\,\beta\,$,\,
as summarized in Table\,\ref{tab:1}.
For comparison, we also show the Yukawa couplings of the light Higgs boson $h^0$,
$\,G_{hff}^{}=-\xi_{h}^{f}\fr{m_f^{}}{v}$,\, in the parentheses of this table.
The trilinear gauge couplings of $H^0$ take the form
$\,G_{HVV}^{}= \cosba \,2M_V^2/v$\, with $\,V=W^\pm,\,Z^0$,\,
while the $h^0VV$ couplings are given by
$\,G_{hVV}^{}= \sin(\beta\!-\!\alpha) \,2M_V^2/v$\,.
%

%%%%%%%%%%%%%%%%%%%%%%%% Sec.2.2 %%%%%%%%%%%%%%%%%%%%%%%%%
\vspace*{2.5mm}
\subsection{\hspace*{-2mm}Constraints from Theory and Existing Experiments}
\vspace*{2mm}
\label{sec:2.2}

\begin{figure}[t]
%\vspace*{-8mm}
\begin{centering}
\includegraphics[height=6.8cm,width=7.7cm]{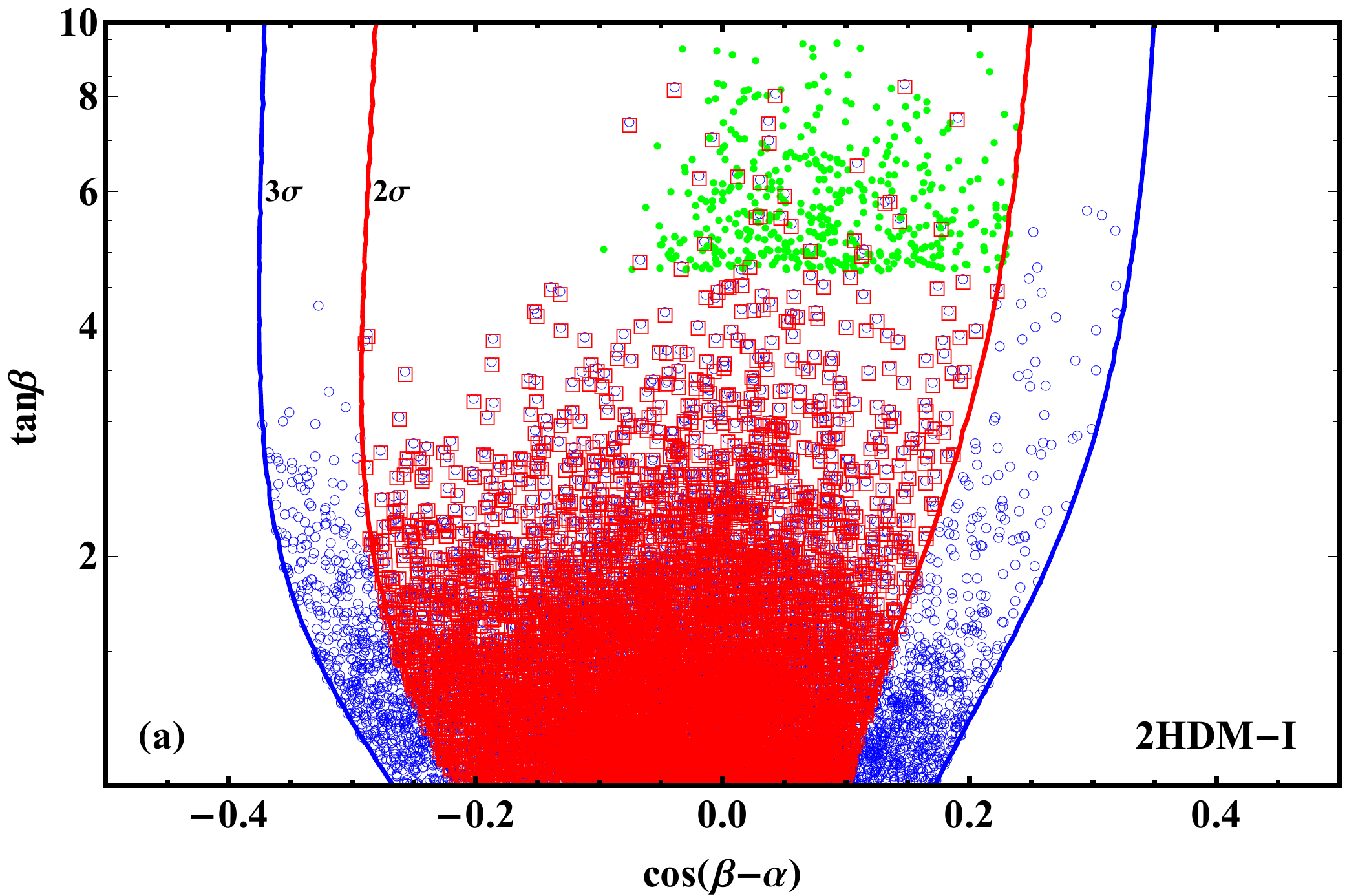}
\hspace*{-1mm}
\includegraphics[height=6.8cm,width=7.7cm]{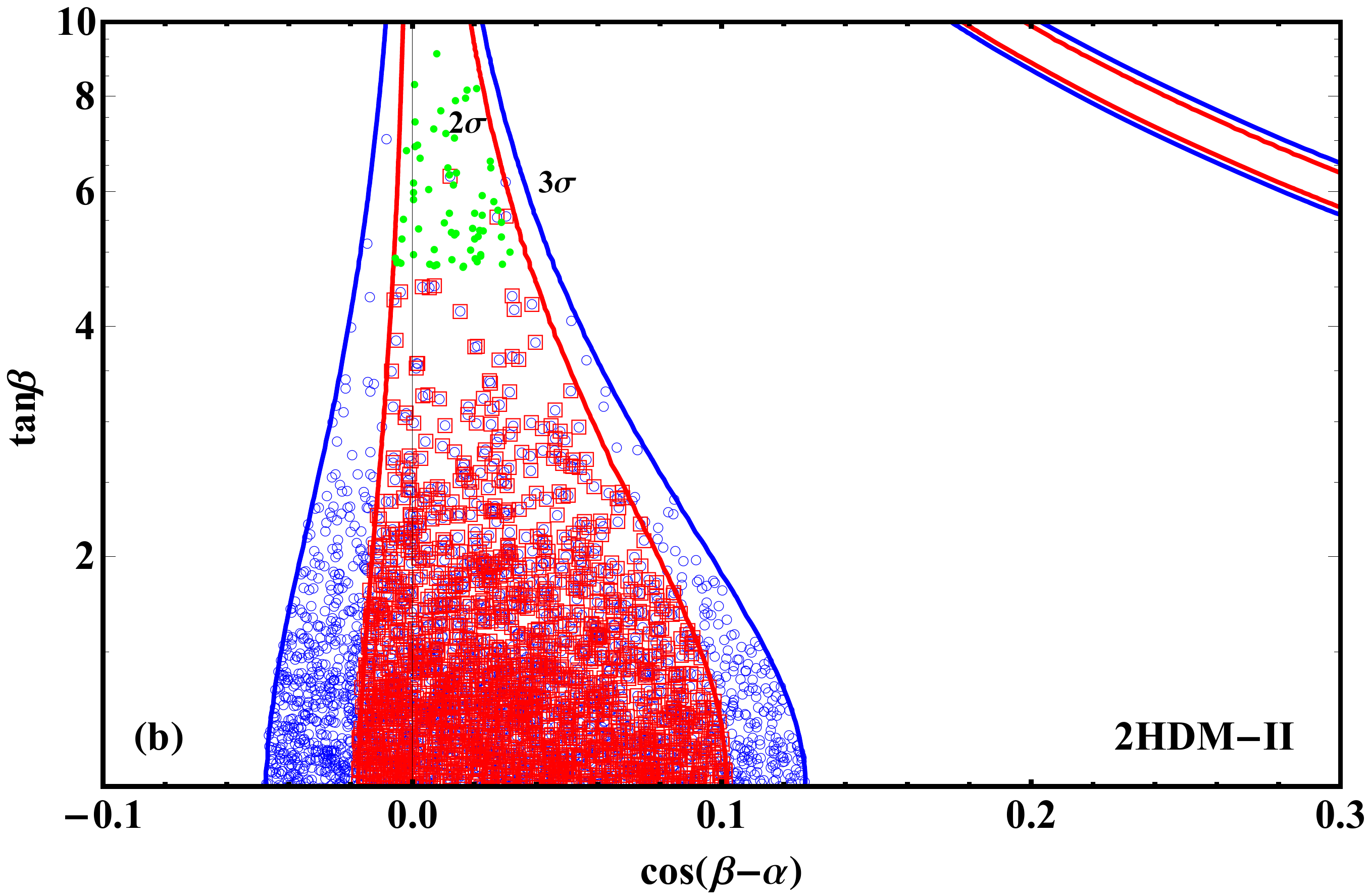}
\vspace*{-5mm}
\caption{Constraints on the 2HDM parameter space by
the global fit of the light Higgs boson $h(125\text{GeV})$.
Plot-(a) shows the limits for 2HDM-I and plot-(b) for 2HDM-II,
where the red (blue) curves give the $2\sigma$\,($3\sigma$) contours
(without theory and precision constraints).
The region with red (blue) dots depict the allowed parameter space
by including both the theory constraints and the Higgs global fit
at $2\sigma$\,($3\sigma$) level. The dotted regions present
a {\it uniform parameter scan} which favors relatively low $\tanb$\,.\,
The green dots are generated at $2\sigma$ level for the special parameter region
obeying $\,M_{12}^2=M_H^2\cos^2\!\alpha\cot\!\beta$\,
and $\tan\!\beta >5$\,. %(which has a very low likelihood).
}
\label{fig:ST}
\label{fig:22}
\end{centering}
\vspace*{-1mm}
\end{figure}

Requiring the Higgs potential \eqref{eq:V}
bounded from below, we have the stability conditions,
\begin{eqnarray}
\lambda_{1,2}^{}>0 \,, \quad~
\lambda_{3}^{}\!+\!\sqrt{\lambda_{1}^{}\lambda_{2}^{}}>0 \,, \quad~
\lambda_{3}^{}\!+\!\sqrt{\lambda_{1}^{}\lambda_{2}^{}}+\lambda_{4}^{}
>|\lambda_{5}^{}| \,.
\hspace*{10mm}
\end{eqnarray}
Furthermore, the high energy behaviors of scattering amplitudes involving
longitudinal weak gauge bosons should obey
the perturbative unitarity\,\cite{unitarity1}.
According to the equivalence theorem\,\cite{ET},
such scattering amplitudes are well approximated by
the corresponding Goldstone boson scattering amplitudes.
The $s$-wave unitarity condition $\,|\textrm{Re}(a_0^{})|<\fr{1}{2}\,$ imposes the
following constraints on the quartic Higgs couplings,
\beqs
\label{eq:UB}
\begin{eqnarray}
\label{eq:UB1}
\left| {{3}}(\lambda_{1}^{}\!+\!\lambda_{2}^{})
\pm\sqrt{{{9}}(\lambda_{1}^{}\!-\!\lambda_{2}^{})^2
+{{4}}(2\lambda_{3}^{}\!+\!\lambda_{4}^{})^2\,} \right|
&<& 16\pi\,,
\\
\label{eq:UB2}
\left|(\lambda_{1}^{}\!+\!\lambda_{2}^{})\pm\sqrt{(\lambda_{1}^{}\!-\!\lambda_{2}^{})^2
\!+\!4\lambda_{4,5}^2\,}\right| &<& 16\pi\,,
\\[0.6mm]
\label{eq:UB3}
\left|\lambda_{3}^{}\!+\!2\lambda_{4}^{}\!\pm\! 3\lambda_{5}^{} \right| \,<\, 8\pi,~~~~~
\left|\lambda_{3}^{}\!\pm\! \lambda_{4,5}^{}\right| &<& 8\pi\,.
\end{eqnarray}
\eeqs
From Eq.(\ref{eq:lambda}), we see that a large $\tan\!\beta$
can strongly enhance the coupling $\,\lambda_1^{}\,$ due to the factor
$\,1/\!\cos^2\!\beta\sim\tan^2\!\beta\,$.\,
Without fine-tuning the masses and mixing angles,
this can easily violate the perturbative unitarity bounds for large $\tan\!\beta$.\,
We will discuss this in more detail later when we show the results
of the parameter scan in Fig.\,\ref{fig:22}.

The existing electroweak precision data will constrain the one-loop contributions
induced by the Higgs-gauge couplings via oblique corrections \cite{STU}.\,
Around the alignment limit, we can expand the 2HDM contributions to the
oblique parameters\,\cite{STUHaber}\cite{stu1} as follows,
\beqs
\label{eq:STU}
\begin{eqnarray}
S &=& \frac{1}{\,\pi M_Z^2\,}\left[ \mathcal B_{22}(M_Z^2;M_{H}^2,M_{A}^2)\!-\!
\mathcal B_{22}(M_Z^2;M_{H^\pm}^2,M_{H^\pm}^2)\right]
\!+{O}\!\left(\cos^2(\beta\!-\!\alpha)\right)\!,
\\[1mm]
T &=& \frac{1}{\,16\pi M_W^2s_W^2\,}
\left[F(M_{H^\pm}^2,M_{A}^2)\!+\!F(M_{H^\pm}^2,M_{H}^2)\!-\!F(M_{A}^2,M_{H}^2)\right]
\!+{O}\!\left(\cos^2(\beta\!-\!\alpha)\right) \!,
\\[1.8mm]
U &=& -S+\frac{1}{\pi M_W^2}
\left[ \mathcal B_{22}^{}(M_W^2;M_{H^\pm}^2,M_{A}^2)
\!+\! \mathcal B_{22}^{}(M_W^2;M_{H}^2,M_{H^\pm}^2)
\!-\! 2\mathcal B_{22}^{}(M_W^2;M_{H^\pm}^2,M_{H^\pm}^2)\right]
\hspace*{8mm}
\nn\\
&& + \,{O}\!\left(\cos^2(\beta\!-\!\alpha)\right),
\end{eqnarray}
\eeqs
where $\,F(x,y)=\fr{1}{2}(x\!+\!y)\!-\!\fr{xy}{\,x-y\,}\!\ln\!\fr{x}{y}$\,,\,
and the function $\,\mathcal{B}_{22}^{}$\, is given by
\beqs
\begin{eqnarray}
\mathcal{B}_{22}^{}(z;x,y)&=&
\frac{z}{\,24\,} \!\left\{ \ln (xy)-
\!\!\left[\frac{3(x\!-\!y)}{z}\!-\!\frac{3(x^2\!-\!y^2)}{z^2}\!+\!\frac{(x\!-\!y)^3}{z^3}
  \right]\!\ln\!\frac{x}{y} - 6F\!\(\frac{x}{z},\frac{y}{z}\)\hspace*{15mm}
  \right.
\nn\\[1.5mm]
&& -\!\left[\frac{10}{3}\!-\!\frac{8(x\!+\!y)}{z} \!+\! \frac{2(x\!-\!y)^2}{z^2}\right]
\!-\!\left[\frac{(x\!-\!y)^2}{z^2} \!-\! \frac{2(x\!+\!y)}{z} \!+\! 1
\right]\!f\!\(\frac{x}{z},\frac{y}{z}\)\!,~~~~~~
%\\
%F(x,y) &=& \frac{\,x\!+\!y\,}{2}-\frac{xy}{\,x\!-\!y\,}\ln \frac x y \,,
%
\\[1.5mm]
f(x,y) &=&
\left\{
\begin{array}{ll}
-2 \sqrt{\!\Delta}\!
\dis\left(\!\arctan\!\frac{\,x\!-\!y\!+\!1\,}{\sqrt{\!\Delta}}
-\arctan\!\frac{\,x\!-\!y\!-\!1\,}{\sqrt{\!\Delta}} \!\right)\!,
\hspace*{8mm}
&  (\Delta >0)\,,
\\[2mm]
0 & (\Delta =0)\,,
\\[2mm]
\sqrt{\!-\Delta} \dis\ln\!
\frac{\,x\!+\!y\!-\!1\!+\!\sqrt{\!-\Delta}\,}
     {\,x\!+\!y\!-\!1\!-\!\sqrt{\!-\Delta}\,} \,,
& (\Delta <0)\,,
\\
\end{array}
\right.
\end{eqnarray}
\eeqs
where $\,\Delta(x,y) =2(x\!+\!y)\!-\!(x\!-\!y)^2\!-\!1$\,.\,
The leading order contributions to the oblique corrections \eqref{eq:STU}
only involve the masses of new Higgs bosons $(H^0,\,A^0,\,H^\pm)$.\,
This is because the couplings of trilinear vertices involving two new Higgs bosons
($\,W^\pm$-$H^\mp$-$H^0$,\,
$W^\pm$-$H^\mp$-$A^0$,\, $Z^0$-$H^0$-$A^0$,\, and $\,Z^0$-$H^+$-$H^-$)
either contain the factor $\,\sin(\beta\!-\!\alpha)$\,
or have no $(\alpha,\beta)$-dependence,
while the cubic vertices with only one new Higgs boson
($\,W^\pm$-$H^\mp$-$h^0$, $\,Z^0$-$A^0$-$h^0$, and $\,H^0$-$V$-$V\,$)
are suppressed by $\,\cos(\beta\!-\!\alpha)$\,.
Besides, the other cubic vertices $\,h^0$-$V$-$V\,$ ($V=W^\pm\!,Z^0$)
have couplings proportional to $\,\sin(\beta\!-\!\alpha)$, and lead to
the suppression factor $\,\sin^2(\beta\!-\!\alpha)-1=-\cos^2(\beta\!-\!\alpha)$\,
after subtracting the corresponding SM contributions.

\begin{figure}[t]
\begin{centering}
\includegraphics[height=6cm,width=7.7cm]{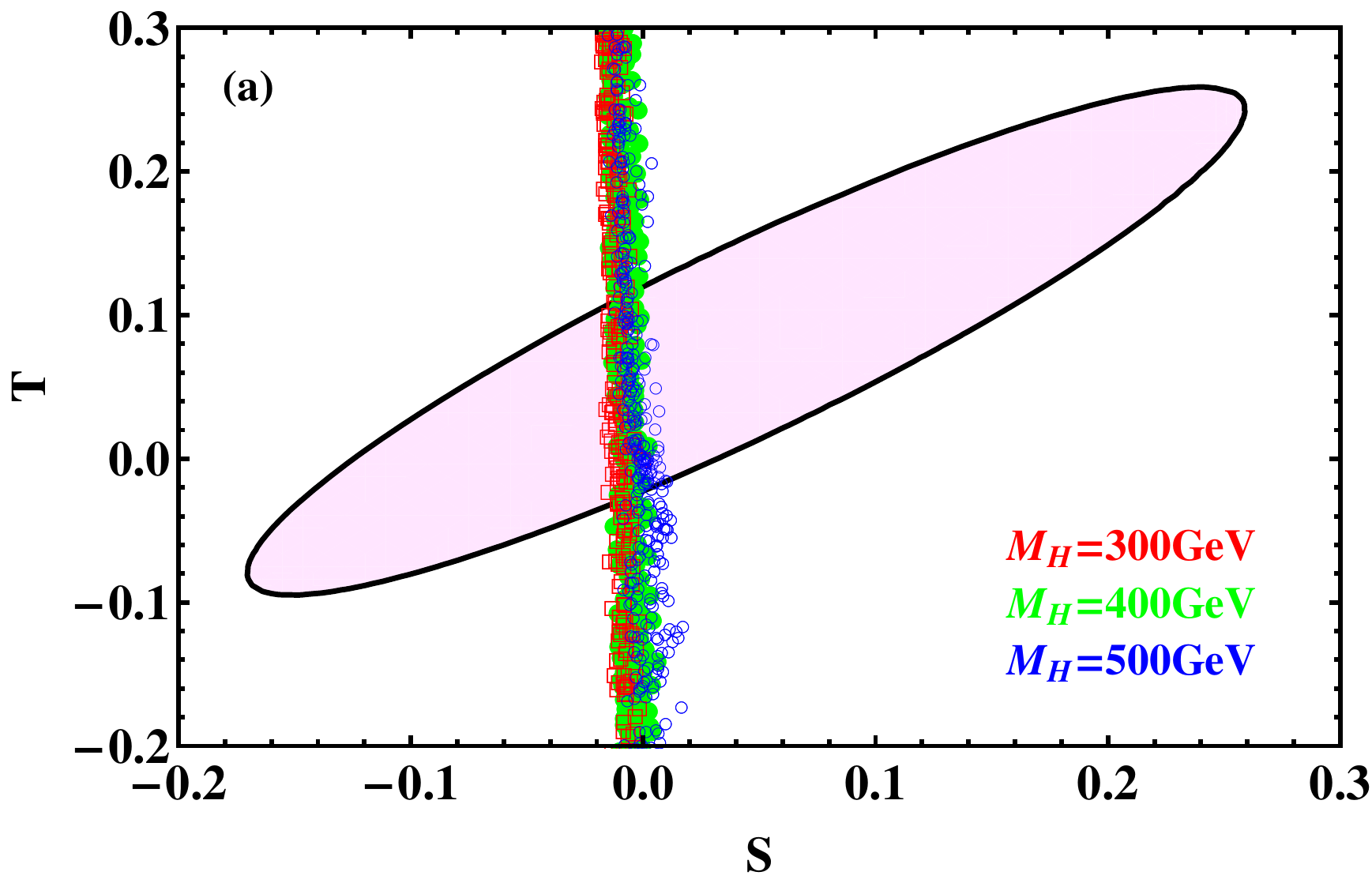}
\hspace*{-2mm}
\includegraphics[height=6cm,width=7.7cm]{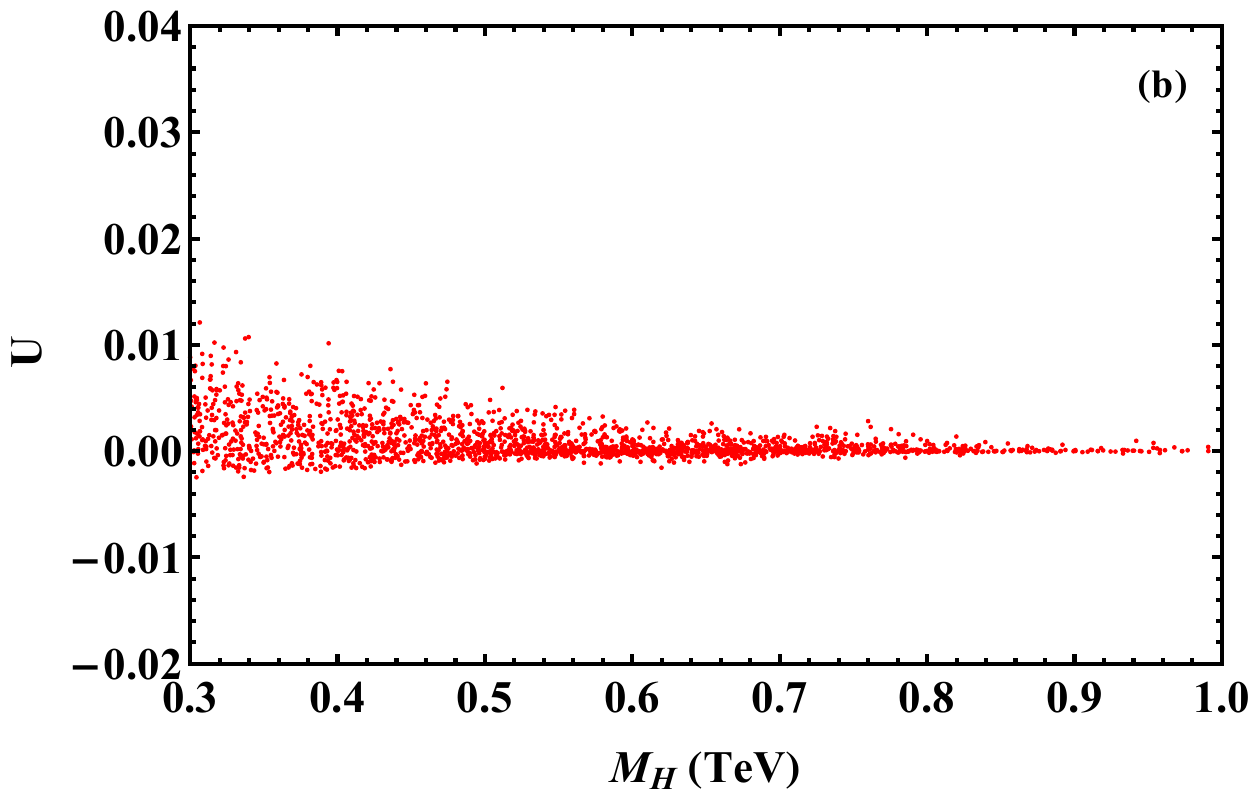}
\vspace*{-3mm}
\caption{Electroweak precision constraints on the 2HDM.
Plot-(a) shows the $S-T$ contour at 95\%\,C.L.\
(by setting $U=0$), in comparison with the 2HDM predictions
for $\,\MH =(300,400,500)$GeV which correspond to (red,\,green,\,blue)
dotted regions, respectively.
Plot-(b) presents the 2HDM prediction for $U$ parameter
over the mass-range $\,\MH=(0.3\!-\!1$)\,TeV.
}
\label{fig:ST}
\label{fig:1}
\end{centering}
\end{figure}

In Fig.\,\ref{fig:1}(a)
we present the 2HDM predictions of $\,(S,\,T)$\, by scanning the 2HDM parameter space,
where the (red, green, blue) dotted regions correspond to $H^0$ mass $\,\MH=(300,400,500)$GeV,
respectively. As a comparison, we also show the 95\%\,C.L.\ contour
from the precision constraints\,\cite{stu2} in the same plot (with $U=0$).
In Fig.\,\ref{fig:1}(b), we present the $\,U$ parameter prediction of the 2HDM
over the mass-range $\,\MH=(0.3-0.8$)TeV.\,
We find that in the 2HDM,  the oblique contribution to $\,T\,$
is much larger than $\,S\,$,\, and the $\,S\,$ and $\,U$ parameters are fairly small
in the relevant parameter region,
namely, $\,|T|\gg |S| \sim |U|< 2\!\times\!10^{-2}\,$.\,
Hence, Fig.\,\ref{fig:1} shows that the nontrivial constraint mainly comes from the $T$ parameter.
Since the current electroweak precision data constrain the $T$ parameter
to be quite small, especially for small $|S|\lesssim  2\!\times\!10^{-2}$
as restricted by the ellipse contour in Fig.\,\ref{fig:1}(a),
this requires the mass of $H^\pm$ to be fairly degenerate with that of $H^0$ or $A^0$.
(For the numerical analyses in Fig.\,\ref{fig:1}, we have used the
exact one-loop formulas for the oblique parameters\,\cite{stu1}.
We also compared our results with Ref.\,\cite{stu1} and the 2HDMC code\,\cite{2HDMC}
for consistency checks.)

\vspace*{1mm}

Next, we further derive the existing constraints on the 2HDM parameter space
by making a global fit for the LHC Run-1 and Run-2 measurements on the light Higgs boson
$h^0$(125GeV).\,  Since this Higgs state $h^0$(125GeV) is fairly SM-like,
the new physics corrections to $h^0$ couplings are tightly constrained
and other new Higgs states need to be significantly heavier.
Hence, we may regard the $h^0$ global fit bounds as indirect constraints
on the new Higgs states, in contrast to the constraints from their direct searches.

In our analysis, we perform the Higgs fit by minimizing $\chi^2$ with the inputs of signal strengths
$\mu_i^{}$ from the LHC experimental fits (including their errors and correlations).
For the Run-1 Higgs data of the LHC\,(7+8TeV), we have used the
combined analysis of ATLAS and CMS\,\cite{Run1-sum-h}.
For the current Higgs data from the Run-2 of the LHC\,(13TeV),
we include the ATLAS measurements on $\,h\!\to\!\gaga \,$ \cite{Atlas2-gaga-ZZ}\cite{tthATLAS},
$\,h\!\to\! ZZ^*\!\to\! 4\ell\,$ \cite{Atlas2-gaga-ZZ},
$\,h\!\to\! WW^*\!\to\!\ell\nu\ell\nu\,$ \cite{Atlas2-WW},
and $\,h\!\to\! b\bar{b}\,$ \cite{tthATLAS}\cite{tthATLASbb}\cite{Atlas2-bb} channels,
and the CMS measurements on $\,h\!\to\!\gaga\,$ \cite{CMS2-gaga}\cite{CMS2x-gaga},
$\,h\!\to\! ZZ^*\!\to\!4\ell\,$ \cite{CMS2-ZZ},
$\,h\!\to\! WW^*\!\to\!\ell\nu\ell\nu\,$ \cite{CMS2-WW},
$\,h\!\to\! \tau\tau\,$ \cite{CMS2-tautau},
and $\,h\!\to\! b\bar{b}\,$ \cite{tthbb}\cite{CMS2-bb} channels.
We make use of these Run-1 and Run-2 Higgs data from
ATLAS and CMS collaborations, and perform a global fit to derive the current
LHC constraints on the 2HDM parameter space.
We present the $2\sigma$ contours (red curves)
and $3\sigma$ contours (blue curves) in the
$\,\cos(\beta\!-\!\alpha)-\tanb\,$ plane for 2HDM-I in Fig.\,\ref{fig:22}(a) and
for 2HDM-II in Fig.\,\ref{fig:22}(b).
We further incorporate the theoretical requirements from the Higgs stability and
perturbative unitarity, and present the combined constraints on the
allowed parameter regions marked by red dots (in box shape) at the $2\sigma$ level
and by blue dots (in circle shape) at the $3\sigma$ level.

\vspace*{1mm}

In Fig.\,\ref{fig:22},
we are making a {\it uniform parameter scan} according to Eq.\eqref{eq:2HDM-scan},
which shows that the likelihood for the red and blue dots becomes smaller
when $\tan\!\beta$ increases.\footnote{We thank Yun Jiang
for discussing the 2HDM parameter scan in Ref.\,\cite{Bernon:2015qea}
and for comparing with their analysis \cite{Bernon:2015qea}.}\,
We see that without special fine-tuning of the parameter space,
the current fit favors the relatively low $\tan\!\beta$ region,
i.e., $\,\tan\!\beta\lesssim 9\,$
for 2HDM-I and $\,\tan\!\beta \lesssim 7\,$ for 2HDM-II.
This is largely due to the unitarity bounds
\eqref{eq:UB1}-\eqref{eq:UB2} in which the Higgs coupling $\lambda_1^{}$
is enhanced by $\,1/\!\cos^2\!\beta\sim\tan^2\!\beta$
[cf.\ Eq.\eqref{eq:lambda-1}].
The other Higgs couplings \eqref{eq:lambda-3}-\eqref{eq:lambda-5}
are also enhanced by the factor $\,\sin\!2\alpha/\!\sin\!2\beta\propto\tan\!\beta\,$ and
$\,2M_{12}^2/\!\sin\!2\beta\sim M_{12}^2\tan\!\beta\,$ for large $\tan\!\beta$.\,
If we select a special parameter space obeying the condition
\beqa
\label{eq:cond-M12}
M_{12}^2=M_H^2\cos^2\!\alpha\cot\!\beta\,,\,
\eeqa
we find that Eq.\eqref{eq:lambda-1} reduces to
%
%\beqa
$\lambda_1^{} =
({\sin^2\!\alpha}/{\cos^2\!\beta})
({M_h^2}/{v^2})$,\, %\,\sim\, {M_h^2}/{v^2}$,\,
%\eeqa
%
and the enhancement factor
$\,1/\!\cos^2\!\beta\sim\tan^2\!\beta\,$ can be removed
by taking the alignment limit
$\,\cos(\beta-\alpha)\!\to\! 0$\,.
The alignment limit gives
$\,\alpha = \beta -\frac{\pi}{2}\,$,\,
under the convenition\,\cite{Haber2015},
$\,\beta\in [0,\,\frac{\pi}{2}]\,$ and $\,\beta-\alpha\in [0,\,\pi]\,$.\,
In this limit, the above condition \eqref{eq:cond-M12} becomes
$\,M_{12}^2=\frac{1}{2}\sin 2\beta\,M_H^2\,$,\,
and the Higgs couplings \eqref{eq:lambda-1}-\eqref{eq:lambda-5}
reduces to
\beqs
\label{eq:lambda-x}
\beqa
&&\lambda_1^{}=\lambda_2^{} = \frac{\,M_h^2\,}{v^2} \,,
\\[1.5mm]
&& \lambda_3^{}= \frac{\,M_h^2+2(M_{H^\pm}^2\!\!-\!M_H^2)\,}{v^2} \,,
\\[1.5mm]
&& \lambda_4^{}= \frac{\,M_A^2\!+\!M_H^2\!-\!2M_{H^\pm}^2\,}{v^2} \,,
\\[1.5mm]
&& \lambda_5^{}= \frac{\,M_H^2\!-\!M_A^2\,}{v^2} \,.
\eeqa
\eeqs
We see that the resultant Higgs couplings above
do not increase with $\tan\!\beta$ and could more easily satisfy the
unitarity bounds \eqref{eq:UB} only if the squared masses of the heavy Higgs bosons
are all nearly degenerate $\,M_H^2\sim M_A^2\sim M_{H^\pm}^2\,$.\,
For illustration, we make a separate scan on the special parameter region
under the condition \eqref{eq:cond-M12}
[in addition to the default condition \eqref{eq:2HDM-scan}]
and for $\,\tan\!\beta>5\,$.\,
This is represented by the green dots in Fig.\,\ref{fig:22}
and indeed reaches larger $\tan\!\beta$ regions for both 2HDM-I and 2HDM-II,
as expected. But we have to keep in mind that the condition \eqref{eq:cond-M12}
only represents a very small region of the generic parameter space defined
in Eq.\eqref{eq:2HDM-scan}, so it has low likelihood and is hard to be reached
by the conventional {uniform} parameter scan.
For consistency check, we have also made comparison with
Ref.\,\cite{Bernon:2015qea} for the global Higgs fit.

\vspace*{1mm}

Fig.\,\ref{fig:22} shows that the current LHC global fit of $h$(125GeV)
shifts the viable parameter space somewhat towards
$\,\cos(\beta\!-\!\alpha)<0\,$ region for 2HDM-I, while it pushes the allowed
parameter range significantly to $\,\cosba >0\,$ side for 2HDM-II.
The reasons are the following.
We note that the current LHC data give tight constraints on the signal strengths
$\mu(gg$F+$t\bar{t}h)$ and $\mu$(VBF+$Vh$).
From the combined Run-1 data\,\cite{Run1-sum-h}, we see that
$h^0\!\to\!WW^*$ channel favors $\mu$(VBF+$Vh)>1$ and
$h^0\!\to\!b\bar{b}$ channel favors $\mu$(VBF+$Vh)<1$,\,
which can be both explained by a reduced $hb\bar{b}$ coupling (relative to its SM value),
since the reduced $hb\bar{b}$ coupling can decrease Br$(h^0\!\!\to\!b\bar{b})$
and enhance branching ratios of all other final states.
Also, $h^0\!\to\!ZZ$ channel prefers $\,\mu(gg$F+$t\bar{t}h)>1\,$ and thus an
enhanced $ht\bar{t}$ coupling.
For the LHC Run-2, we note that ATLAS data\,\cite{Atlas2-gaga-ZZ} significantly favor
$\mu$(VBF)$>1$ in both $h^0\!\!\to\!\gaga,ZZ^*$ channels,
while their combination has little effect on $\,\mu(gg$F).
The $Vh$ production at ATLAS Run-2 also prefers $\,\mu(Vh)<1\,$
via $h^0\!\!\to\!b\bar{b}$ channel \cite{Atlas2-bb}.
Thus, these features can be explained by a reduced $hb\bar{b}$ coupling.
Besides, the CMS Run-2 data\,\cite{CMS2-ZZ} mildly favor $\,\mu$(VBF+$Vh)<1$\,
via $h^0\!\!\to\!ZZ^*$ and thus an enhanced $hb\bar{b}$ coupling,
while $h^0\!\!\to\!\ga\ga$ channel at the CMS Run-2
slightly prefers $\,\mu(gg$F+$t\bar{t}h)>1\,$
\cite{CMS2-gaga} and so an enhanced $ht\bar{t}$ coupling.
Finally, inspecting the $hb\bar{b}$ and $ht\bar{t}$ couplings in Table\,\ref{tab:1}
and expanding them around the alignment limit
(with $\,\beta-\alpha \equiv \frac{\pi}{2}-\delta\,$
and $\delta$ as a small deviation),
we find that up to the first order of $\delta$,\,
the $hb\bar{b}$ coupling equals
$\,1\!+\!\delta\!\cot\!\beta\,$ in 2HDM-I and
$\,1\!-\delta\tan\!\beta\,$ in 2HDM-II,
while the $ht\bar{t}$ coupling equals $\,1\!+\!\delta\!\cot\!\beta\,$ in both 2HDM-I,II.
Hence, a reduced $hb\bar{b}$ coupling requires
$\,\delta<0\,$ and thus $\,\cos(\beta\!-\!\alpha)<0\,$ in 2HDM-I,
pushing the viable parameter space towards the left-hand-side
in Fig.\,\ref{fig:22}(a);
while for 2HDM-II, this requires $\,\delta>0\,$ and thus
$\,\cos(\beta\!-\!\alpha)>0\,$,\,  shifting the parameter space
towards the right-hand-side in Fig.\,\ref{fig:22}(b).
Also, a mildly enhanced $ht\bar{t}$ coupling requires $\,\delta>0\,$
for both 2HDM-I,II, so its effect will partially cancel that of
the $ht\bar{t}$ coupling for 2HDM-I, and add together with the
$ht\bar{t}$ effect for 2HDM-II.
This explains that the 2HDM-II has larger asymmetry over
$\,\cos(\beta\!-\!\alpha)\,$ [Fig.\,\ref{fig:22}(b)] than that
of the 2HDM-I [Fig.\,\ref{fig:22}(a)].

%-----------------------------------------
%
\begin{table}[t]
	\centering
	%\vspace*{3mm}
	\begin{tabular}{l||l|c|c|c|c|c}
		\hline\hline
		&&&&&&\\[-2.5mm]
	~~Experiments &  ~~~$M_H^{}$\,(GeV) & ~300~ & ~400~ & ~500~ & ~600~ & ~800~
		\\[-3mm]
		&&&&&&
		\\
		\hline
		&&&&&& \\[-3.5mm]
		ATLAS Run-2~\, & $H\!\!\to\! ZZ(4\ell )$
		& 0.18
		& 0.22
		& 0.11
		& 0.12
		& 0.056
		\\[1mm]
		 & $H\!\!\to\! ZZ(\ell\ell\nu\nu)$
		 & 2.0
		 & 0.36
		 & 0.22
		 & 0.82
		 & 0.28
		\\[1mm]
		  & $H\!\!\to\! WW$
		  & 4.1
		  & 1.3
		  & 0.8
		  & 0.49
		  & 0.32
		\\[1mm]
		  & $H\!\!\to\! \tau\tau$
		  & 0.55
		  & 0.27
		  & 0.14
		  & 0.052
		  & 0.024
		\\[1mm]
				\hline
		CMS Run-2~\, & $H\!\!\to\! hh(bb\tau\tau)$
		& 4.4
		& 1.2
		& 0.31
		& 0.43
		& 0.23
		\\[1mm]
		& $H\!\!\to\! \tau\tau$
		& 0.25
		& 0.13
		& 0.079
		& 0.081
		& 0.043		
		\\[1mm]	
		\hline
		ATLAS Run-1~\, & $H\!\!\to\! ZZ$
		& 0.26
		& 0.066
		& 0.043
		& 0.021
		& 0.012
		\\[1mm]
		& $H\!\!\to\! hh$
		& 2
		& 0.83
		& 0.18
		& 0.08
		& 0.05
		\\[1mm]
		\hline
		CMS Run-1~\, & $H\!\!\to\! VV$
		& 0.3
		& 0.26
		& 0.15
		& 0.1
		& 0.07
		\\[1mm]
		& $H\!\!\to\! hh(bb\gamma\gamma)$
		& 1.1
		& 1.1
		& 0.45
		& 0.37
		& 0.12
		\\[1mm]
		\hline\hline
	\end{tabular}
\caption{Current upper limits (95\%\,C.L.) from the LHC Run-1 and Run-2
         direct searches on the production cross sections (in pb) of the heavier Higgs boson
         $H^0$ in various decay channels, where the decay branching fractions of the
         final states $VV$ or $hh$ are not included.}
%\vspace*{-1mm}
\label{tab:direct}
\label{tab:22}
\end{table}
%

%\vspace*{3mm}
\begin{figure}[t]
\hspace*{-4mm}
\begin{centering}
\includegraphics[height=6cm,width=7.9cm]{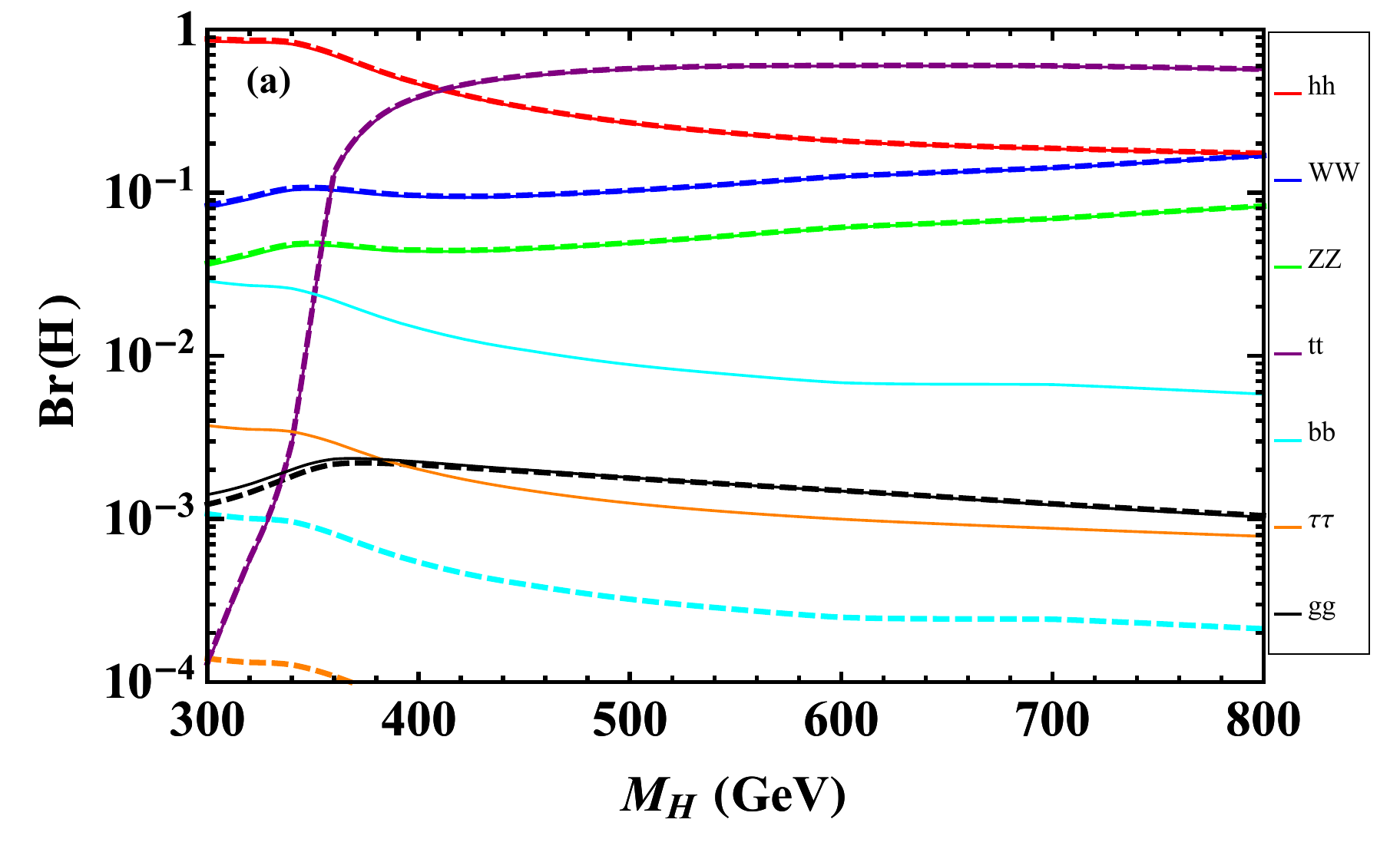}
\hspace*{-4.5mm}
\includegraphics[height=6cm,width=7.9cm]{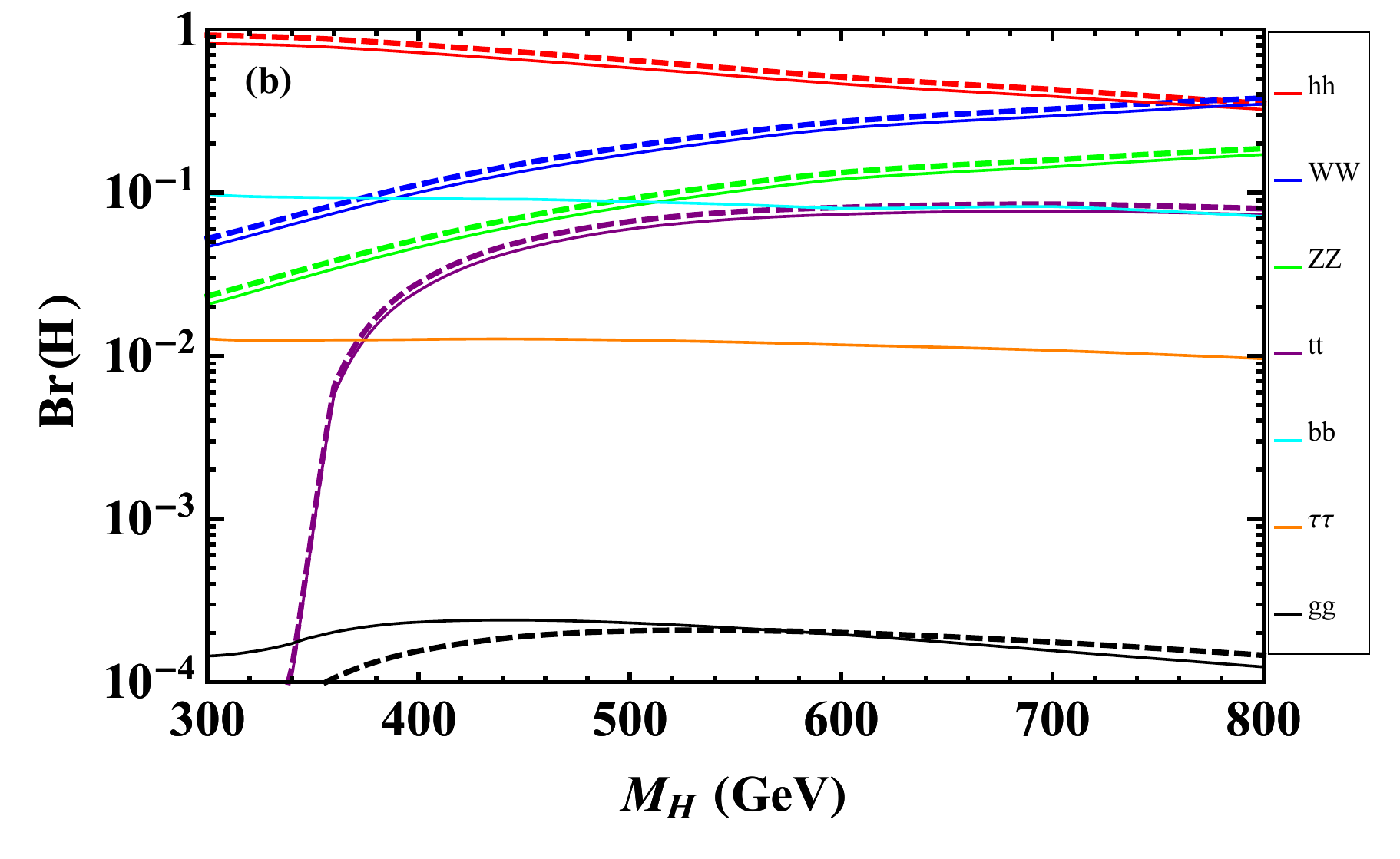}
%[height=6.9cm,width=8.9cm]{figs/figbr.pdf}
\vspace*{-5mm}
\caption{Decay branching fractions of the heavy Higgs boson $H^0$ as functions
of its mass $M_H^{}$ for $\tan\beta = 2$\, [plot-(a)] and
$\tan\beta = 5$\, [plot-(b)].\,
where we set $(M_A^{},\, M^2_{12}) = (800\textrm{GeV}, -(200\textrm{GeV})^2)$\,
and $\,\cos(\beta-\alpha) = 0.1$\, for illustration.
The dashed (solid) curves represent the results of 2HDM-I (2HDM-II) in each plot.
}
\label{fig:Br}
\label{fig:3new}
\label{fig:n3}
\end{centering}
\vspace*{-1mm}
\end{figure}

\vspace*{1.5mm}

Finally, for this study,
we will consider the upper bounds from the existing LHC Run-1 and Run-2 searches
on a heavier neutral Higgs state $H^0$ with decays in various channels.
These will put additional constraints on the 2HDM parameter space through
various $H^0$ couplings.
The LHC Run-1 searches include
$\,H^0\!\to\! hh \!\to\! bb\gamma\gamma$\,
from CMS\,\cite{CMSHhh}, the combined searches of
$\,H^0\!\to\! hh \!\to\! bb\gamma\gamma$,
$bbbb$, $bb\tau\tau$, $\gamma\gamma WW^*$ from ATLAS\,\cite{Aad:2015xja},
$\,H^0\!\to\! ZZ$\, from ATLAS\,\cite{ATLASHVV}, and the combined searches of
$\,H^0\!\to\! W^+W^-/ZZ$\, from CMS\,\cite{CMSHVV}.		
The LHC Run-2 searches include
% $\,H^0\!\to\! hh \!\to\! bbbb$\, from ATLAS\,\cite{ATLAS:2016ixk},
$\,H^0\!\to\!WW$\, from ATLAS\,\cite{HWWlvlvATLAS13},
$\,H^0\!\to\! ZZ\,$ from ATLAS\,\cite{HZZ4lATLAS13}\cite{HZZ2l2vATLAS13},
$\,H^0\!\to\!\tau\tau$\, from ATLAS\,\cite{tataATLAS}
and from CMS\,\cite{tataCMS},
and $\,H^0\!\to\! hh \!\to\! bb\tau\tau$\, from CMS\,\cite{CMS:2017orf}.
With these, we summarize in Table\,\ref{tab:22}
the current upper limits (95\%\,C.L.) from the LHC Run-1 and Run-2
direct searches on the production cross sections of the heavier Higgs boson
$H^0$ in various decay channels, where the numbers do not contain
the decay branching fractions of the final states $VV$, $hh$, and
$\tau\tau$\,.\,

\vspace*{1mm}

For the heavy Higgs state $H^0$,\, we analyze its decay branching fractions
in Fig.\,\ref{fig:Br}.
We present the branching fractions over the mass range $M_H^{}=(300-800)$GeV, for
$\,\tan\!\beta = 2$\, [plot-(a)] and
$\,\tan\!\beta = 5$\, [plot-(b)].
where we input $(M_A^{},\, M^2_{12}) = (800\textrm{GeV}, -(200\textrm{GeV})^2)$\,
and $\,\cos(\beta-\alpha) = 0.1$\, for illustration.
The solid curves represent the branching fractions of 2HDM-I,
and the dashed curves stand for 2HDM-II.

\vspace*{1mm}

\begin{figure}[t]
%\vspace*{-3mm}
\begin{centering}
\includegraphics[height=6cm,width=7.7cm]{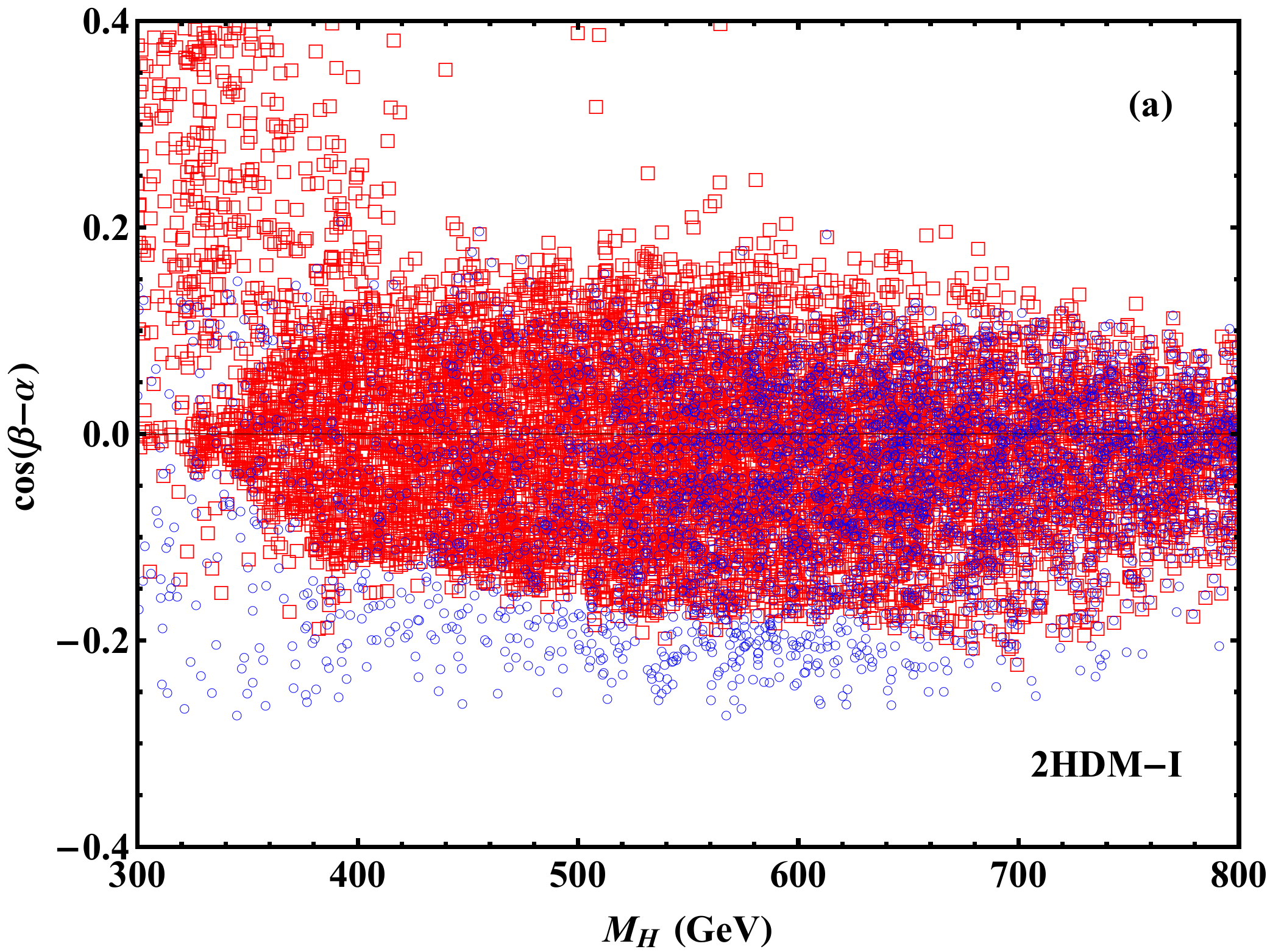}
\hspace*{-2mm}
\includegraphics[height=6cm,width=7.7cm]{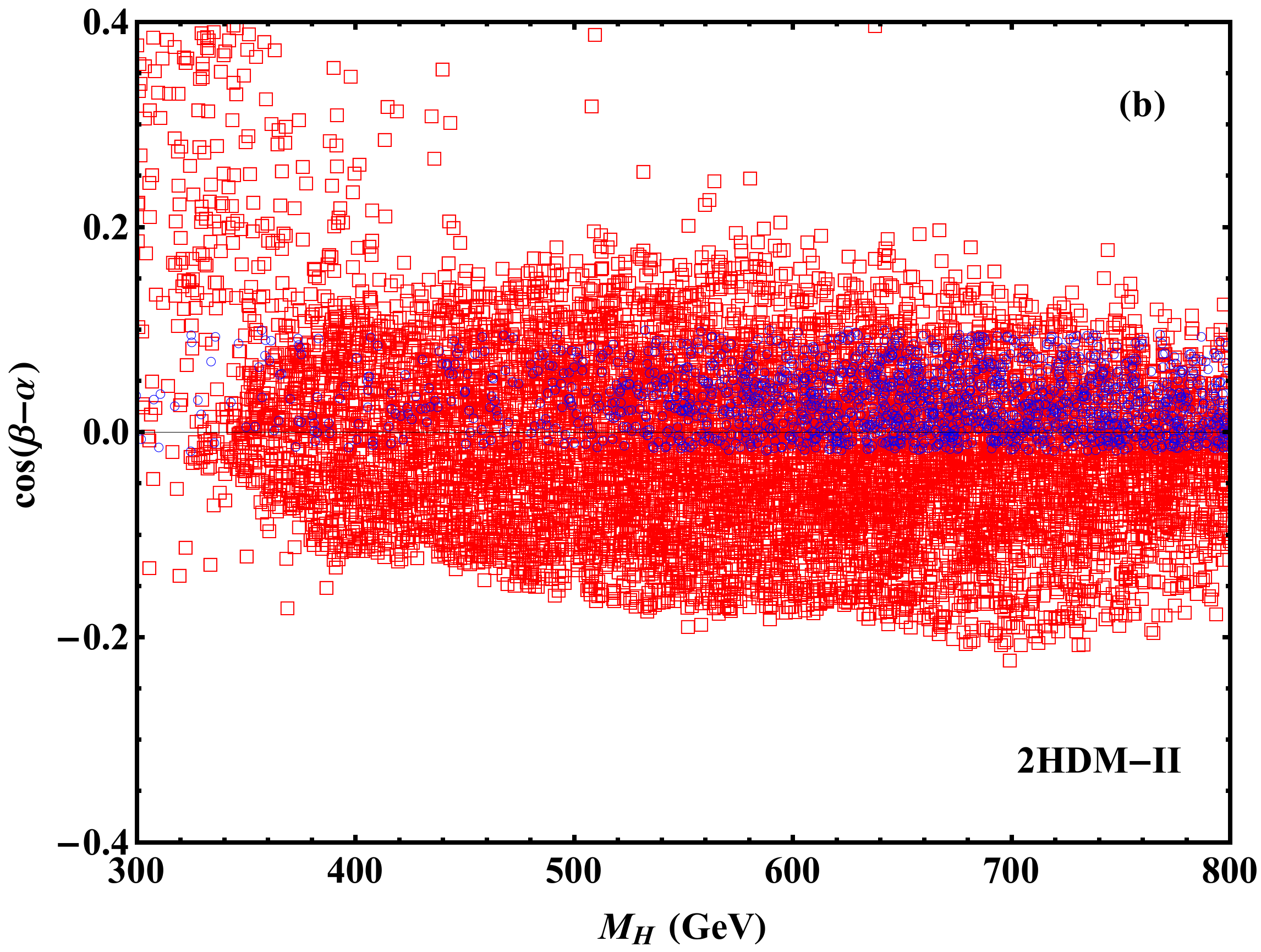}
\vspace*{-2mm}
\caption{Current experimental constraints on the 2HDM parameter space in
the $\,\MH \!- \cos(\beta\!-\!\alpha)$\, plane
for 2HDM-I in plot-(a) and for 2HDM-II in plot-(b).
The blue points (circle shape) satisfy the theoretical requirements,
the electroweak precision limits ($2\sigma$), and
the LHC bounds ($2\sigma$) by the Higgs global fit of $h$(125GeV) data.
The red points (square shape) present
the existing LHC direct search limits ($2\sigma$) on the heavier Higgs boson $H^0$
(combined with the theoretical requirements).
}
\label{fig:cba}
\label{fig:2}
\label{fig:33}
\label{fig:n4}
\end{centering}
%\vspace*{-3mm}
\end{figure}

Then, in Fig.\,\ref{fig:33}, we present the current experimental constraints
on the 2HDM parameter space in the plane of $\,\MH-\cos(\beta\!-\!\alpha)$.\,
The parameter region with blue dots (circle shape) satisfy the theoretical conditions,
the electroweak precision limits ($2\sigma$), and
the LHC bounds ($2\sigma$) from the Higgs global fit of $h$(125GeV) data.
The red dots (square shape) present the parameter region obeying
the existing LHC direct search limits ($2\sigma$) on the heavier Higgs boson $H^0$
in combination with the theoretical constraints.
The electroweak precision tests mainly bound the oblique parameter $\,T\,$
as shown in Fig.\,\ref{fig:1}, and prefer
the masses $M_{H^\pm}^{}$ and $M_A^{}$ to be fairly degenerate for
$\,\MH\lesssim 500\,$GeV.\,
The present LHC global Higgs fit prefers $h$(125GeV) to be quite SM-like,
and favors the 2HDM parameter space around the alignment limit
(cf.\ Fig.\,\ref{fig:22}).
Fig.\,\ref{fig:33} shows that
the allowed region of $\,\cos(\beta\!-\!\alpha)$\, (with blue dots)
in 2HDM-I is more shifted to $\,\cosba <0\,$ as in plot-(a),
while the region with blue dots in 2HDM-II is largely excluded on
the $\,\cosba <0\,$ side, as in plot-(b).
These features are consistent with Fig.\,\ref{fig:22}.
The current LHC direct search limits on the heavier Higgs state $H^0$ are reflected
by the red dotted regions in Fig.\,\ref{fig:33}.
They are comparable to the bounds imposed by the LHC $h$(125GeV) global fit (combined
with the electroweak precision limits) for 2HDM-I,
but they are significantly weaker for the case of 2HDM-II.

%

%%%%%%%%%%%%%%%%%%%%%%%% Sec.2.3 %%%%%%%%%%%%%%%%%%%%%%%%%
\vspace*{2.5mm}
\subsection{\hspace*{-2mm}H$^0$ Production in Di-Higgs Channel and Benchmarks}
\vspace*{2mm}
\label{sec:2.3}

For the present study, we focus on the productions and decays of the heavy
Higgs state $H^0$.\,
We have summarized the $H^0$ Yukawa couplings for 2HDM-I and 2HDM-II
in Table\,\ref{tab:1}. The gauge couplings of $H^0$ take
the form $\,G_{HVV}^{}= \cosba \,2M_V^2/v$\,,\, ($V=W,\,Z$),\,
which differs from the SM Higgs-gauge coupling by a factor
$\,\cos(\beta-\alpha)\,$.
With these, we determine the $ggH$ vertex by rescaling SM contributions inside
the loop accordingly. The ratios of the decay widths with respect to
the SM results are given as follows,
\begin{eqnarray}
\frac{\Gamma(H\!\to\! VV)}{\,\Gamma(H\!\to\! VV)_{\textrm{sm}}} &\,=\,&
\cos^2(\beta-\alpha)\,,~~~~~~~
\frac{\Gamma(H\!\to\! ff)}{\,\Gamma(H\!\to\! ff)_{\textrm{sm}}}\,=\,
\(\!\xi_H^f\!\)^2,
\nn\\
\frac{\Gamma(H\!\to\! gg)}{\,\Gamma(H\!\to\! gg)_{\textrm{sm}}\,} &\,=\,&
\left|\sum_{f=t,b}\xi_H^f
\frac{\,A_{1/2}^H(\tau_f^{})\,}{A_{1/2}^H(\tau_t^{})}\right|^2\!,
\end{eqnarray}
where $\,\tau_f^{} = M_H^2/4m_f^2$\, and
\beqs
\beqa
A_{1/2}^{}(x) &=& \frac{2}{\,x^2\,}\left[ x+\!(x\!-\!1)f(x)\right],
\\[1mm]
f(x) &=&
\left\{\begin{array}{ll}
\arcsin^{2}\!\!\sqrt{x} \,, &~~~~ (x\leqq 1)\,,
\\[1.5mm]
\dis\! -\frac{1}{4}\!
\left(\!\ln\!\frac{\,1\!+\!\sqrt{1\!-\!x^{-1}}\,}
                {\,1\!-\!\sqrt{1\!-\!x^{-1}}\,}\!-i \pi\!\right)^{\!\!2}\!,
&~~~~ (x>1)\,.
\end{array}\right.
\eeqa
\eeqs
Around the alignment limit the decay width $\,\Gamma(H\!\to\! VV)$\,
is suppressed by $\,\cos^2(\beta\!-\!\alpha)$,\,
while for down-type fermions the partial width
$\,\Gamma(H\!\to\! ff)\,$ is enhanced by a factor $\,\tan^2\!\beta\,$.

\vspace*{1mm}

The main production mechanism of the neutral Higgs boson $H^0$ at the LHC
is the gluon fusion production $\,gg\!\to\! H^0$.\,
In the 2HDM-I, all Yukawa couplings rescale by a common factor
$\,(\sin\alpha/\!\sin\beta)\,$ with respect to the corresponding SM values.
Hence, the gluon fusion production is still dominated by the top-loop,
and the cross section is rescaled by the factor $\,(\sin\alpha/\!\sin\beta)^2$.\,
In the 2HDM-II, the up-type and down-type Yukawa couplings
have different rescalings from their SM values,
where the rescaling of down-type Yukawa couplings is proportional to
$\,1/\!\cos\!\beta\propto\tanb$\,.\,
For $\,\tanb \gg 1$\,,\,
the bottom-loop in the gluon fusion process may have visible contribution,
but it becomes negligible for $\,\tanb = \mathcal{O}(1)$\,.\,
Fig.\,\ref{fig:22}(b) shows that the current constraints strongly favor
$\,\tanb = \mathcal{O}(1)$\, for the 2HDM-II, so the bottom-loop contribution
is negligible\,\cite{2HDM}.
We take the four-flavor scheme in the present study, and the relevant
$b$-related production process is the bottom-pair associated production
$\,gg\!\to\!Hb\bar{b}$ \cite{Hbb98}\cite{Hdecay},
which is also negligible for the 2HDM-I and
for the 2HDM-II [with small $\,\tanb =\mathcal{O}(1)$\,].

\vspace*{1mm}

We can deduce the coupling of the cubic scalar vertex $Hhh\,$
from the Higgs potential,
\begin{eqnarray}
\label{eq:Hhh_coup1}
G_{Hhh}^{}\,
&\!=\!&  \, \frac{\,\cos(\beta\!-\!\alpha)\,}{v}
\!\left[\!\(\frac{6 M_{\!12}^{2}}{\sin\! 2\beta} \!-\! M_{H}^{2} \!-\! 2M_{h}^{2} \!\)
\!\!\(\!\cos\!2(\beta\!-\!\alpha)-\frac{\,\sin\!2
(\beta\!-\!\alpha)\,}{\tan\!{2\beta}}\!\)\! -\frac{2M_{12}^{2}}{\,\sin\!2\beta\,}
\right]~~~~~~~~~~
\nonumber\\
&\!=&   -\frac{1}{\,v\,}\!
\(\!\frac{8 M_{\!12}^{2}}{\,\sin\! 2\beta\,} - M_{H}^{2} - 2M_{h}^{2}\!\)
\!\cos(\beta\!-\!\alpha) +\mathcal{O}\!\!\left(\cos^2(\beta\!-\!\alpha)\right) \!,
\end{eqnarray}
where we expand the formula around the alignment limit in the second line.
For the mass-range $\MH >2\Mh$,\, the tree-level decay width of $\,H\!\!\to\! hh\,$ is
\beqa
\Gamma(H\!\!\to\! hh) \,=\,
\frac{\,G_{Hhh}^2\,}{\,32\pi M_H^{}\,}\sqrt{1\!-\!\frac{\,4M_h^2\,}{M_H^2}\,}\,.
\eeqa
The di-Higgs decay width is also suppressed by
$\,\cos^2(\beta\!-\!\alpha)$\, in the alignment limit,
but it may receive enhancement from other mass-parameters
$M_{12}^2$ and $M_H^2$ in the Higgs potential.

\begin{figure}[t]
%\vspace*{4mm}
\begin{centering}
\hspace*{-3mm}
\includegraphics[height=6cm,width=7.8cm]{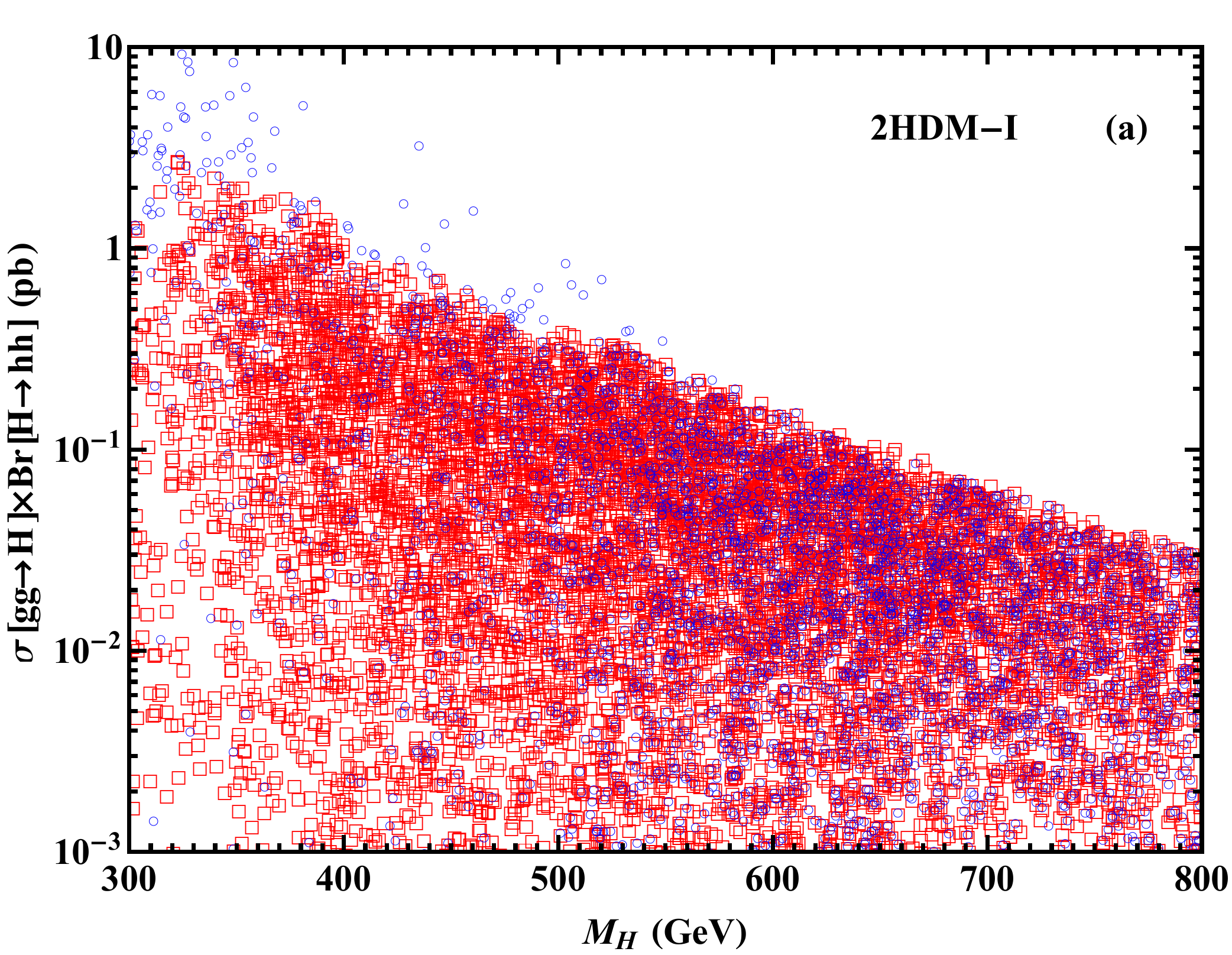}
\hspace*{-1.5mm}
\includegraphics[height=6cm,width=7.8cm]{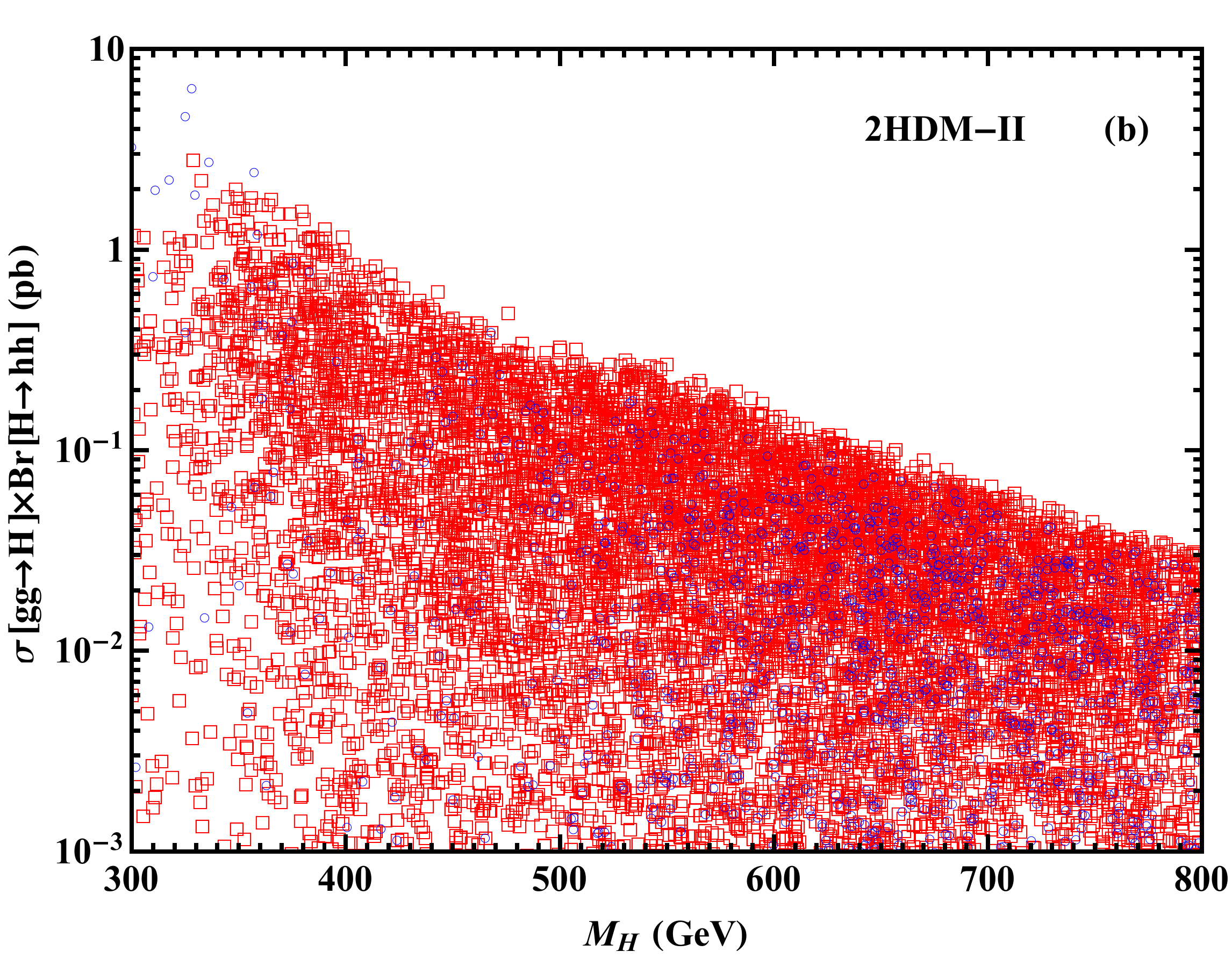}
\vspace*{-7mm}
\caption{$\sigma(gg\!\to\!H)\!\times\!\text{Br}(H\!\!\to\! hh)\,$ as
a function of Higgs mass $\,\MH\,$ at the LHC(14TeV),
for 2HDM-I in plot-(a) and for 2HDM-II in plot-(b).
The blue points (circle shape) satisfy the theoretical requirements,
the electroweak precision limits ($2\sigma$), and
the LHC bounds ($2\sigma$) by the Higgs global fit of $h$(125GeV) data.
The red points (square shape) present
the LHC direct search limits ($2\sigma$) on the heavy Higgs boson $H^0$
(combined with the theoretical requirements).
}
\label{fig:ggHhh}
\label{fig:3}
\label{fig:44}
\label{fig:n5}
\end{centering}
\end{figure}

To set viable benchmarks for searching the new Higgs boson $H^0$
via resonant di-Higgs production
in $\,hh\to WW^*WW^*$\, channel, we will implement the theory constraints and
the current experimental bounds on the 2HDM parameter space,
as we have analyzed in Sec.\,\ref{sec:2.2}.
The requirements of vacuum stability and perturbative unitarity strongly favor
the small $\,\tan\!\beta\,$ region
[except the alignment limit $\,\cos(\beta-\alpha)\sim 0\,$] \cite{llc}\cite{Bernon:2015qea}.
For small $\,\tan\!\beta\,$,\, the bottom-loop contribution in the gluon fusion
production can be safely ignored\,\cite{2HDM}.
Besides, the bottom-pair associated production is subdominant,
especially for the experimental searches aiming at gluon fusion production without
making extra $b$-tagging. According to the analysis in Sec.\,\ref{sec:3},
we generate events for $\,pp\!\to\! H^0b\bar{b}\,$ in the four-flavor scheme.
With the selection cuts aiming at gluon fusion production, in particular
the $b$-veto which helps to suppress the top-related backgrounds,
we find $H^0b\bar{b}$ contribution unimportant for the current study.
Hence, in the following we will focus on the gluon fusion production,
\begin{eqnarray}
\label{eq:CX-HX}
\sigma(gg\!\to\! H^0) \,=\,
\frac{\Gamma(H^0\!\!\to\! gg)}{\,\Gamma(H^0\!\!\to\! gg)_{\text{sm}}^{}\,}
\sigma(gg\!\to\! H^0)_{\text{sm}}^{} \,,
\end{eqnarray}
which includes only the top-loop contribution.
The NLO QCD corrections in the 2HDM are assumed to be the same as
in the SM which are already included in
$\,\Gamma(H^0\!\!\to\! gg)_{\text{sm}}$\,.

In Fig.\,\ref{fig:44}, we present the allowed cross sections of
$\,gg\!\to\! H^0\!\!\to\! hh$\, as a function of the heavy Higgs mass
$\MH$ at the LHC(14TeV) for the 2HDM-I  [plot-(a)]
and 2HDM-II [plot-(b)].
We note that the existing constraints in Fig.\,\ref{fig:33}
will set upper bounds on the resonant $H^0$ production cross sections
at the on-going LHC Run-2 and the HL-LHC.
In Fig.\,\ref{fig:44}, we plot the red dots (square shape)
to show the viable parameter region allowed by
the LHC limits ($2\sigma$) of the existing $H^0$ direct searches
combined with the theoretical requirements.
For comparison, the blue dots (circle shape) represent the parameter space
obeying the theoretical requirements,
the indirect electroweak precision limits ($2\sigma$), and
the LHC bounds ($2\sigma$) from the global fit of $h$(125GeV).
Inspecting Fig.\,\ref{fig:44}, we see that the allowed $H^0$ production cross sections
in the 2HDM-I and 2HDM-II are not much different over the wide mass range
of $\MH =(300-1000)$\,GeV. But, the distribution of the blue dots for 2HDM-II
[plot-(b)] is relatively sparser than that for 2HDM-I [plot-(a)], due to
the stronger constraints on the 2HDM-II by the current LHC global fit of $h$(125GeV)
(cf.\ Fig.\,\ref{fig:33}).
As a side remark, for the neutral pseudo-scalar Higgs boson $A^0$,
it has no $A$-$h$-$h$ vertex and $A$-$V$-$V$ vertex ($V\!=W,Z$) at tree level.
Hence its dominant decay channel is $A\!\to b\bar{b}$
(for $M_A< 2m_t^{}$) or $A\!\to t\bar{t}$ (for $M_A\gtrsim 2m_t^{}$)
\cite{MSSM}\cite{Hdecay},
where the final state fully differs from that of our diHiggs channel
$\,H\!\to hh\!\to 4W$\, and thus does not affect our current LHC analysis.

\vspace*{1mm}

Based upon our analyses of the existing indirect and direct
experimental bounds on the 2HDM (combined with theoretical constraints),
we will systematically study the direct probe of the
heavy Higgs boson via
$\,gg\!\to\! H^0\!\!\to\! hh \!\to\! WW^*WW^*$\, channel at the LHC(14TeV)
in the following Section\,\ref{sec:3}.
For this, we set up three benchmark scenarios for the mass $\MH$ and
the cross section
$\,\sigma(gg\!\to\! H^0)\!\times\! \textrm{Br}(H^0\!\!\to\! hh \!\to\! WW^*WW^*)$
as follows,
\begin{eqnarray}
\label{eq:benchmark}
(\MH ,\,\sigma\!\times\!\textrm{Br}) \,=\,
(300\text{GeV},\,60\text{fb}),~
(400\text{GeV},\,40\text{fb}),~
(500\text{GeV},\,12\text{fb}),~~~~~~~~
\end{eqnarray}
which will be denoted by (H300, H400, H500) for short.

\vspace*{1mm}

For an illustration, we further present three explicit parameter samples to realize
the benchmarks \eqref{eq:benchmark}. It is consistent with all the
theoretical and experimental constraints discussed above.
We show this sample in Table\,\ref{tab:3new}
for the 2HDM-I and 2HDM-II, respectively.
For the parameter samples in Table\,\ref{tab:3new}, we have also explicitly computed
their oblique corrections and obtain the following results corresponding to the benchmarks
(H300, H400, H500),
\beqs
\label{eq:ST-BM}
\beqa
\text{2HDM-I:}~~~ &&
S = (-0.014,\,-0.010,\,-0.0054),~~~~
T = (0.051,\,0.096,\,0.084); \hspace*{13mm}
\\
\text{2HDM-II:}~~~ &&
S = (-0.015,\,-0.011,\,-0.0073),~~~~
T = (0.049,\,0.061,\,0.079); \hspace*{13mm}
\eeqa
\eeqs
and the parameter $\,U=O(10^{-4})$\, is negligible.
As a consistency check, we also recompute the oblique corrections
by using the 2HDMC code\,\cite{2HDMC}, which gives
$\,S = (-0.013,\,-0.010,\,-0.0054)$,\,
$T = (0.051,\,0.097,\,0.085)$\, for the 2HDM-I,
and
$\,S = (-0.015,\,-0.010,\,-0.0071)$,
$T = (0.049,$ $\,0.061, 0.077)$\, for the 2HDM-II.
We find that these agree well with our above results \eqref{eq:ST-BM}.

\begin{table}[t]
\centering
\vspace*{1.5mm}
\begin{tabular}{c||c|c|c|c|c|c|c|c|c}
\hline\hline
2HDM-I & $\tan\!\beta$ & $\cos(\beta\!-\!\alpha)$
& $M_A^{}$ & $M_{H^\pm}^{}$ & $M_{12}^2$ & $G_{\!Hhh}^{}$
& Br($hh$)& Br($t\bar t$)
& $\sigma\!\times\!\text{Br}$\,(fb)
\\
\hline
H300 & $2.10$ & $~~0.025$ &572 & 582 & 36032 & 24.6 & 0.594
& 0.006 & 60
\\
\hline
H400 & 1.29 & $-0.102$ & 601 & 625 & 59620 & 127 & 0.081 & 0.847 & 40
\\
\hline
H500 & 1.11 & $-0.127$ & 537 & 589 & $-6686$ & 172 & 0.034 & 0.904 & 12
\\
\hline\hline
2HDM-II & $\tan\!\beta$ & $\cos(\beta\!-\!\alpha)$
& $M_A^{}$ & $M_{H^\pm}^{}$ & $M_{12}^2$ & $G_{\!Hhh}^{}$
& Br($hh$)& Br($t\bar t$)
& $\sigma\!\times\!\text{Br}$\,(fb)
\\
\hline
H300 & $2.07$ & $0.051$ & 652 & 660 & 32816 & 42.2 & 0.406 & 0.001 & 60
\\
\hline
H400 & 1.00 & $0.083$ & 636 & 650 & 71662 & 128 & 0.077 & 0.87 & 40
\\
\hline
H500 & 1.52 & $0.088$ & 634 & 661 & $94627$ & 183 & 0.110 & 0.80 & 12
\\
\hline\hline
\end{tabular}
\caption{Explicit parameter samples to realize the benchmarks
\eqref{eq:benchmark} for the 2HDM-I and 2HDM-II, respectively.
Here we denote $\text{Br}(hh)=\text{Br}(H^0\!\!\to\!hh)$ and
$\text{Br}(t\bar{t})=\text{Br}(H^0\!\!\to\!t\bar{t})$.
All the mass parameters and the cubic Higgs coupling $G_{Hhh}^{}$ are
in the unit of GeV.}
\label{tab:3new}
\end{table}
%

%%%%%%%%%%%%%%%%%%%%%%%%%%%%%%%%%%%%%%%%%%%%%%
%%=============== Section 3 ================%%
%%%%%%%%%%%%%%%%%%%%%%%%%%%%%%%%%%%%%%%%%%%%%%

\vspace*{3mm}
\section{\hspace*{-2mm}Analyzing Signals and Backgrounds at the LHC}
\label{sec:3}

In this section, we perform systematical Monte Carlo analysis for the resonant neutral
Higgs $H^0$ signal $\,gg\!\to\! H^0\!\!\rightarrow\! h^0h^0\!\rightarrow\! WW^{*}WW^{*}$\,
and its main backgrounds at the LHC\,(14TeV).
We set up the signal process model by FeynRules\,\cite{Alloul:2013bka}
with $ggH^0$ and $H^0h^0h^0$ vertices.
We generate the events by MadGraph5 package\,\cite{Alwall:2014hca}
at parton level, and then process them by Pythia\,\cite{Sjostrand:2006za}
for hadronization and parton shower.
Finally, we use Delphes\,3\,\cite{deFavereau:2013fsa} for detector simulations.
Here, for the resonant diHiggs production,
we include the new vertices $ggH^0$ and
$H^0h^0h^0$ in the MadGraph model file.
For the $ggH^0$ vertex, we use the precise form factor from the top-loop,
which only depends on the masses $M_H^{}$ and $M_t^{}$.
We include the same $K$-factor for the NLO QCD corrections as in the
SM-type Higgs production $\,gg\!\to\! H^0$ \cite{Djouadi-SMh2005}.
As consistency checks, we have used the SusHI package\,\cite{SusHI}
to recompute the resonant production cross section
$\,\sigma(gg\!\to\! H^0)$ and get full agreement.
We also note that the cross section of the resonant on-shell $H^0$
production (followed by the cascade decay $H^0\!\!\to\! h^0h^0$)
overwhelms the nonresonant SM contribution\,\cite{Hhh-2HDM}.

\vspace*{1mm}

We study two major decay channels of the final state $WW^{*}WW^{*}$:
(i).~$\ell^{\pm}\nu \ell^{\pm}\nu\, 4q\,$ with same-sign di-leptons (SS2L);
(ii).~$\ell^{\pm}\nu \ell^{\mp}\nu \ell^{\pm}\nu \,2q\,$ with tri-leptons (3L).
For $W$ boson, its branching fractions of leptonic decays
$\,W\!\!\to\! e\nu,\,\mu\nu,\,\tau\nu$\,
are 10.8\%, 10.6\%, and 11.3\%, respectively,
while its hadronic decay $\,W\!\!\to\! q\bar{q}'\,$ has branching fraction 67.6$\%$
\cite{Agashe:2014kda}. For this analysis, we include the detected
$e$ and $\mu$ from $\tau$ decays as well.
These two decay channels of $WW^{*}WW^{*}$ have
branching fractions about 9.6\% and 9.2\%, respectively.
Although they are quite small (less than 10\%), we note that
requiring the detection of the same-sign di-leptons or the tri-leptons
in the final state can significantly reduce the QCD backgrounds and enhance
the signal sensitivity.

%%%%%%%%%%%%%%%%%%%%%%%% Sec.3.1 %%%%%%%%%%%%%%%%%%%%%%%%%

%\vspace*{2.5mm}
\subsection{\hspace*{-2mm}Final State Identification}
\vspace*{2mm}
\label{sec:3.1}
\label{sec:objiden}

To analyze the signal sensitivity, we apply the ATLAS procedure
to identify the final states in both SS2L and 3L decay channels.
Jets, leptons, and transverse missing energy are selected by the following cuts,
\beqa
 p_T^{}(\ell)>10\,\textrm{GeV},~~
 %p_T(\mu)>25\,\textrm{GeV},~~
 p_T^{}(j)>25\,\textrm{GeV},~~
 \slashed{E}_T>10\,\textrm{GeV},~~
 |\eta(j)|,|\eta(\ell)|<2.5\,.
\eeqa
Electrons with $\,1.37<|\eta(\ell)|<1.52$\, are rejected
in order to remove the transition region of electromagnetic calorimeter of ATLAS.
After the trigger, we require the leading lepton passing the trigger requirement
$\,p_T^{}(\ell)>25$\,GeV.
The $b$-taging algorithm based on $p_T^{}$ of jet is implemented
in Delphes\,\cite{ATL-PHYS-PUB-2015-022}.

The reconstructed objects in the final state
have to be well separated spatially to prevent the potential double-counting.
We implement the following criteria\,\cite{criteria}:
(i).~any electron overlapped with a muon with $\,\Delta R(e,\mu)<0.1\,$ is removed;
(ii).~for any electron pair with $\,\Delta R(e,e)<0.1$,\, the electron with lower
$p_T^{}$ is removed; (iii).~any electron within $\,\Delta R(j,e)<0.3$\, is removed;
(iv).~any muon within $\,\Delta R(\mu,j)<0.04+10/p_{\mu,T}^{}\textrm{(GeV)}$\, is removed.

%%%%%%%%%%%%%%%%%%%%%%%% Sec.3.2 %%%%%%%%%%%%%%%%%%%%%%%%%

\vspace*{2.5mm}
\subsection{\hspace*{-2mm}Analysis of Same-Sign Di-lepton Decay Channel}
\vspace*{2mm}
\label{sec:3.2}
\label{sec:2lss_analysis}

With the identification of the final state particles as in Sec.\,\ref{sec:objiden},
our analysis of the same-sign di-leptons (SS2L) channel further requires
the sub-leading lepton obeying $\,p_T^{}(\ell)>25\,$GeV\,
to reduce the fake backgrounds (as will be described in the following) and
\begin{equation}
\label{eq:2LSSBES}
n_{\ell}^{} =2\, (\textrm{same-sign}), \quad
n_j^{}\geqq 3\,.
\end{equation}
The jets arising from the off-shell $W$ boson decays could be soft and
the requirement of $\,n_j^{} \geqq 3\,$ provides an optimal significance.
The above defines the basic event selection for the SS2L channel.

The main prompt backgrounds that contribute to the same-sign di-leptons include
$\,W^{\pm}W^{\pm}$,\, $W^{\pm}h$\,(with $h\!\to\! W^{\pm}W^{\mp})$,\,
$Zh$\,(with $Z\!\to\!\ell\ell$, $h\!\to\!WW$),\,
$t\bar{t}W$,\, $t\bar{t}Z$\,(with $Z\!\to\!\ell\ell)$,\,
$t\bar{t}h$\,(with $h\!\to\! W^{\pm}W^{\mp})$,\,
$ZZ$\,(with $Z\!\to\!\ell\ell)$\, and \,$WZ$\,(with $W\!\to\! \ell\nu$\,
and $\,Z\!\to\!\ell\ell\,)$.\,
The background processes with a pair of top quarks (namely, $t\bar{t}W$,\,
$t\bar{t}Z$, and $t\bar{t}h$) can be efficiently rejected by $b$-veto.
With the basic event selection and $b$-veto, the $t\bar{t}Z$ background becomes negligible.
The diboson backgrounds, $WZ$, $ZZ$, and $Zh$, can be suppressed by requiring exactly
two same-sign leptons as in Eq.(\ref{eq:2LSSBES}).
Hence, we can safely ignore $ZZ$ and $Zh$ backgrounds given their small cross sections,
and only include $WZ$ channel for the background estimate.

\begin{table}[t]
\begin{center}
\begin{tabular}{c|c|c}
\hline
\hline
Signals  &
$\sigma\!\times\!\textrm{BR}$\,(fb)   & $\sigma\!\times\!\textrm{BR}$\,(fb) \\
~\& Backgrounds~ & (before PreS) & (after PreS)
\\
\hline
H300 &5.8   &0.22    \\
H400 &3.8   &0.18    \\
H500 &1.15  &0.044    \\
\hline
$W^{\pm}W^{\pm}$ &29.6  &2.57 \\
$Wh$  & 24.1\,\cite{PhysRevLett.107.152003}     &0.39    \\
$t\bar{t}W$ & 54.6\,\cite{Campbell2012}    &1.61 \\
$t\bar{t}h$ & 12.6\,\cite{1310.1132}   &0.185    \\
$WZ$  & 921\,\cite{PhysRevD.60.113006}     &15.0    \\
\hline
$W$+jets & 1358000   & {1.4}    \\
$t\bar{t}$\,(semi-leptonic)
& {433500}\,\cite{https://twiki.cern.ch/twiki/bin/view/LHCPhysics/TtbarNNLO}~
& {2.28}    \\
\hline
$Z$+jets & 141200    & 0.30    \\
$t\bar{t}$\,(leptonic) & {104436}  & {3.80}    \\
\hline
\hline
\end{tabular}
\end{center}
\vspace*{-2mm}
\caption{$\sigma\!\times\!\textrm{Br}$ for the signal process (with three benchmarks) and
for the major backgrounds in the SS2L decay channel before and
after pre-selections (PreS), where the last two categories denote backgrounds
with fake leptons and charge-misidentifications.
For each background (with a cited reference), the cross section includes
the QCD corrections ($K$-factor),
while the rest of cross sections are computed
by MadGraph5 at the leading order.}
\label{xs_2lss_preselection}
\label{tab:3}
\label{tab:33}
\end{table}

\begin{figure}[t]
\vspace*{-15mm}
 \centering
\includegraphics[height=5.0cm,width=0.48\textwidth]{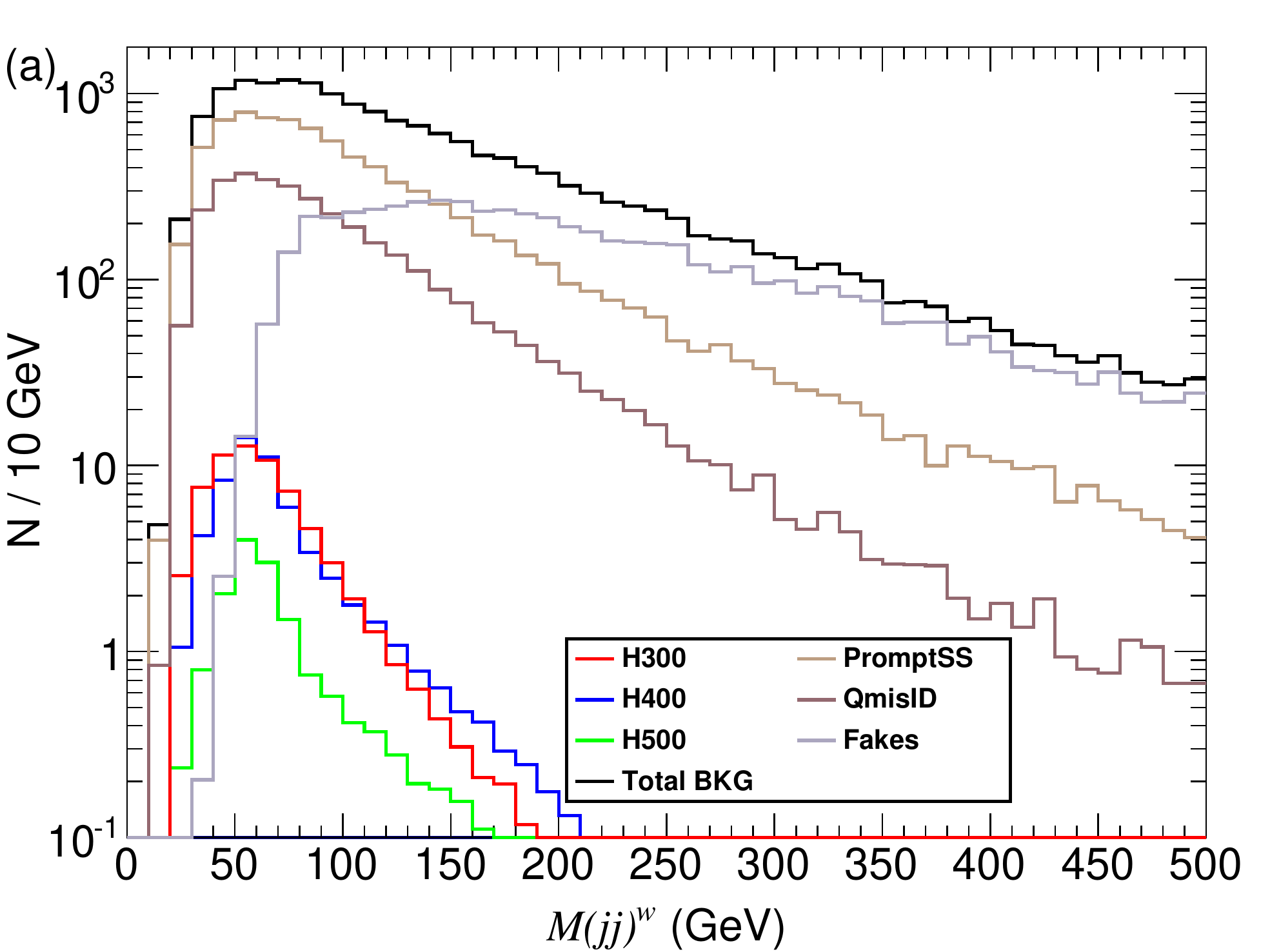}
%\label{subf:m_jj_w1}
\hspace*{-3mm}
\includegraphics[height=5.0cm,width=0.48\textwidth]{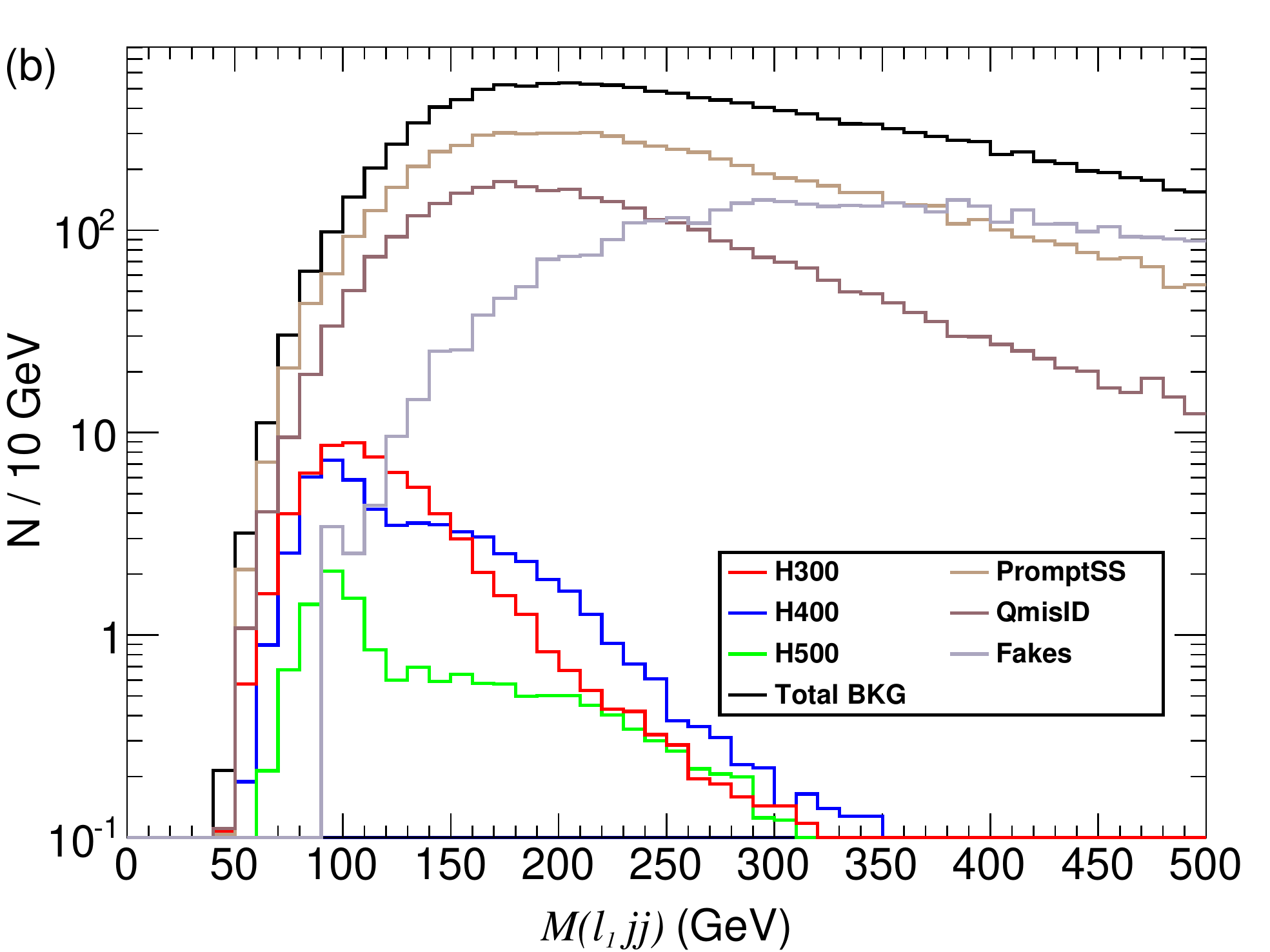}
%\label{subf:m_l1jj}
\\[0mm]
\includegraphics[height=5.0cm,width=0.48\textwidth]{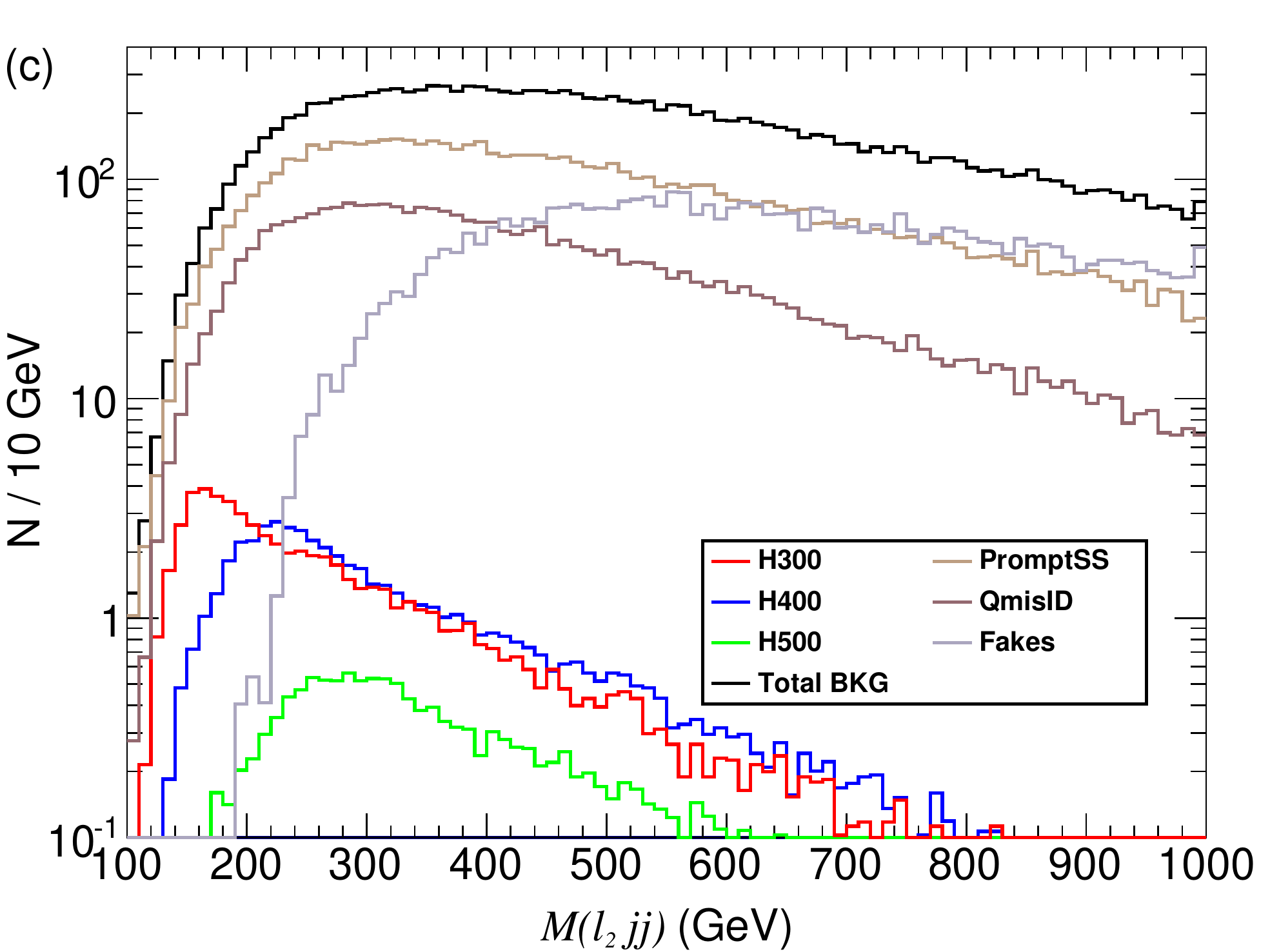}
%\label{subf:m_l2jj}
\hspace*{-3mm}
\includegraphics[height=5.0cm,width=0.48\textwidth]{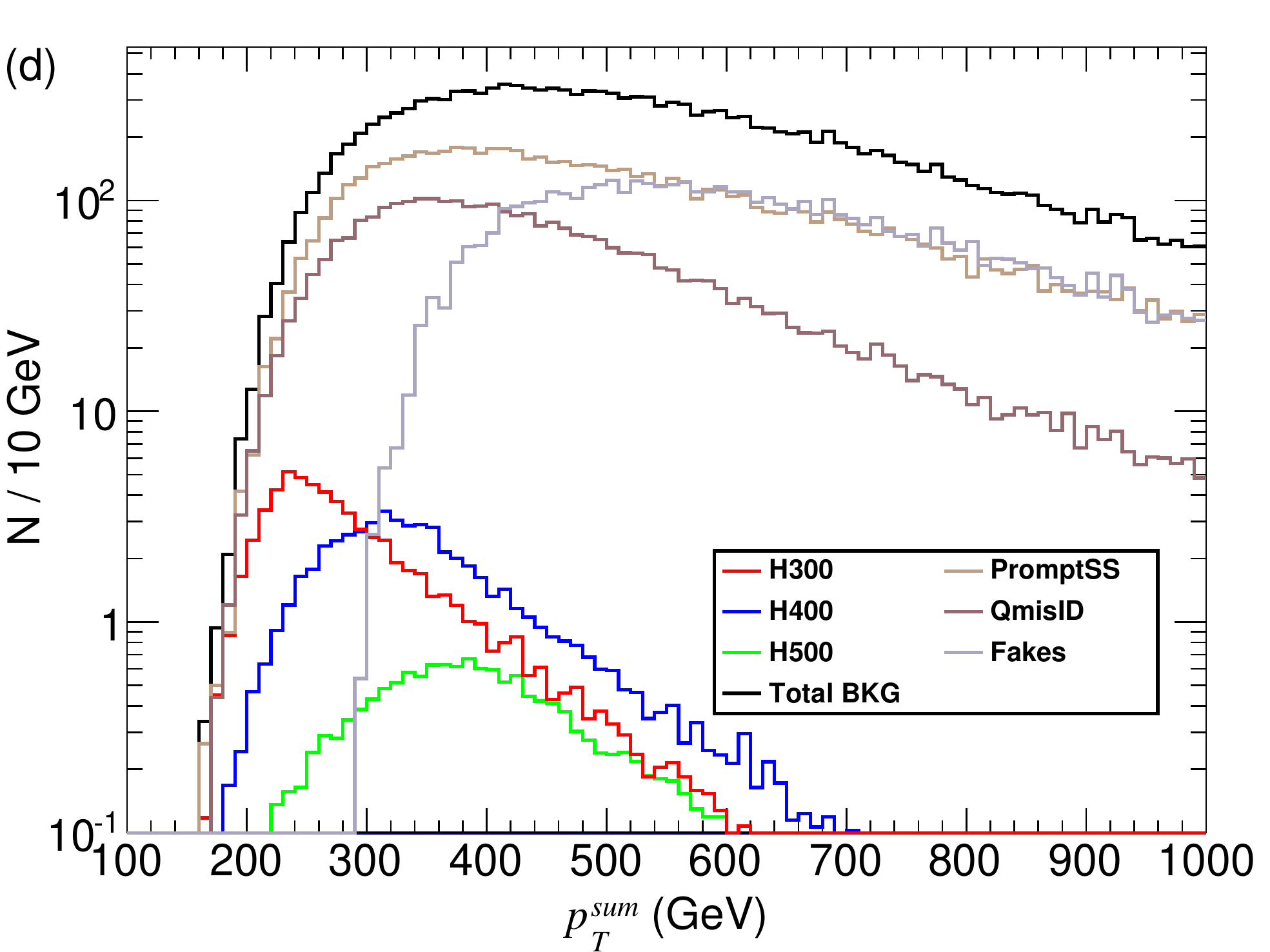}
%\label{subf:pt_sum}
\\[0mm]
\includegraphics[height=5.0cm,width=0.48\textwidth]{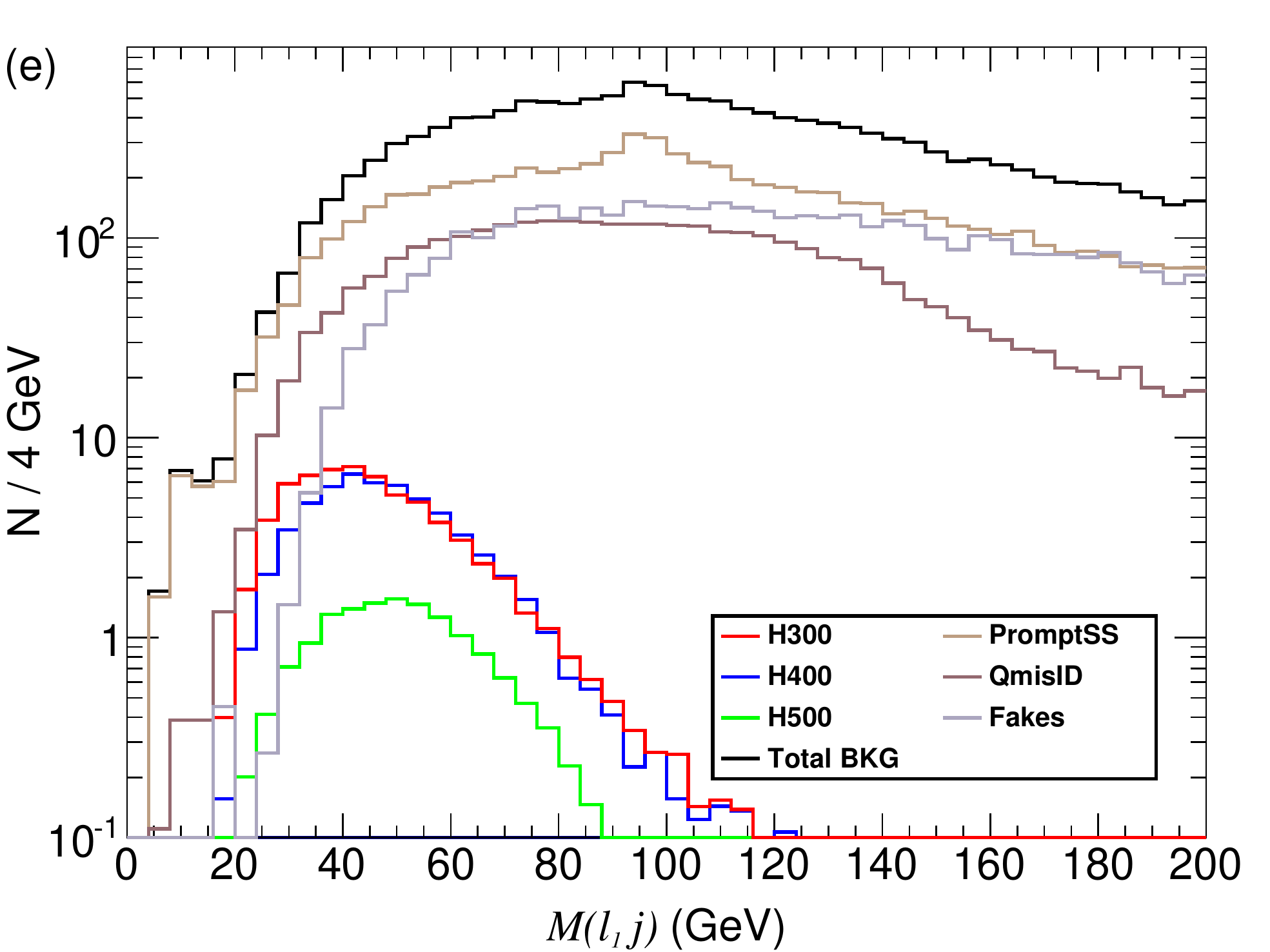}
%\label{subf:m_l1j}
\hspace*{-3mm}
\includegraphics[height=5.0cm,width=0.48\textwidth]{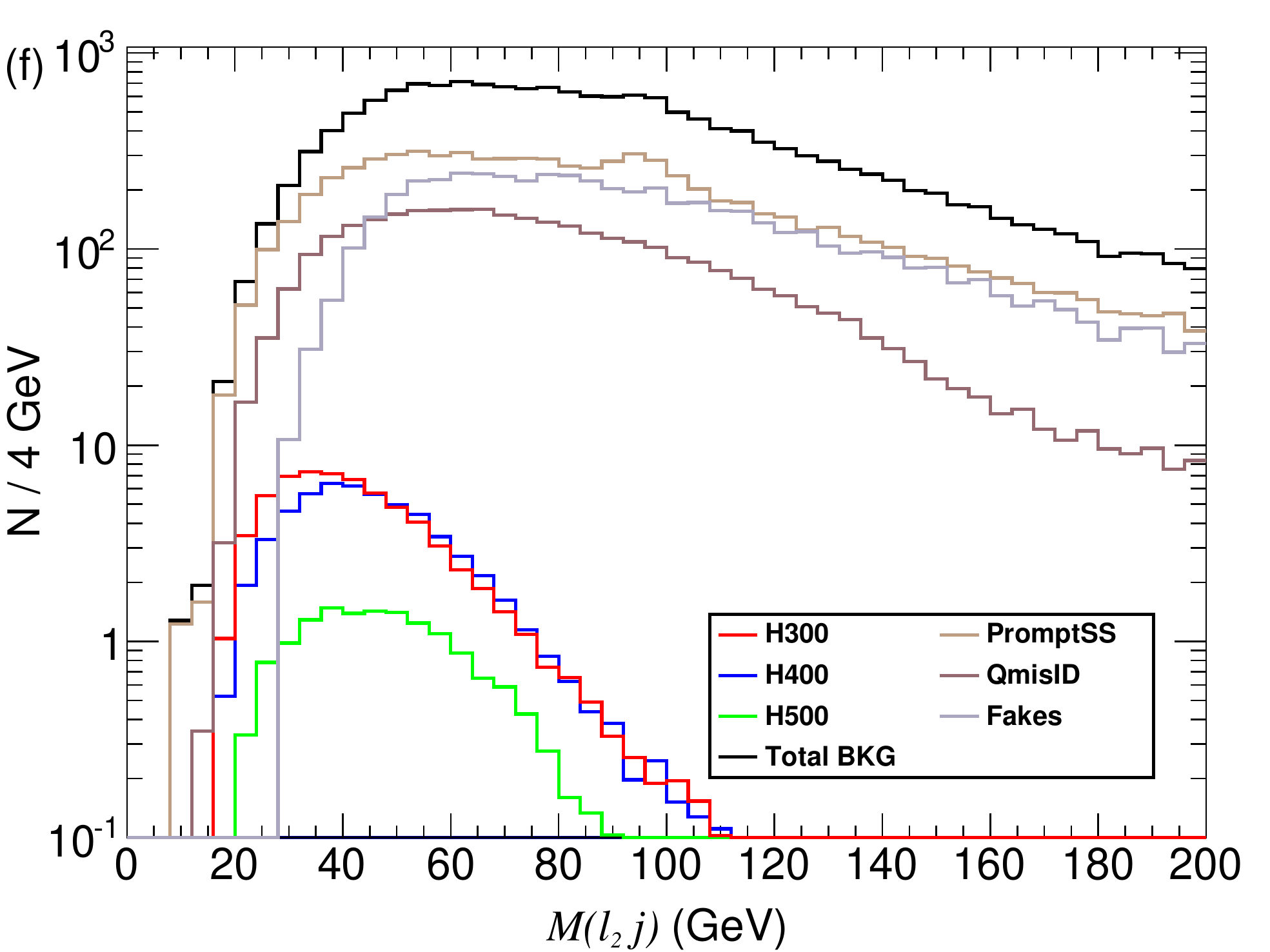}
%\label{subf:m_l2j}
\\[0mm]
\includegraphics[height=5.0cm,width=0.48\textwidth]{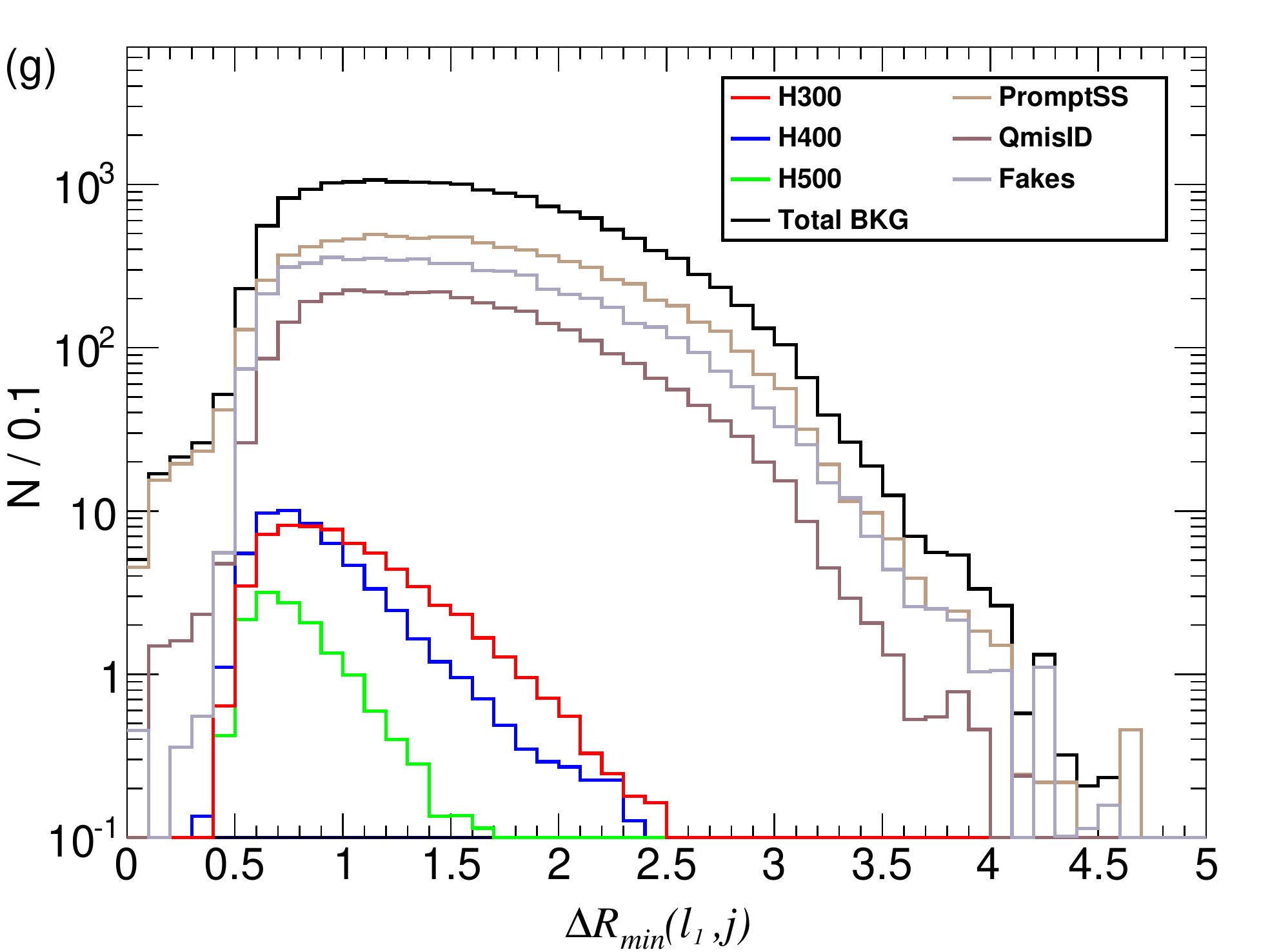}
%\label{subf:mindR_l1j}
\hspace*{-3mm}
\includegraphics[height=5.0cm,width=0.48\textwidth]{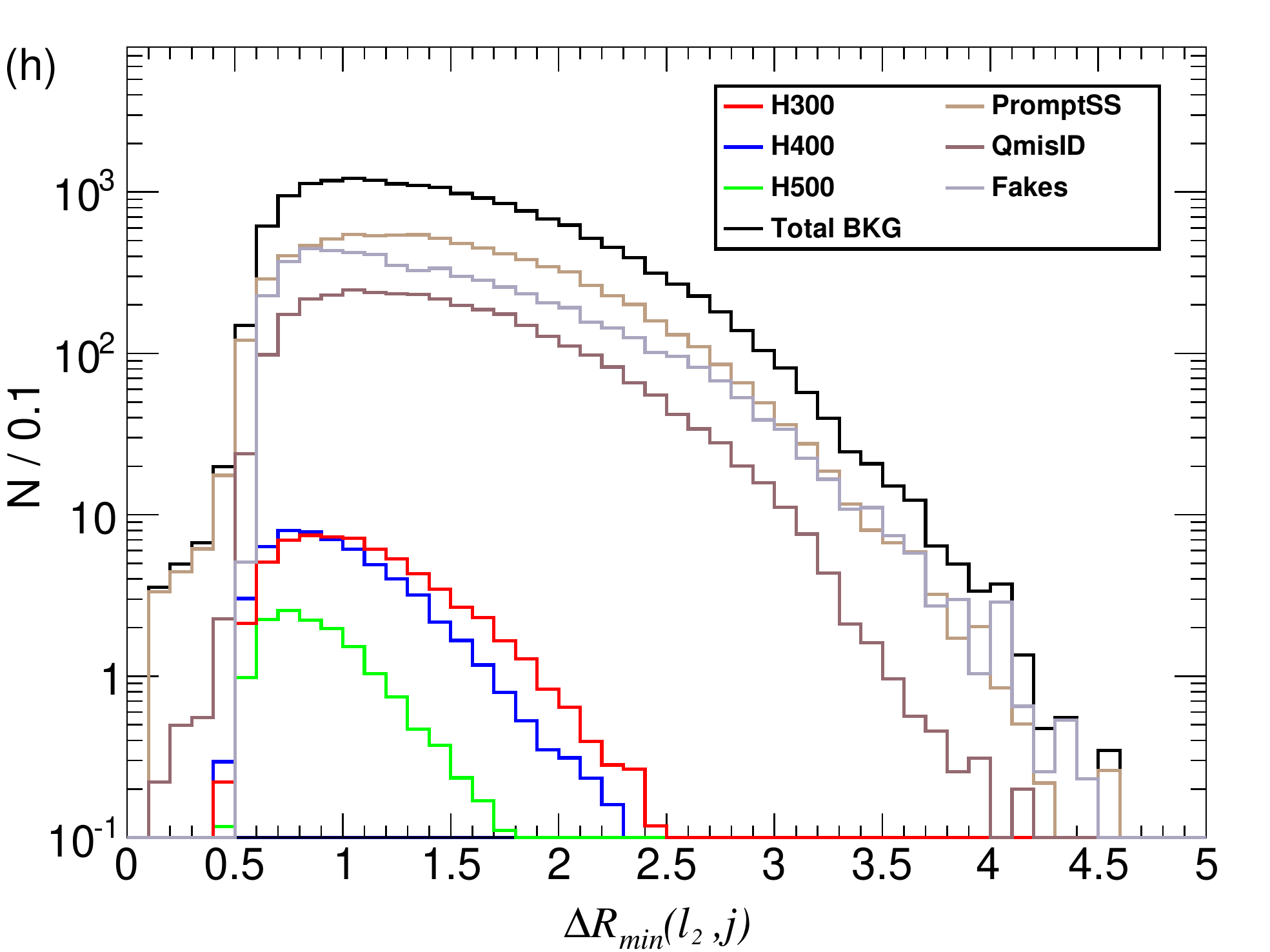}
\label{subf:mindR_l2j}
\vspace*{-1mm}
\caption{Distributions of different kinematical variables in the SS2L decay channel:
(a) invariant mass of the two closet jets;
(b) invariant mass of the leading lepton and the two closet jets;
(c) invariant mass of the sub-leading lepton and the remaining jet(s),
(d) linear sum of $p_T^{}$ from two leptons, jets and transverse missing energy,
(e) invariant mass of the leading lepton and the closet jet,
(f) invariant mass of the sub-leading lepton and the closet jet,
(g) minimum $\,\Delta R\,$ distance between the leading lepton and jet,
(h) minimum $\,\Delta R\,$ distance between the sub-leading lepton and jet.
For these plots, we input an integrated luminosity of 300\,fb$^{-1}$.}
\label{kine-1}
\label{fig:4}
\label{fig:55}
\label{fig:n6}
\vspace*{-13mm}
\end{figure}

For the SS2L channel, backgrounds with fake leptons from jets or charge misidentifications
(QmisID) can also be significant.
Jet faking leptons mainly come from $W$+jets final state and semi-leptonic mode of
$t\bar{t}$ final state (which both have large cross sections).
For the samples of fake electrons, we assign a weight to each event in the following way.
(i).~We generate $W+$jets background (including two or more jets) and
the $t\bar{t}$ background (in the semi-leptonic decay mode).
(ii).~For each selected event of lepton+jets, we loop over
all possible jets that could fake an electron with a probability
$\,P=0.0048\times\exp[-0.035\!\times\! p_T^{}(\text{GeV})]$\,
as a function of jet $p_T^{}$ \cite{ATL-PHYS-PUB-2013-004}.
(iii).~We sum over all fake rates and divide it by two to account for the same sign fakes
with the selected leptons.
(iv).~We randomly choose one jet to be the fake electron according to the fraction
of fake rates, and rescale the jet's energy to its 40\%
as that of the fake electron\,\cite{ATL-PHYS-PUB-2013-004}.
Since fake electrons are usually soft, we find that the selection cut
$\,p_T^{}(e)>25$\,GeV\, helps to significantly suppress these backgrounds
and makes them comparable to other prompt backgrounds.
With the upgrade Level-0,1 Muon Trigger for ATLAS detector\,\cite{ATL-PHYS-PUB-2016-026},
the contribution of fake muons with $\,p_T^{}(\mu)>25\,$GeV is small and can also be safely ignored.
The QmisID mainly comes from the $Z$+jets and the pure leptonic mode of $t\bar{t}$\,,\,
with one charge misidentified lepton. Pseudo-events are generated in a way similar to
that for the fake electrons, with a weight 0.0026 assigned
to each event\,\cite{ATL-PHYS-PUB-2016-026}.
These backgrounds can be significantly suppressed by $Z$-veto, i.e.,
$\,|M(\ell\ell)\!-\!M(Z)|>10$\,GeV.\,
With these, we find that the contribution due to QmisID is negligible.

\begin{figure}[t]
\centering
\includegraphics[height=6.5cm,width=0.49\textwidth]{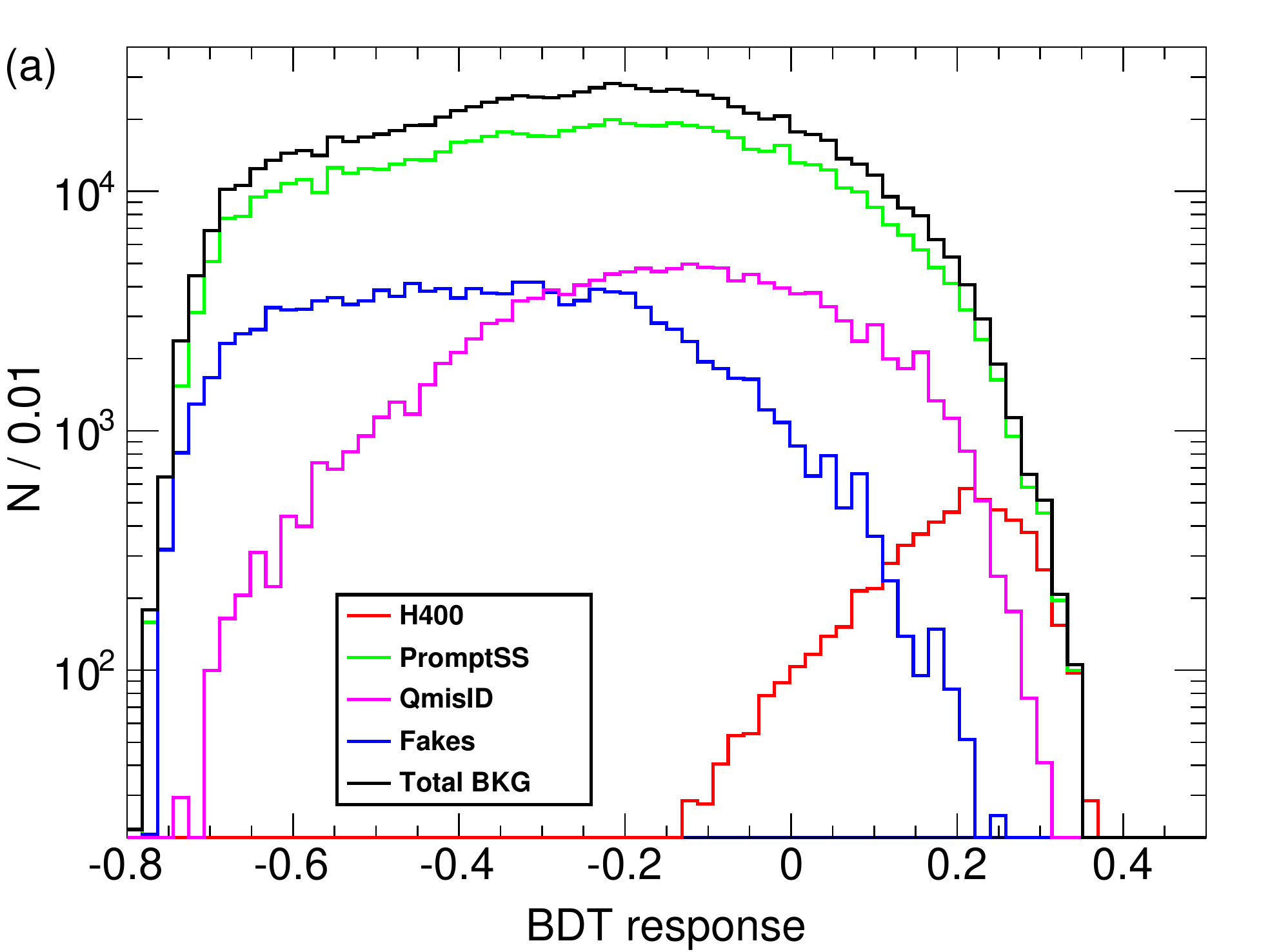}
%\label{subf:BDT_H300}
\hspace*{-3mm}
\includegraphics[height=6.5cm,width=0.49\textwidth]{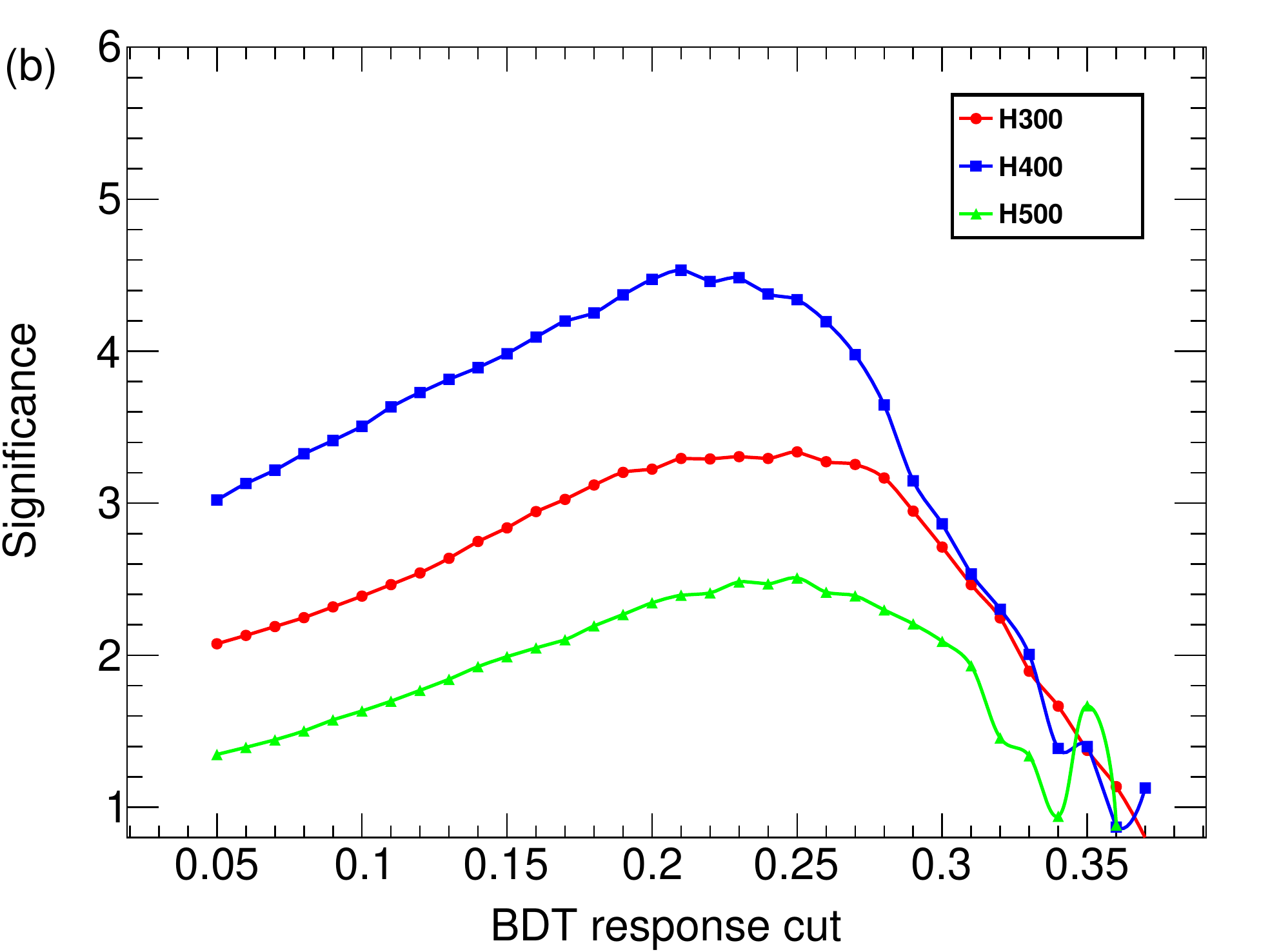}
%\label{subf:BDTcut_H300}
\vspace*{-2mm}
\caption{BDT optimization for Higgs signals over backgrounds.
Plot-(a) depicts the distributions of the output BDT response for the
Higgs signal of $\,\MH =400$\,GeV and for different backgrounds. %with all components normalized to 1.
Plot-(b) presents the significance $\SZZ$ as a function of
the BDT cut for $\,\MH =(300,\,400,\,500)$\,GeV shown by (red,\,blue,\,green) curves.
The optimal BDT cuts are (0.25,\,0.21,\,0.25) for the three curves
of $\,\MH =(300,\,400,\,500)$\,GeV, which correspond to the maximal significance
$\,\SZZ =(3.3,\,4.5,\,2.5)$. We input an integrated luminosity of
(300,\,1000,\,3000)fb$^{-1}$ for $\,\MH =(300,\,400,\,500)$\,GeV.
}
\label{BDT_H300}
\label{fig:5}
\label{fig:66}
\label{fig:n7}
\end{figure}

After all pre-selection cuts, including the basic event selection,
$b$-veto, and $Z$-veto, we obtain $\sigma\!\times\!\textrm{Br}$
as shown in the last column of Table\,\ref{xs_2lss_preselection}.
We will make further optimization to increase the significance.
In Fig.\,\ref{kine-1}, we present the distributions of kinematic variables
after all pre-selections, each of which has its own advantage to
help the discrimination of signals from backgrounds.
Some variables are insensitive to the heavy Higgs mass $\MH$.\,
The invariant mass of the pair of closest jets, $M_{jj}^{W}$,
well represents the mass scale of $W$ boson.
The invariant mass of the leading lepton and the two closest jets,
$\,M(\ell_{1}^{}jj)$,\, reflects the mass scale of the light Higgs boson
$h$(125GeV). To represent $\MH$,\, one obvious choice is the $p_T^{}$ sum
of the selected leptons, jets and transverse missing energy.
For larger Higgs mass $\MH$,\, the intermediate $W$ bosons become more boosted.
The $\,\Delta R\,$ distance between the two leptons and their closest jets,
$\,\Delta R_{\min}^{}(\ell_{i}^{},j)$,\, tend to be smaller,
while their corresponding invariant masses, $M(\ell_{i}^{}j)$, are larger.
In summary, $\,\Delta R_{\min}^{}(\ell_{i}^{},j)$\, has stronger power of
separating the signal from backgrounds for the higher $M_H$ case,
while all invariant-masses and $p_{T}^{\textrm{sum}}$
play a better role for the case of lower $\MH$.\,

\vspace*{1mm}

\begin{table}[t]
\vspace*{4mm}
\begin{center}
\small
\begin{tabular}{c|c|cccccc}
			\hline
			\hline
			{\footnotesize Benchmark} & $\mathcal{L}$
			& $t\bar{t}W$ &$W^{\pm}W^{\pm}$ &$t\bar{t}h$ &Wh &WZ &$t\bar{t}$\,{\footnotesize (leptonic)} \\
			\hline
			H300 &300 &2.8$\pm$0.4    &1.7$\pm$0.3    &1.4$\pm$0.1    &12.2$\pm$0.4   &16.5$\pm$2.2
			& {7.5$\pm$0.8}  \\
			\hline
			H400 &1000 &26.9$\pm$2.2    &19.6$\pm$1.7    &15.1$\pm$0.8    &49.3$\pm$1.6    &125.1$\pm$10.9
			& {45.0$\pm$3.3}  \\
			\hline
			H500 &3000 &30.6$\pm$4.0    &25.5$\pm$3.3    &20.7$\pm$1.7    &45.5$\pm$2.7    &139.3$\pm$19.9
			& 35.7$\pm$4.7  \\
			\hline
			{\footnotesize Benchmark}  & $\mathcal{L}$
			& $W\!+$jets
			& $t\bar{t}$\,{\footnotesize (semileptonic)}
			& Full BKG
			& Signal & $S/\!\sqrt{S\!+\!B}$ & $\SZZ$ \\
			\hline
			H300 &300   & 0$\pm$0 &0$\pm$0  & 42.1$\pm$2.5  &23.4$\pm$0.5  & 2.9    &{3.3} \\
			\hline
			H400 &1000  & 1.1$\pm$0.7    & 1.9$\pm$0.8  &284$\pm$12  & 79.7$\pm$1.4    & 4.2  & {4.5} \\
			\hline
			H500 &3000  & 1.6$\pm$1.6  & 3.7$\pm$2.2  &302$\pm$21  & 44.2$\pm$1.1    & 2.4  & {2.5} \\
			\hline
			\hline
		\end{tabular}
\end{center}
\vspace*{-3.5mm}
\caption{BDT optimization in the same-sign di-lepton (SS2L) channel for the three benchmarks
with different integrated luminosity $\mathcal{L}$ (in the unit of fb$^{-1}$).
We present, after the optimal BDT cuts,
the selected event numbers for the signals, backgrounds,
and median significance $\mathcal{Z}$\,.}
\label{event_yield_2lss_BDT}
\label{tab:4}
\label{tab:44}
\end{table}

To maximize the background rejection, we train a boosted decision tree (BDT)
discriminant, implemented in a TMVA package,
using all the input variables\,\cite{Hocker:2007ht} for each benchmark.
By scanning the median significance $\SZZ$ \cite{Z0}
(cf.\ Eq.\,\eqref{eq:Z>5})
as a function of the BDT response cuts,
we select the BDT-optimized working point before
the statistical fluctuations becoming important.
As an illustration, Fig.\,\ref{BDT_H300}(a) shows
the distribution of the output BDT response
with an integrated luminosity $\,\mathcal{L}=1000\,\text{fb}^{-1}$
for the case of $\,\MH=400$\,GeV.
In Fig.\,\ref{BDT_H300}(b), we describe the median significance $\SZZ$
as a function of the BDT response for three sample Higgs masses
$\,\MH=(300,400,500)$GeV, which correspond to the (red, blue, green) curves.
With the optimal BDT cut at (0.25,\,0.21,\,0.25) for
$\,\MH=(300,400,500)$GeV, we achieve the final significance
$\,\SZZ =(3.3,\,4.5,\,2.5)$ in each case accordingly.
The results for the optimizations of the three benchmarks are summarized in
Table\,\ref{event_yield_2lss_BDT}.
In each case, the BDT optimization helps to suppress all backgrounds significantly.
For instance, the event rates of $\,W^{\pm}W^{\pm}$\, and \,$WZ$\,
get reduced by the kinematical variables that reflect $\Mh$ and $\MH$.\,
The annoying fake leptons backgrounds are also largely suppressed.
As shown in this table, the significance increases from the BDT optimization
is stable under the statistical fluctuations.

%%%%%%%%%%%%%%%%%%%%% Sec.3.3 %%%%%%%%%%%%%%%%%

\vspace*{2.5mm}
\subsection{\hspace*{-2mm}Analysis of Tri-Lepton Decay Channel}
\vspace*{2mm}
\label{sec:3.3}
\label{sec:3l_analysis}

The analysis of the tri-lepton (3L) channel is similar to that of the SS2L channel.
With the final state identification in Sec.\,\ref{sec:objiden},
we require the sub-leading electron to satisfy
$\,p_{T}^{}(e)>$20\,GeV,\, and the sub-leading muon to obey
$\,p_{T}^{}(\mu)>25$\,GeV.
These will help to significantly suppress the fake electrons and fake muons.
Furthermore, we implement a stronger cut $\,\slashed{E}_T>$\,20\,GeV,
since the third neutrino in the 3L channel can contribute to the reconstructed
transverse missing energy with less back-to-back cancellation
than the case of the SS2L channel. The events need to further satisfy,
\begin{equation}
n_{\ell}^{} = 3~(\textrm{total charge}=\pm1),\quad
n_j^{} \geqq 2\,.
\hspace*{10mm}
\end{equation}
The requirement on the total charge can enhance the $Z$-veto efficiency.
Among the three leptons, $\ell_0^{}$\, denotes the lepton with the opposite charge
to the others, $\ell_1^{}$\, for the one closest to $\,\ell_0^{}$,\,
and $\,\ell_2^{}\,$ for the remaining one.
Thus, we define the $Z$-veto by
$\,|M(\ell_0^{}\ell_1^{})-M_Z^{}|>20$\,GeV\, and
$\,|M(\ell_0^{}\ell_2^{})\!-\!M_Z^{}|>10$\,GeV.

\begin{table}[t]
\begin{center}
\begin{tabular}{c|c|c}
\hline\hline
Signals  &
$\sigma\!\times\!\textrm{BR}$\,(fb)   & $\sigma\!\times\!\textrm{BR}$\,(fb) \\
~\& Backgrounds~ & (before PreS) & (after PreS)
\\
\hline
H300 &5.5    &0.15     \\
H400 &3.7    &0.13     \\
H500 &1.1   &0.034     \\
\hline
$WZ$  &921            &1.82     \\
$Wh$  &11.5           &0.11     \\
$ZZ$ &152\,\cite{PhysRevD.60.113006}            &0.09     \\
$WWW$ &12.1\,\cite{0804.0350}           &0.39     \\
$WWZ$ &3.3\,\cite{0804.0350}            &0.09     \\
$t\bar{t}W$ &26.1     &0.30     \\
$t\bar{t}h$ &12.1     &0.11     \\
$t\bar{t}Z$ &29.6     &0.19     \\
\hline
$Z$+jets &141200      &0.05     \\
$t\bar{t}$\,(leptonic) & {104436}  & {2.64} \\
\hline
\hline
\end{tabular}
\end{center}
\vspace*{-4mm}
\caption{$\sigma\!\times\!\textrm{Br}$ for the signal process
(with three benchmarks) and the major backgrounds
in the 3L decay channel before and after the pre-selections (PreS), where the last category denotes
backgrounds with fake leptons.}
\label{xs_3l_preselection}
\label{tab:5}
\label{tab:55}
\vspace*{4mm}
\end{table}

The background estimation for the 3L channel also has several differences.
Besides those discussed in the analysis of SS2L channel,
the prompt backgrounds include $ZZ$ final state (which might not be negligible here)
and tri-boson processes $WWW$ and $WWZ$.\,
For the fake backgrounds, only jet-faked leptons from the final state $Z$+jets
and the final state $t\bar{t}$ in pure leptonic mode need to be considered
due to their large cross sections.
We generate samples with fake electrons in the same way as described for
the analysis of SS2L channel. The fake backgrounds from $Z$+jets can be
largely suppressed by $Z$-veto.
We show $\,\sigma\!\times\!\text{Br}\,$ after the pre-selections
in Table\,\ref{xs_3l_preselection}.

\begin{figure}[t]
\vspace*{-5mm}
 \centering
\includegraphics[height=5.2cm,width=0.48\textwidth]{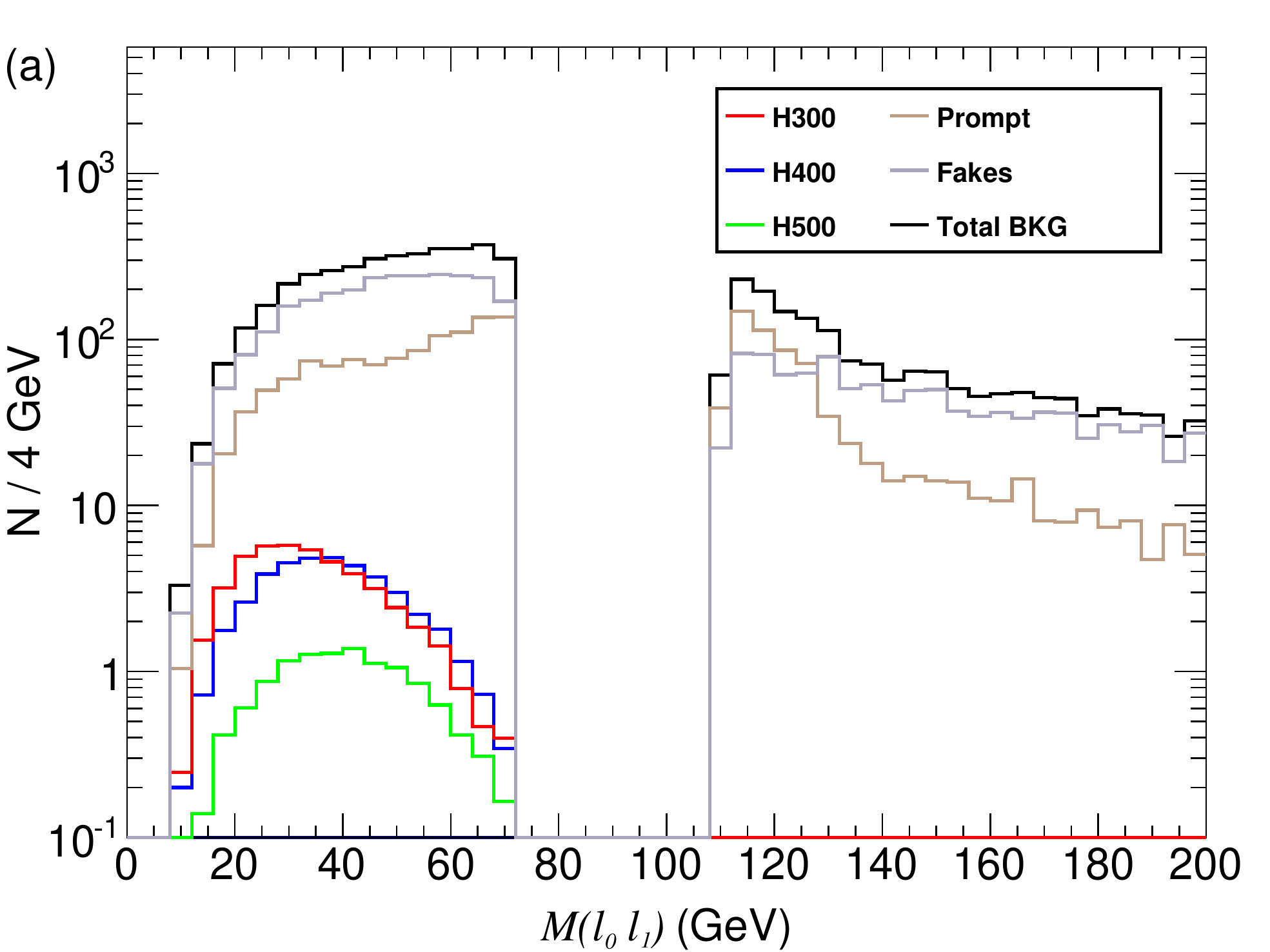}
%\label{subf:3l_m_l0l1}
\hspace*{-3mm}
\includegraphics[height=5.2cm,width=0.48\textwidth]{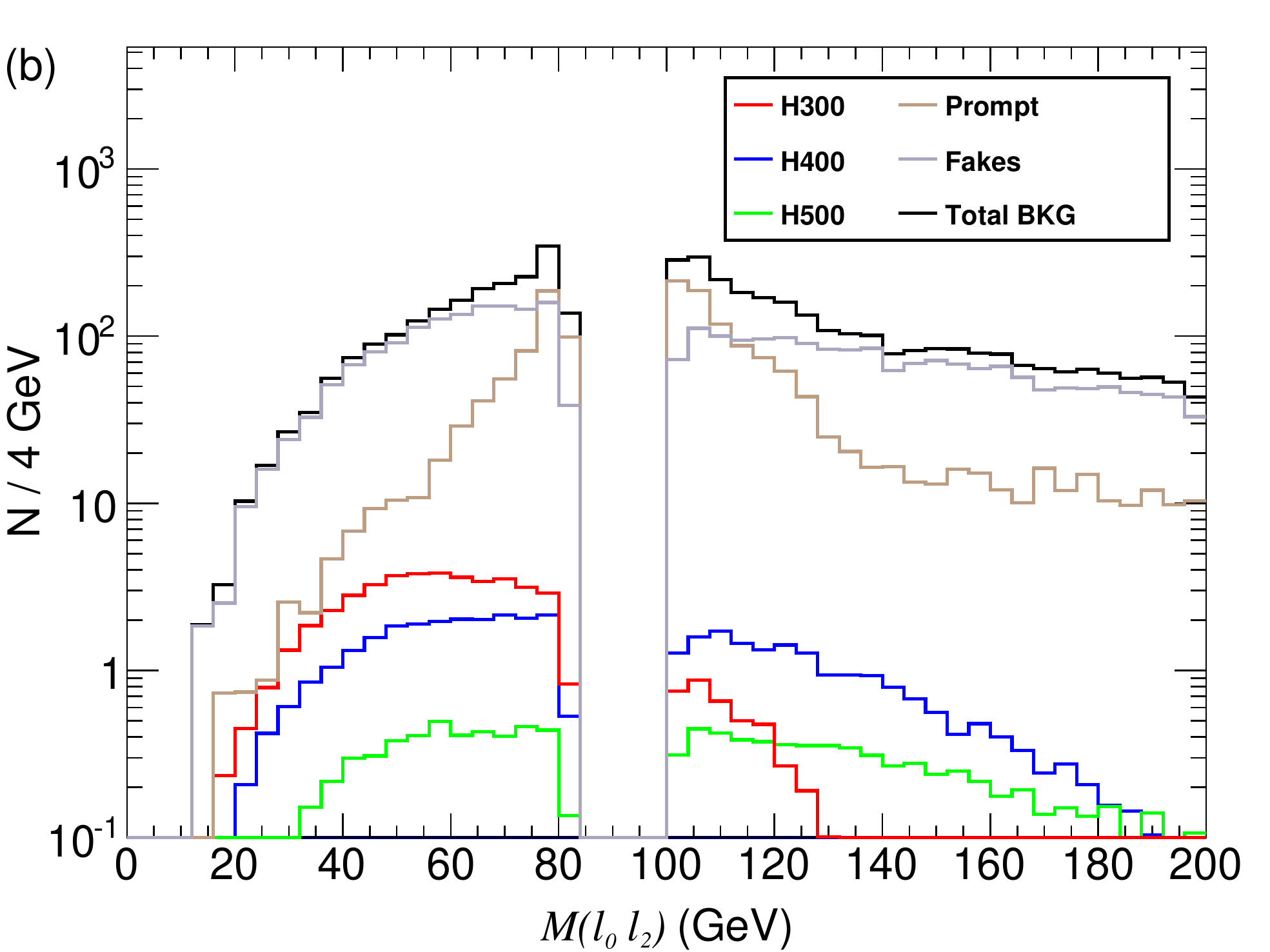}
%\label{subf:3l_m_l0l2}
\\[1mm]
\includegraphics[height=5.2cm,width=0.48\textwidth]{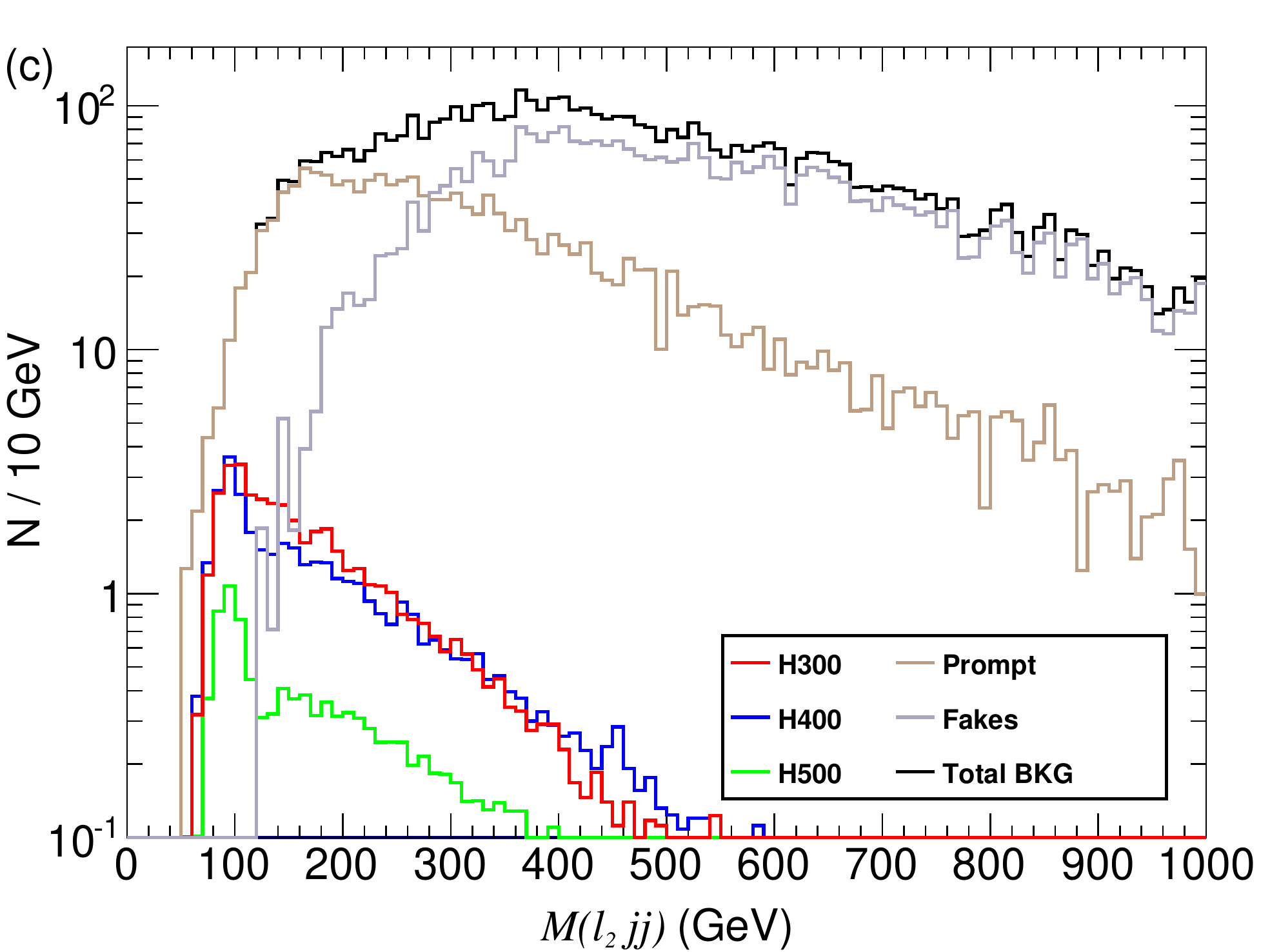}
%\label{subf:3l_m_l2jj}
\hspace*{-3mm}
\includegraphics[height=5.2cm,width=0.48\textwidth]{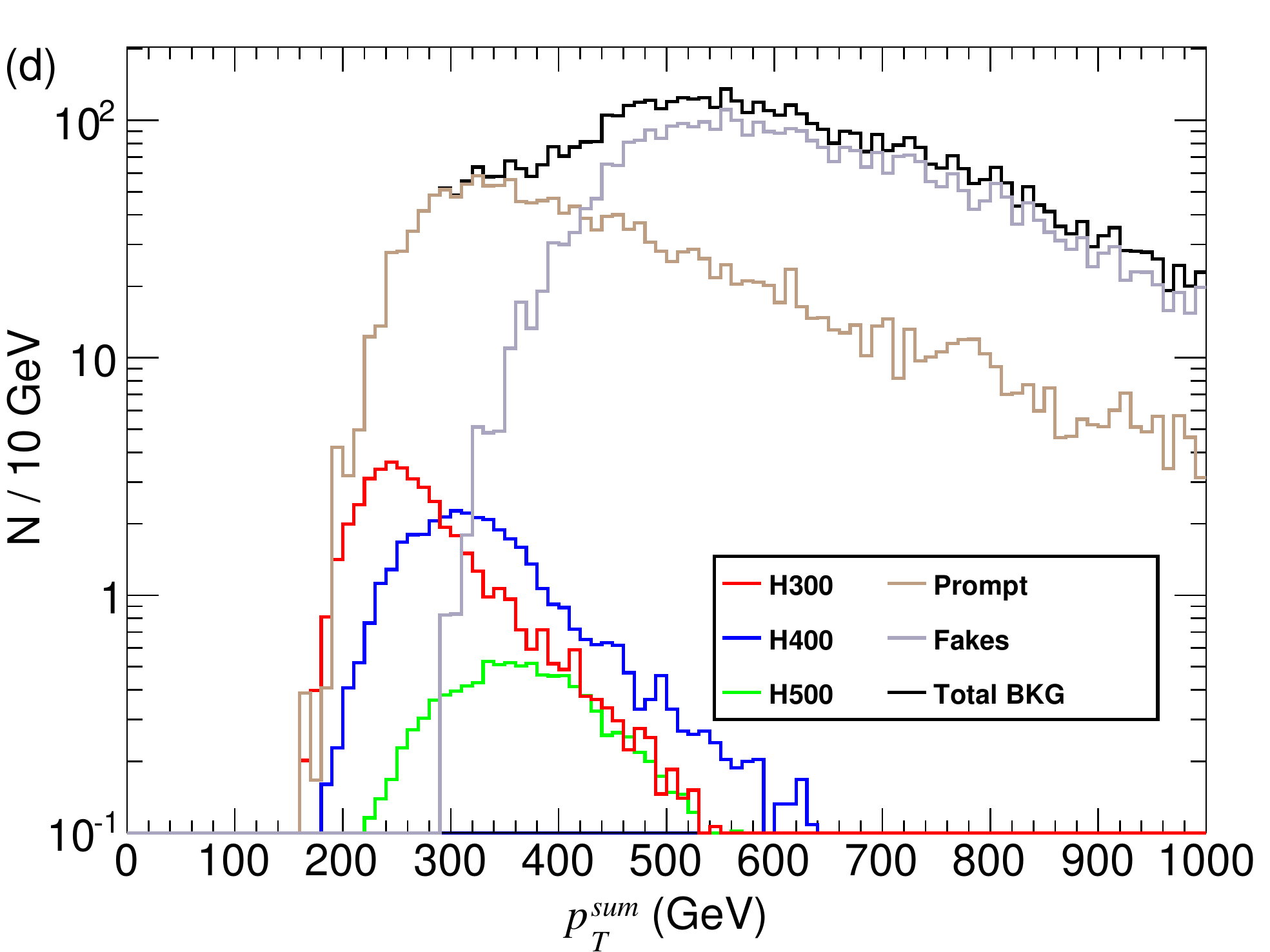}
%\label{subf:3l_pt_sum}
\\[1mm]
\includegraphics[height=5.2cm,width=0.48\textwidth]{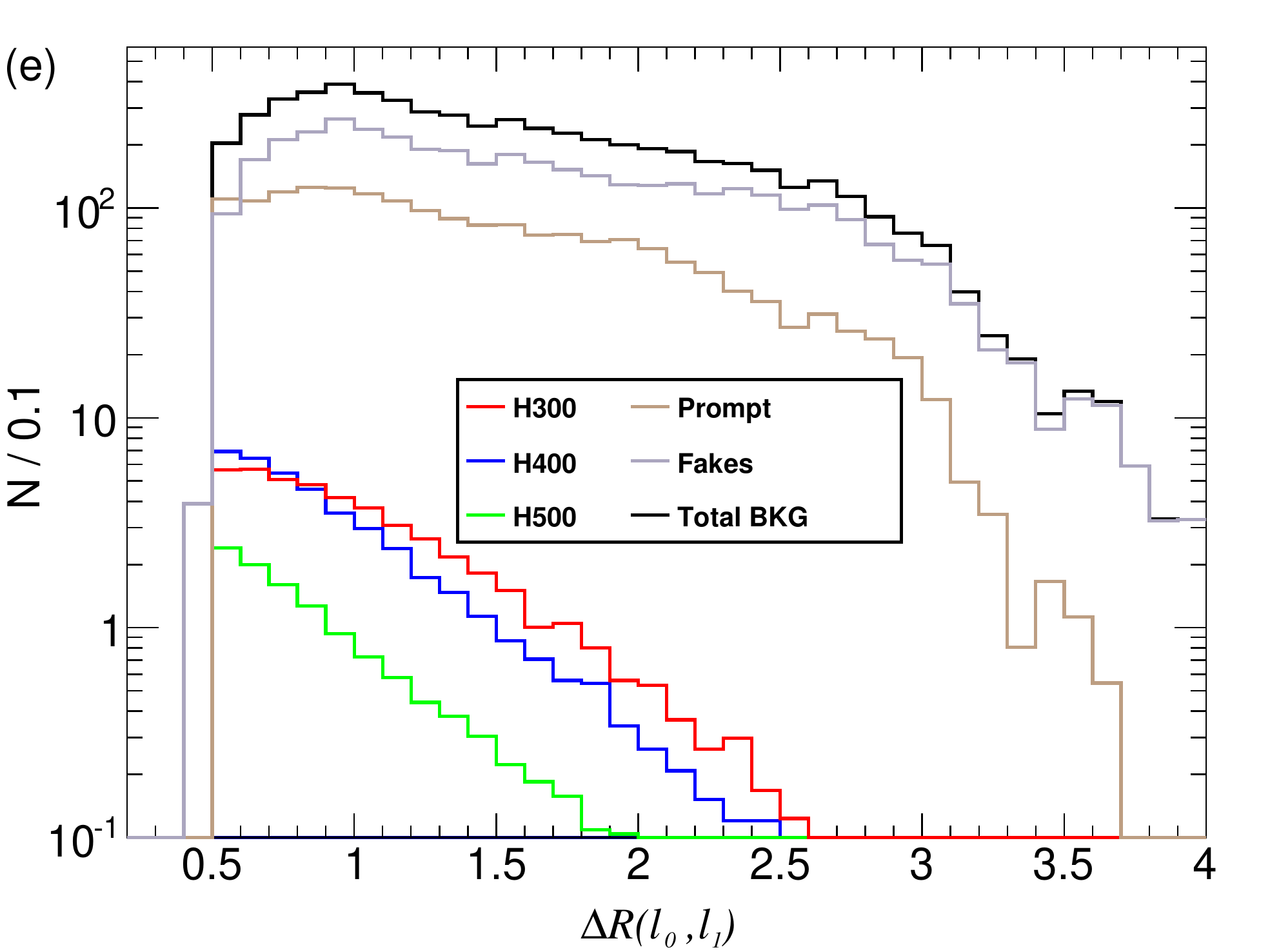}
%\label{subf:3l_dR_l0l1}
\hspace*{-3mm}
\includegraphics[height=5.2cm,width=0.48\textwidth]{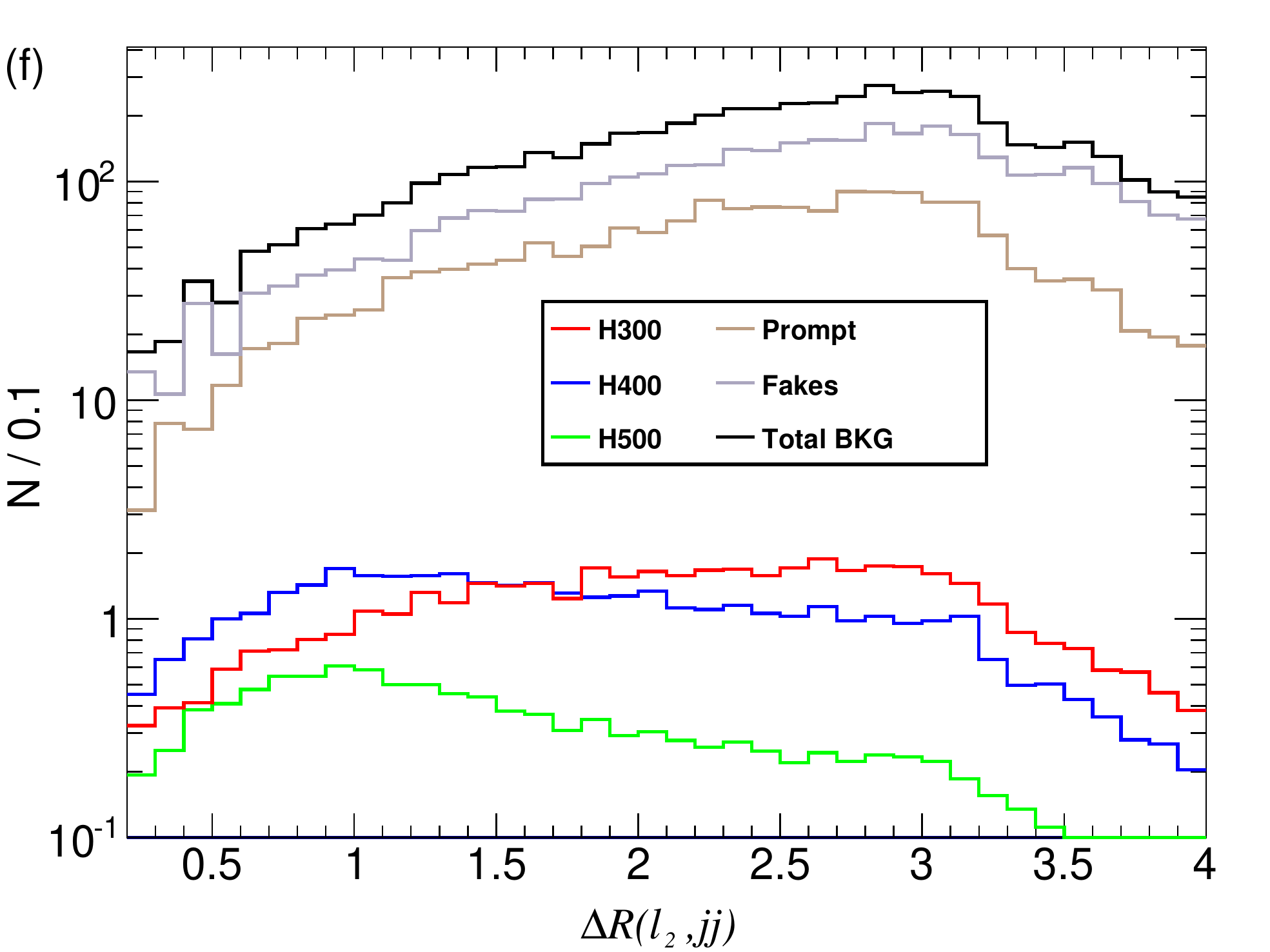}
%\label{subf:3l_dR_l2jj}
\vspace*{-2mm}
\caption{Distributions of different kinematical variables in the 3L decay channel:
(a) invariant mass of $\ell_0$ and $\ell_1$,
(b) invariant mass of $\ell_0$ and $\ell_2$,
(c) invariant mass of $\ell_2$ and selected jets,
(d) linear $p_T$ sum of two leptons, selected jets and transverse missing energy,
(e) $\Delta R$ distance of $\ell_0$ and $\ell_1$,
(f) $\Delta R$ distance of $\ell_2$ and selected jets system.
For these plots, we input an integrated luminosity of 300fb$^{-1}$.}
\label{fig:kine-3l}
\label{fig:6}
\label{fig:77}
\label{fig:n8}
\end{figure}
\begin{table}[t]
\begin{center}
\small
\begin{tabular}{c|c|ccccccc}
\hline
\hline
Process & $\mathcal{L}$  &$t\bar{t}W$ & $ZZ$ & $t\bar{t}h$ & Wh & WZ & $t\bar{t}Z$  \\
\hline
H300 &300 &4.6$\pm$0.6    &0.8$\pm$0.2    &3.3$\pm$0.4    &8.3$\pm$0.3    &7.7$\pm$1.5    &0.06$\pm$0.02  \\
\hline
H400 &1000 &18.0$\pm$2.2    &2.6$\pm$0.5   &10.4$\pm$1.1    &23.8$\pm$1.0    &23.7$\pm$4.7    &0.10$\pm$0.06    \\
\hline
%H500 &3000 & 17.2$\pm$3.7   &1.6$\pm$0.7    &9.4$\pm$1.8    &17.2$\pm$1.4    &8.5$\pm$4.9 & 0.10$\pm$0.10   \\
H500 &3000 &9.4$\pm$2.7 &0.80$\pm$0.46  &3.6$\pm$1.2    &9.5$\pm$1.1 &2.8$\pm$2.8 &0$\pm$0 \\
\hline
\hline
Process & $\mathcal{L}$   &$WWW$ &$WWZ$ &$t\bar{t}$\,{\footnotesize (leptonic)}
			& Full Bkgnds & Signal & $S/\!\sqrt{S\!+\!B}$ & $\SZZ$ \\
\hline
H300 &300   &2.5$\pm$0.2      &0.48$\pm$0.05   & 0.4$\pm$0.3 &{28.2$\pm$1.7} &28.8$\pm$0.6 &3.8 & 4.8 \\
\hline
H400 &1000  &9.5$\pm$0.7   &1.7$\pm$0.2  & {0.6$\pm$0.6} &90.4$\pm$5.6  &62.3$\pm$1.3   & 5.0 & 6.0 \\
\hline
%H500 &3000  &10.8$\pm$1.2    &1.2$\pm$0.2   &0$\pm$0  & {66.2$\pm$6.7}  &34.6$\pm$1.0  & 3.4 & 3.9 \\
H500 &3000 &5.9$\pm$0.9    &0.7$\pm$0.2    &0$\pm$0   &32.7+-4.4 &28.1$\pm$0.8 &3.6 &4.4 \\
\hline
\hline
\end{tabular}
\end{center}
\vspace*{-4mm}
\caption{BDT optimization in the tri-lepton (3L) channel for the three benchmarks
with different integrated luminosity $\mathcal{L}$ (in the unit of fb$^{-1}$).
We present, after the optimal BDT cuts, the selected event numbers for the signals and backgrounds
(in which the $Z$+jets background is negligible and not listed here), as well as the median significance
$\mathcal{Z}\,$.}
\label{event_yield_3l_BDT}
\label{tab:6}
\label{tab:66}
\end{table}

In Fig.\,\ref{fig:kine-3l},
we present the distributions of useful kinematical variables after the pre-selection.
Using these we further implement the BDT optimizations in Table\,\ref{tab:6}.
By definition, $\ell_0^{}$ and $\ell_1^{}$ come from the decays of
one Higgs boson $h$, while $\ell_2^{}$ and the two closest jets
arise from the decays of the other Higgs boson $h$.\,
The invariant masses, $M(\ell_0^{}\ell_1^{})$\, and
\,$M(\ell_2^{}jj)$,\, are then controlled by the light Higgs mass $\Mh$.\,
The $\,\Delta R\,$ distances,\, $\Delta R(\ell_0^{}, \ell_1^{})$
and $\Delta R(\ell_2^{}, jj)$,\, tend to be smaller for increasing
$\,\MH$\, and are useful for the optimization.
In summary, $\,\Delta R(\ell_0^{}, \ell_1^{})$ and $\Delta R(\ell_2^{}, jj)$\,
have better powers of background suppression for the case of larger $\MH$,\,
while the rest are important for the case of lower $\MH$.

Using the same procedure, we can implement the BDT optimization in the analysis of
3L channel for the three benchmarks. Our results are summarized in Table\,\ref{event_yield_3l_BDT}.
The dominant prompt background $WZ$ gets efficiently suppressed by
the valuable variables constructed for $\ell_0^{}$ and $\ell_1^{}$.\,
Impressively, the fake leptons backgrounds become almost negligible after the optimal BDT cut.
With these, we have achieved higher sensitivity for the 3L channel
since it has less backgrounds than the SS2L channel.

\vspace*{1.3mm}

In passing, we also compare the multi-variable-analysis via
the BDT approach with the traditional cut-based method,
and show how the BDT approach can improve the sensitivities.
For this comparison, we will use the same set of optimization variables
in the two analyses.
For the decay channels 2LSS and 3L, we list the set of variables as follows,
%
%\begin{itemize}
%\item
\beq
\ba{rl}
\hspace*{-5mm}
\text{[2LSS]:}~ &
M(jj)^{W}, M(\ell_1^{},jj), M(\ell_2^{},jj),
p_T^{\text{sum}}, M(\ell_1^{}j), M(\ell_2j),
\Delta R(\ell_1^{},j), \Delta R(\ell_2^{},j);~~~
\\[2mm]
%\item
\hspace*{-5mm}
\text{[3L]:}~ &
M(\ell_0^{}\ell_1^{}), M(\ell_0^{}\ell_2^{}), M(\ell_2^{}jj),
p_T^{\text{sum}}, \Delta R(\ell_0^{},\ell_1^{}), \Delta R(\ell_2^{},jj).
%\end{itemize}
\ea
\eeq
As an illustration,
we optimize the significance with the two most sensitive variables,
$p_T^{\text{sum}}$ and $\Delta R(\ell_1^{},j)$ for the 2LSS channel,
and $p_T^{\text{sum}}$ and $\Delta R(\ell_0^{},\ell_1^{})$
for the 3L channel. %instead of full optimization with all variables.
We summarize the results in
Table\,\ref{tab:compareBDT} for the benchmarks
(H300,\,H400,\,H500), where we apply the optimal cuts
$\,p_T^{\text{sum}}<(300, 500, 600)$\,GeV and
$\Delta R(\ell_1^{},j)<(2.5, 1.5, 1.0)$ to the 2LSS channel,
and $\,p_T^{\text{sum}}<(300, 500, 600)$\,GeV and
$\Delta R(\ell_0^{},\ell_1^{})<(2.0, 1.5, 1.2)$ to the 3L channel.
We see that the BDT optimization gains about $(16-64)\%$ increase of significance
over the cut-based method.
\begin{table}[t]
%\scriptsize
\begin{tabular}{c|ccc||c|ccccccccccc}
\hline\hline
2LSS  & Cut-based & BDT  & R(BDT/Cut) &
3L & Cut-based & BDT & R(BDT/Cut)
\\
\hline
~H300~ & 2.4 & 2.9 & 1.21 &
~H300~ & 3.2 & 3.8 & 1.19
\\
H400 & 3.6 & 4.2 & 1.16 &
H400 & 3.7 & 5.0 & 1.35
\\
H500 & 1.9 & 2.4 & 1.26 &
H500 & 2.2 & 3.6 & 1.64
\\
\hline\hline
\end{tabular}
\caption{Comparison of the significance $S/\!\sqrt{S\!+\!B\,}$
for the cut-based analysis and the BDT approach. In the 4th and 8th
columns, the quantity R(BDT/Cut) equals the ratio of the significance of
the BDT approach over that of the cut-based analysis.}
\label{tab:compareBDT}
\label{tab:8}
\end{table}
%

%%%%%%%%%%%%%%%%%%%%% Sec.3.4 %%%%%%%%%%%%%%%%%

%\vspace*{2.5mm}
\subsection{\hspace*{-2mm}Combination of the SS2L and 3L Decay Channels}
\vspace*{2mm}
\label{sec:3.4}

According to the above systematical analyses of the signals and backgrounds
in the SS2L and 3L decay channels for the three Higgs benchmarks,
we will study the combined sensitivity of both
the SS2L and 3L channels for detecting the new Higgs boson $H^0$ in this subsection.
To handle the relatively small event number, we will use the median significance
$\,\SZZ\,$\,\cite{Z0}. For estimating the sensitivity of future experiments,
we use the following median significance for the discovery reach
under the background-only hypothesis\,\cite{Z0},
\beqa
\label{eq:Z>5}
\mathcal{Z} \,=\, \sqrt{2\!\left[\!(S\!+\!B)\!\ln\!
\frac{\,S\!+\!B\,}{B} - S\right]\,}
\,\geqq 5\,,
\eeqa
while for the exclusion, we use the formula
under the background with signal hypothesis\,\cite{Z0},
\beqa
\label{eq:Z>2}
\mathcal{Z} \,=\, \sqrt{2\!\left(\!B\ln\!
\frac{\,B\,}{\,S\!+\!B} + S\right)\,}
\,\geqq 2\,.
\eeqa
For the case of $\,B\gg S\,$,\, we can expand the formulas
\eqref{eq:Z>5} and \eqref{eq:Z>2} in terms of $\,S/B\,$ or
$\,S/(S\!+\!B)\,$,\, and find that they both reduce to the form
of $\,\SZZ =S/\!\sqrt{B}\,$ or $\,\SZZ =S/\!\sqrt{S\!+\!B\,}\,$,\, as expected.

Using our results from the SS2L and 3L decay channels as presented in
Table\,\ref{tab:4} and Table\,\ref{tab:6},
we combine the significance $\,\SZZ\,$ from Table\,\ref{event_yield_2lss_BDT}
and Table\,\ref{event_yield_3l_BDT},
\beqa
\SZZ_{\textrm{comb}}^{} \,=\,
 \sqrt{\SZZ_{\textrm{SS2L}}^2\!+\SZZ_{\textrm{3L}}^2\,}\,,\,
\eeqa
which is summarized in Table\,\ref{tab:combined_Z0} for the three Higgs benchmarks
in Eq.\eqref{eq:benchmark}.
In the fourth row of Table\,\ref{tab:2},
we derive the combined significance
$\,\SZZ_{\textrm{comb}}^{}$ of the benchmarks (H300,\,H400,\,H500) under
sample inputs of the corresponding integrated luminosity
$\,\mathcal{L}=(300,\,1000,\,3000)\,$fb$^{-1}$, respectively.
In the second row of this table, we also present
the required minimal integrated luminosity
$\,\mathcal{L}_{\min}^{5\sigma}$\, to reach the significance
$\,\SZZ_{\textrm{comb}}^{}\!=5\,$
for each benchmark, by combining the LHC searches in both SS2L and 3L channels.
We see that given an integrated luminosity of 446\,fb$^{-1}$,\, the LHC\,(14TeV)
can reach a discovery of the heavier Higgs state $H^0$ with mass up to 400\,GeV
via the di-Higgs channel $H^0\!\to\! h^0h^0\!\to\!4W$.
\begin{table}[t]
\vspace*{-6mm}
\begin{center}
\begin{tabular}{c|ccc}
\hline
\hline
~Benchmarks~ & ~H300~ & ~H400~ & ~H500~ \\
\hline
\\[-4.4mm]
$\mathcal{L}_{\min}^{5\sigma}$\,(fb$^{-1}$\!)
& 222 & 446 & 2954
\\[-4.4mm]
\\
\hline
\\[-4.4mm]
$\mathcal{L}$\,(fb$^{-1}$\!) &300 &1000 &3000 \\
$\SZZ_{\textrm{comb}}$ & 5.8 & 7.5 & 5.0 \\
\hline
\hline
\end{tabular}
\end{center}
\vspace*{-3.7mm}
\caption{Combined significance $\SZZ_{\text{comb}}^{}$
% for the three benchmarks of Eq.(\ref{eq:benchmark})
at the LHC\,(14TeV).}
\label{tab:combined_Z0}
\label{tab:2}
\label{tab:77}
\end{table}

\vspace*{1.5mm}

In passing, we have also compared the significance of our diHiggs decay channel
$hh\!\to\!4W$ with another channel
$hh\!\to\!b\bar{b}WW^*$ in \cite{LMPHhh}\cite{bbWW}
which studied the SM extension with a real singlet scalar $S$.
After proper rescaling, we find that for $M_H^{}< 400$GeV, the $4W$ channel
has better sensitivity than the $b\bar{b}WW$ channel\,\cite{LMPHhh}\cite{bbWW}
for detecting the heavy Higgs $H^0$,
while for $M_H^{}\gtrsim 400$GeV, the $b\bar{b}WW$ channel study
in \cite{bbWW} has higher sensitivity.

%%%%%%%%%%%%%%%%%%%%%%%%%%%%%%%%%%%%%%%%%%%%%%
%%============ Section 4 ===================%%
%%%%%%%%%%%%%%%%%%%%%%%%%%%%%%%%%%%%%%%%%%%%%%
\vspace*{3mm}
%\newpage
\section{\hspace*{-2mm}Probing the Parameter Space of 2HDM}
\label{sec:4}

In this section, we analyze the probe of the 2HDM parameter space
by searching the heavy Higgs resonant production in the di-Higgs channel
$\,gg\!\to\! H^0\!\!\to\! hh\!\to\! WW^*WW^*$\, at the LHC\,(14TeV).
For this, we shall combine the analyses of both SS2L and 3L channels
in Sec.\,\ref{sec:3}.
We further take into account both the theoretical constraints and the current
experimental bounds as studied in Sec.\,\ref{sec:2}.

\vspace*{1mm}

\begin{figure}[t]
\vspace*{3mm}
\centering%
{\includegraphics[height=6cm,width=7.2cm]{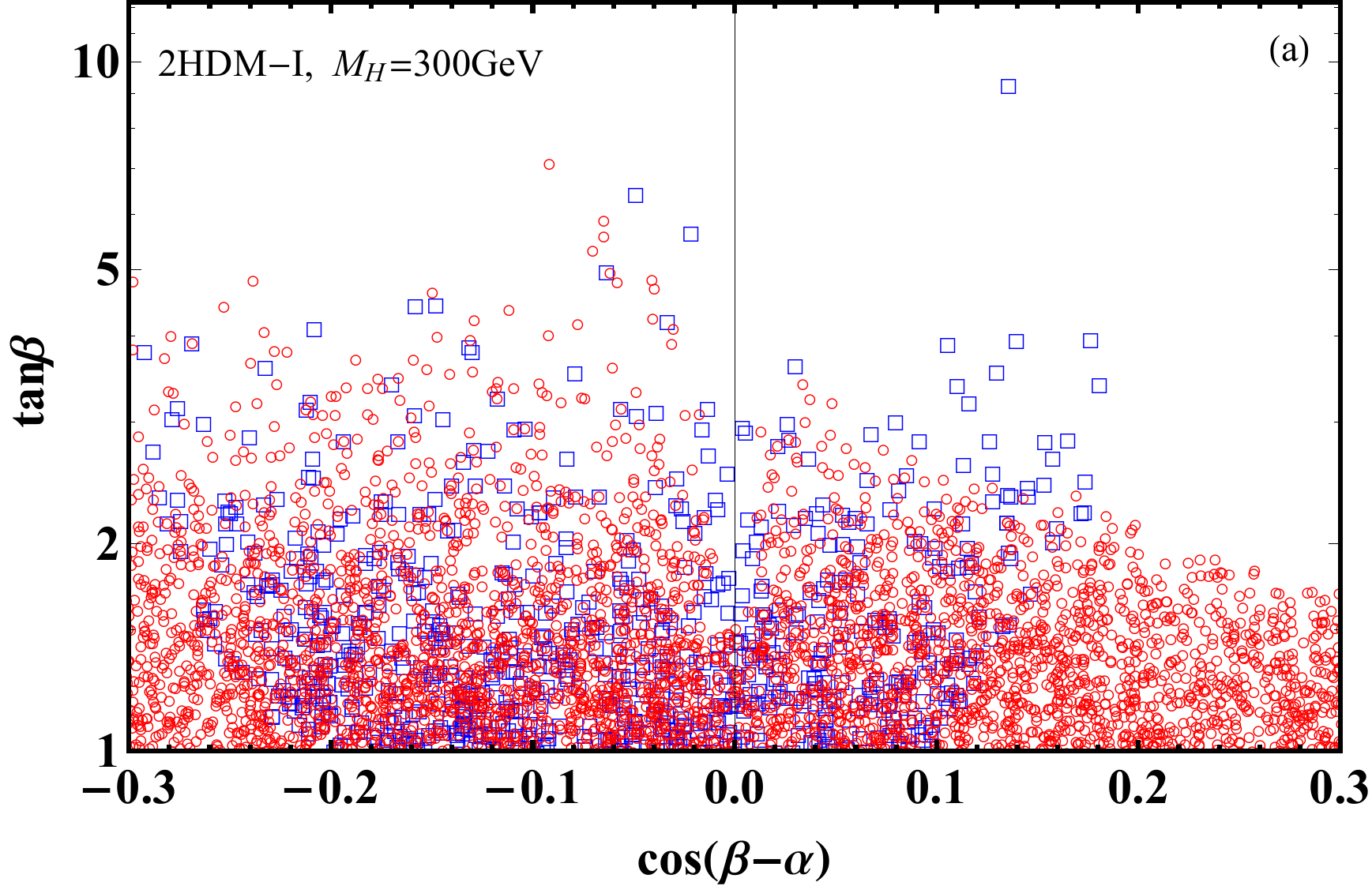}}
\hspace*{1mm}
{\includegraphics[height=6cm,width=7.2cm]{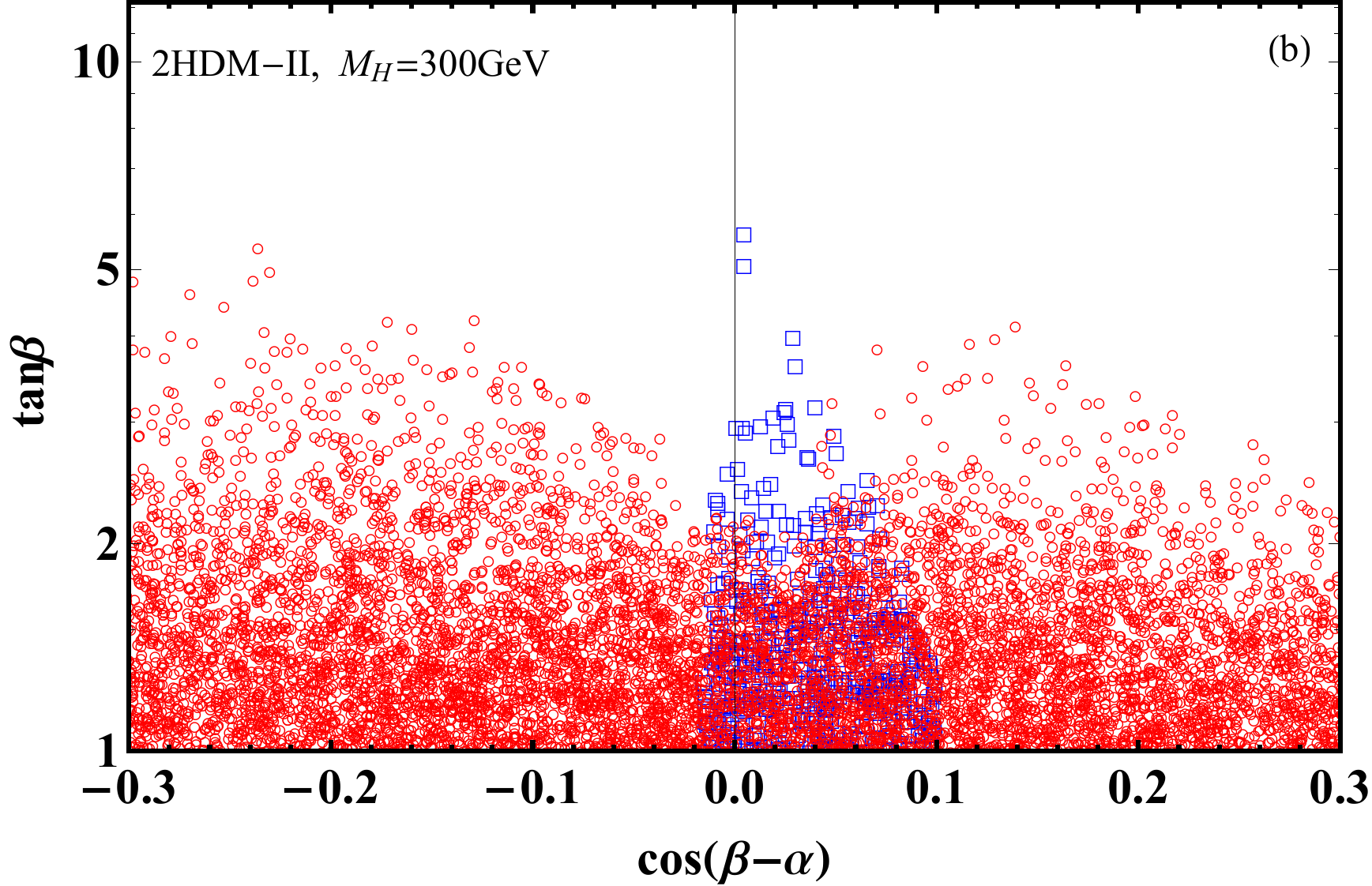}}
\\[2.5mm]
{\includegraphics[height=6cm,width=7.2cm]{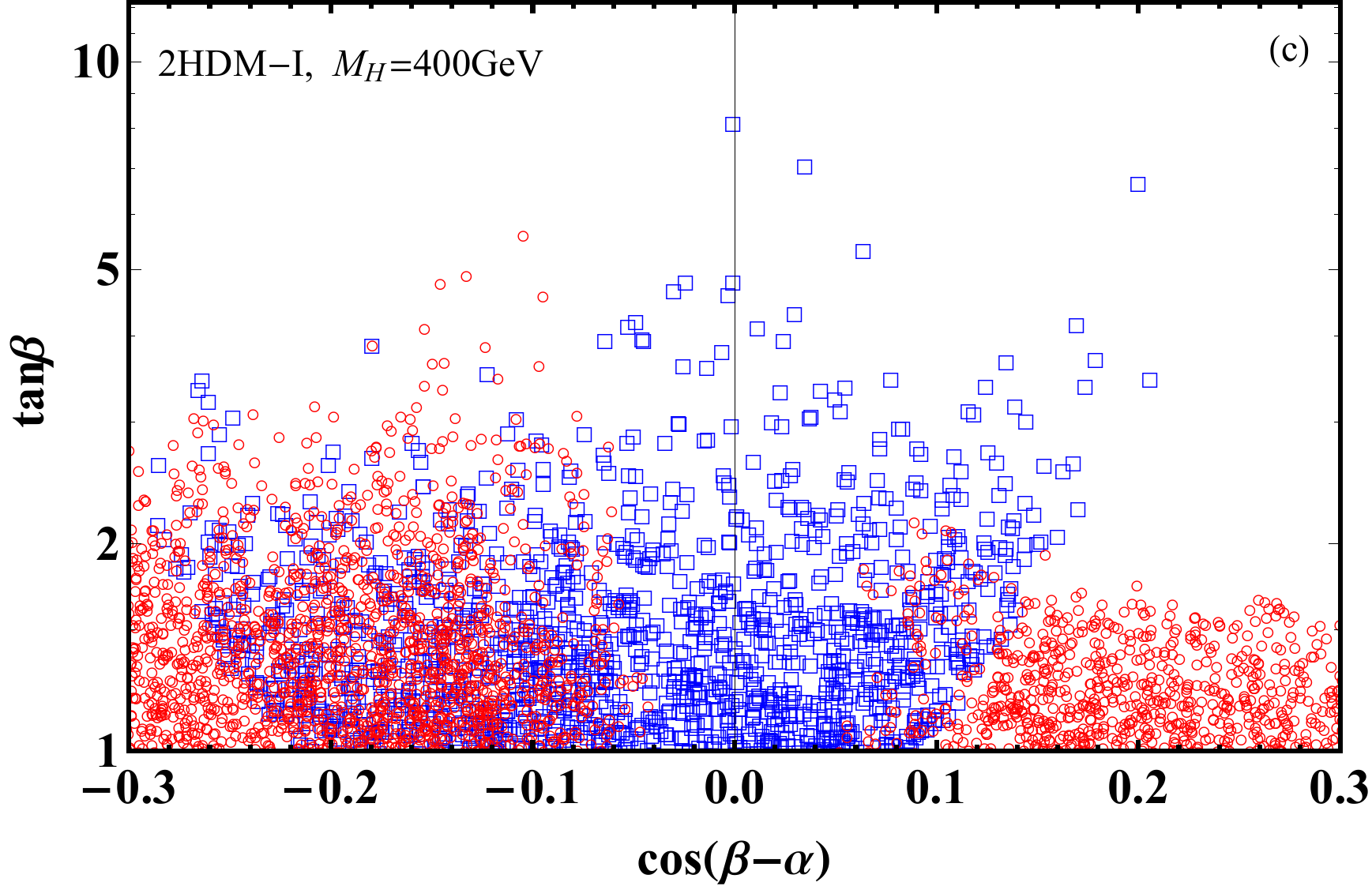}}
\hspace*{1mm}
{\includegraphics[height=6cm,width=7.2cm]{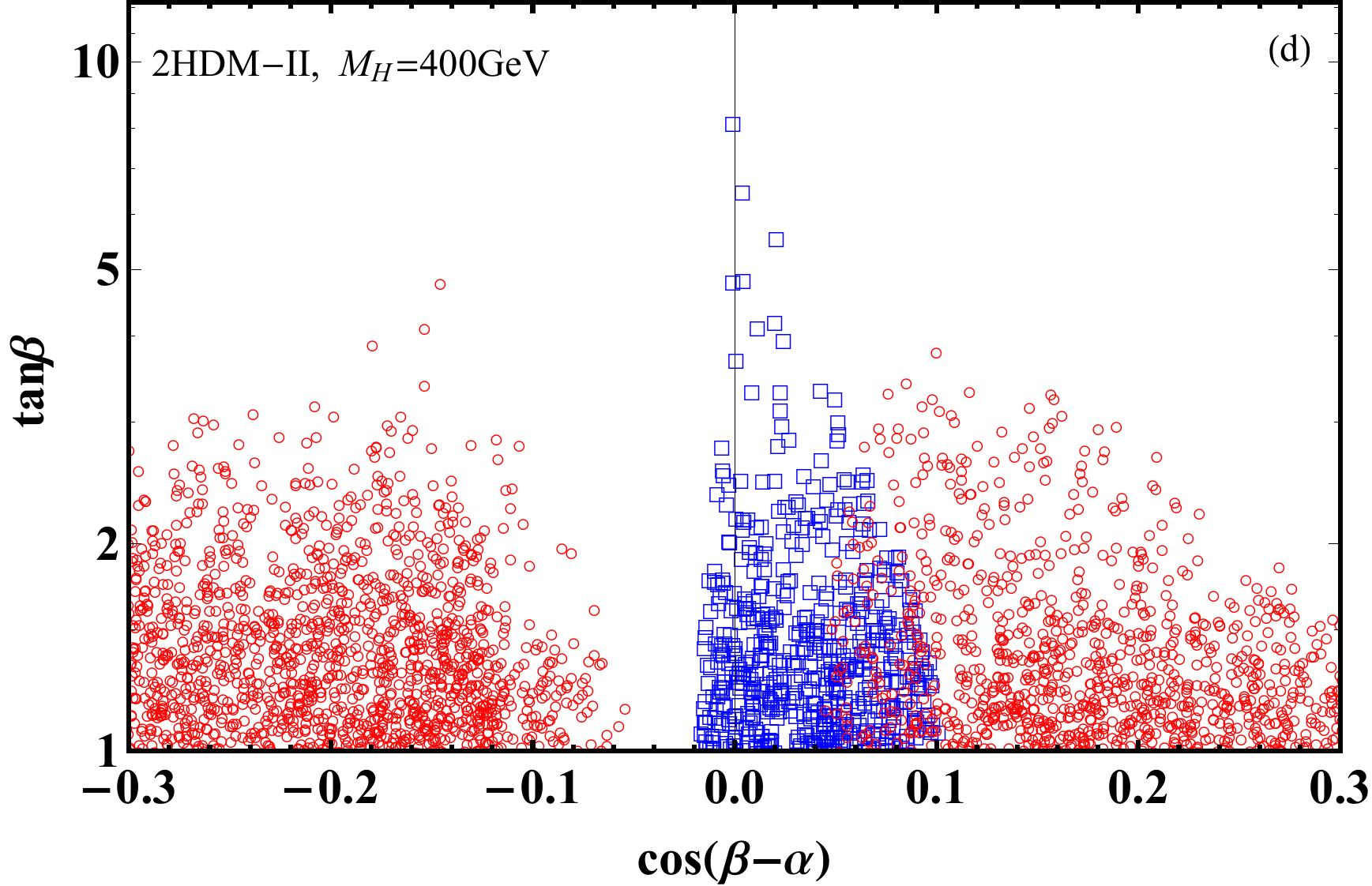}}
\vspace*{-3mm}
\caption{Parameter space in $\cos(\alpha\!-\!\beta)-\ln(\tanb)$ plane
for the sample inputs of heavier Higgs boson mass,
$\,\MH =300$\,GeV [plots (a)-(b)] and
$\,\MH =400$\,GeV [plots (c)-(d)],
and for 2HDM-I [plots (a),(c)] and 2HDM-II [plots (b),(d)].
The blue dots (square shape) satisfy the theoretical constraints,
the electroweak precision limits and the $h(125\text{GeV})$ global fit.
The red dots (circle shape) present the parameter region which can be probed
by the LHC direct searches of the heavier Higgs boson $H^0$, including the existing
heavy Higgs search bounds and our study of
$\,gg\!\to\! H^0\!\!\to\! hh\!\to\! 4W$\, searches at
the LHC Run-2 with $\mathcal{L}=300$\,fb$^{-1}$
(combined with the theoretical constraints).
All these bounds are shown for a significance $\SZZ\geqq 2$\,.
}
\label{fig:PSallLL}
\label{fig:7}
\label{fig:88}
\label{fig:n9}
%\vspace*{5mm}
\end{figure}

In Fig.\,\ref{fig:7},
we present the projection of parameter scan in the plane of
$\,\cos(\alpha\!-\!\beta)-\tan\beta$\,  for the sample inputs of
the heavier Higgs mass $\,\MH =300$\,GeV [plots (a)-(b)] and
$\,\MH =400$\,GeV [plots (c)-(d)], and for
the 2HDM-I [plots (a),(c)] and 2HDM-II [plots (b),(d)], respectively.
The blue dots (square shape) present the allowed parameter region satisfying
the theoretical constraints and the indirect experimental bounds
(including the electroweak precision limits and the LHC global fit of the
SM-like Higgs boson $h$(125GeV),
as we discussed in Sec.\,\ref{sec:2}).
The red dots (circule shape) represent the parameter space which can be probed
by the direct heavy Higgs searches at the LHC
with a significance $\,\SZZ\geqq 2\,$, including
the existing heavy Higgs search bounds and our study of
$\,gg\!\to\! H^0\!\to\! hh\!\to\! 4W\,$ searches
(combined with the theoretical constraints).
For the heavy Higgs searches via $4W$ channel, we derive the expected sensitivity
from the LHC Run-2 with an integrated luminosity
$\,\mathcal{L}=300\,\textrm{fb}^{-1}$.\,

\vspace*{1mm}

From Figs.\,\ref{fig:88}(a) and \ref{fig:88}(b),
we see that for both 2HDM-I and 2HDM-II with $\,\MH =300\,$GeV,
the direct searches of $H^0$ at the LHC Run-1 and Run-2 can substantially
probe the parameter space towards the alignment limit
(represented by the regions with red dots),
and largely cover the viable parameter space allowed by the current indirect searches
(shown by the regions with blue dots).
We note that the red dots can cover sizable regions around the alignment limit
$\,\cosba =0\,$.\, This is because the existing direct searches of $H^0$ at
the LHC Run-2 via $\,H\!\to\!\tau\tau$\, channel\,\cite{tataATLAS}\cite{tataCMS}
already give nontrivial bounds for the case of $\,\MH =300\,$GeV
(cf.\ Table\,\ref{tab:2}), where the Yukawa coupling
$H\tau\tau$ does not depend on $\,\cosba\,$ as shown in Table\,\ref{tab:1}.
The cases with a heavier Higgs mass such as $\,\MH =400\,$GeV become much harder since
Figs.\,\ref{fig:88}(c)-(d) still have significant viable parameter regions
(with blue dots) not covered by the direct heavy Higgs searches
(represented by the red dots).
We also see that the direct heavy Higgs searches can probe the
viable parameter region up to $\,\tanb\simeq 5$\, for 2HDM-I
and $\,\tanb\simeq 3$\, for 2HDM-II.
Given the lower capability of the LHC Run-2 for
detecting $H^0$ in the mass range $\,\MH\gtrsim 400\,$GeV,
we expect that more sensitive direct probes should be achieved at the HL-LHC.
\begin{figure}[t]
%\vspace*{-3mm}
\centering%
\includegraphics[height=5.8cm,width=7.2cm]{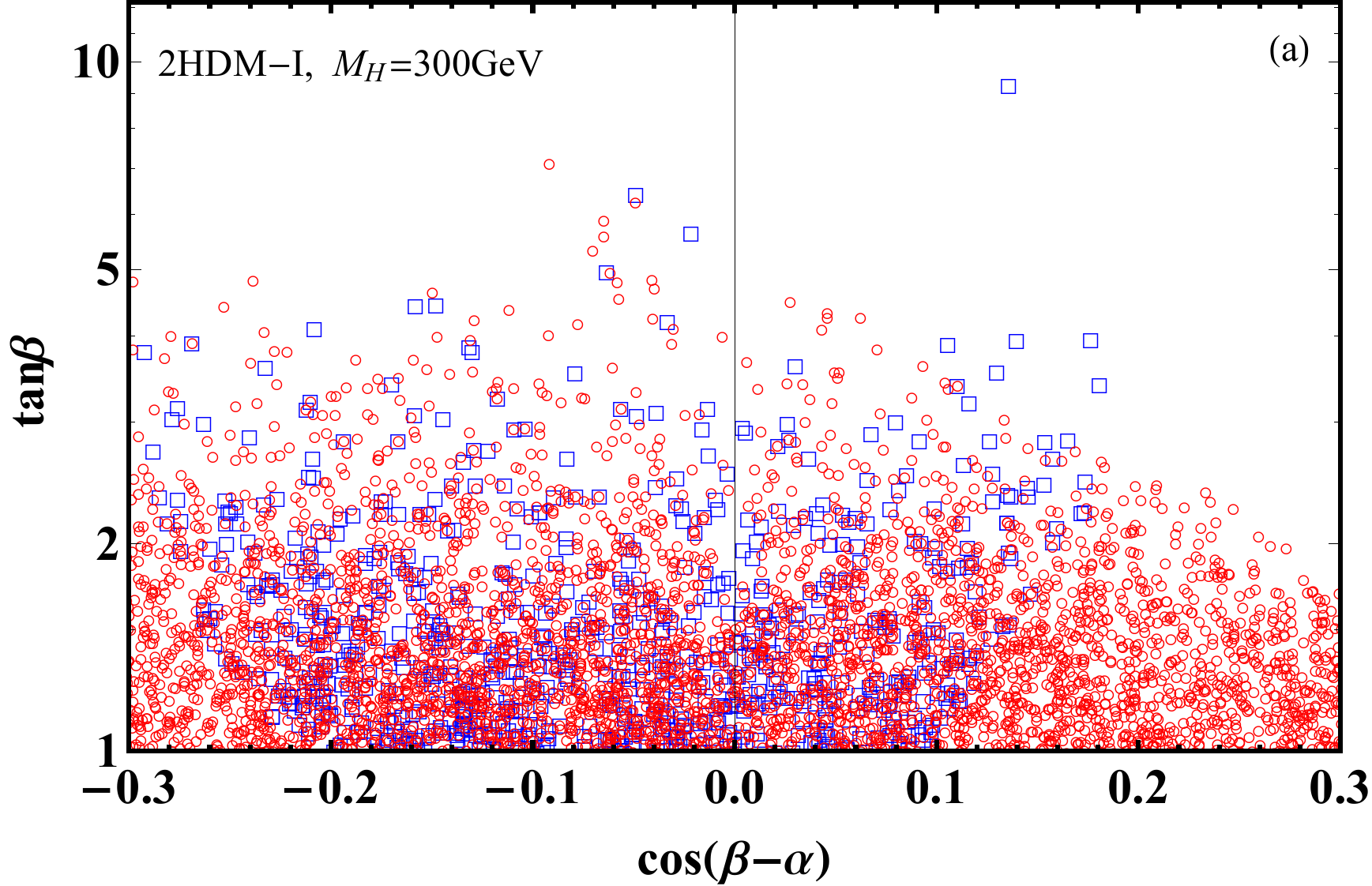}
\includegraphics[height=5.8cm,width=7.2cm]{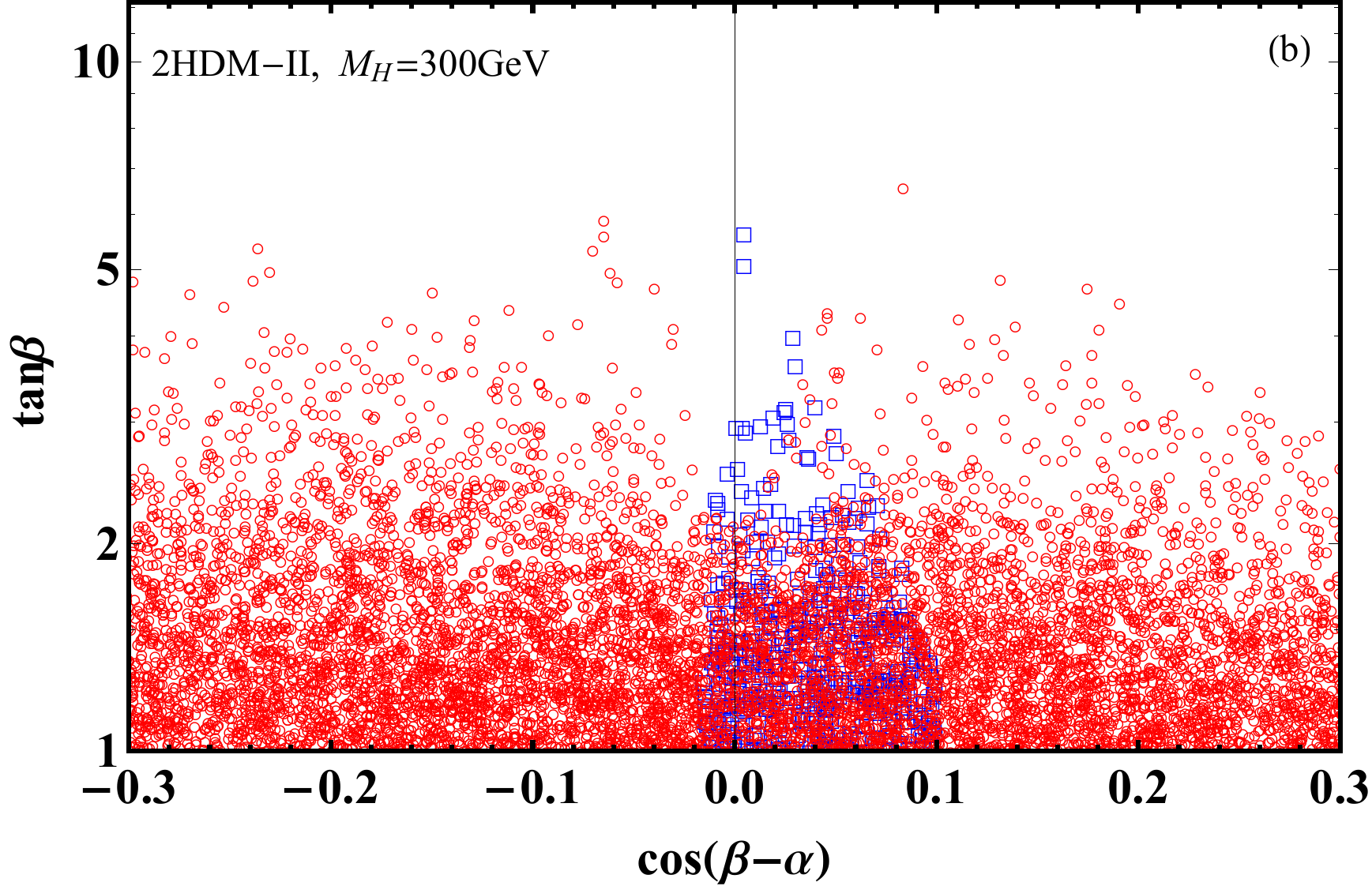}
\\[2mm]
\includegraphics[height=5.8cm,width=7.2cm]{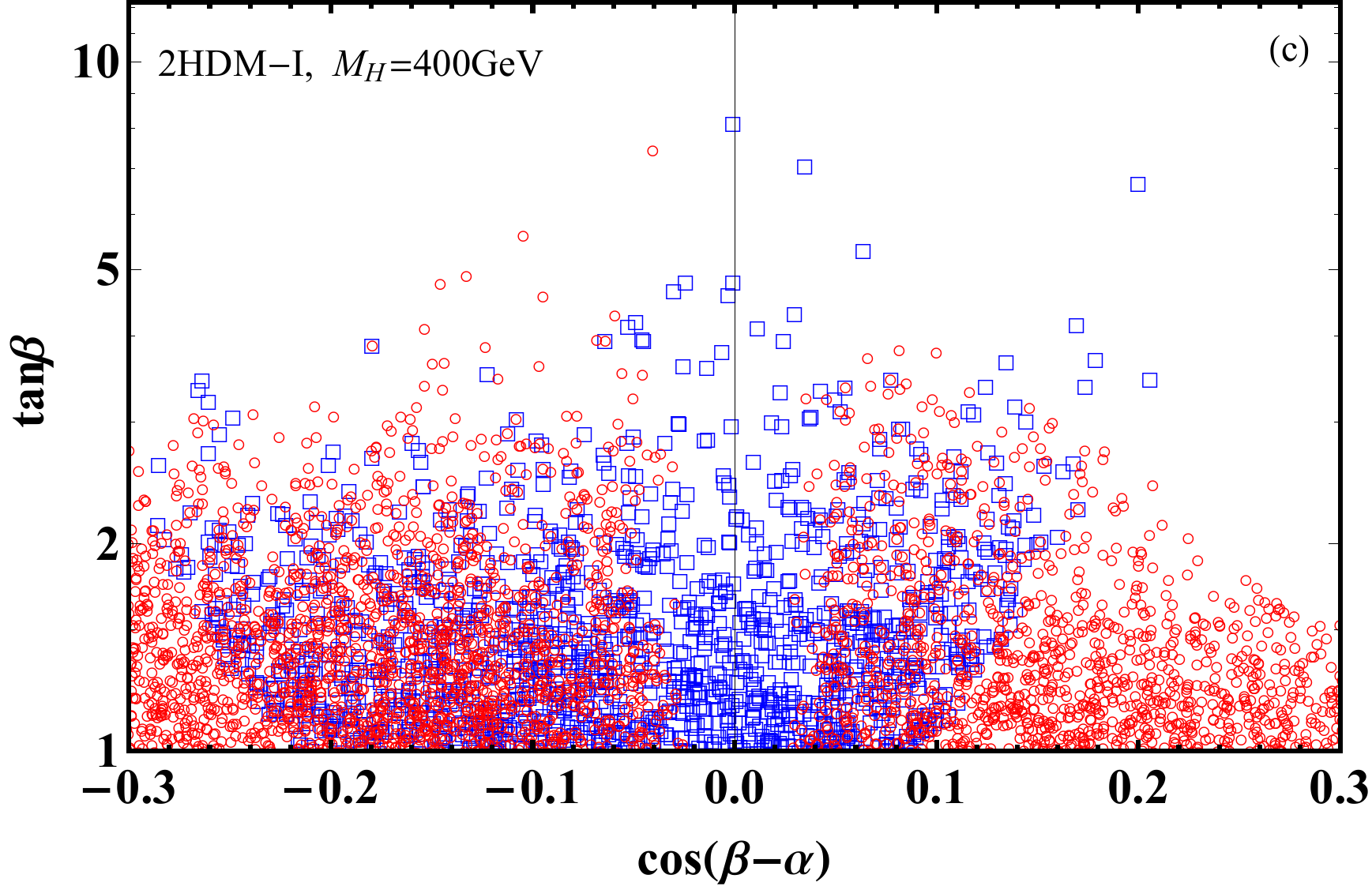}
\includegraphics[height=5.8cm,width=7.2cm]{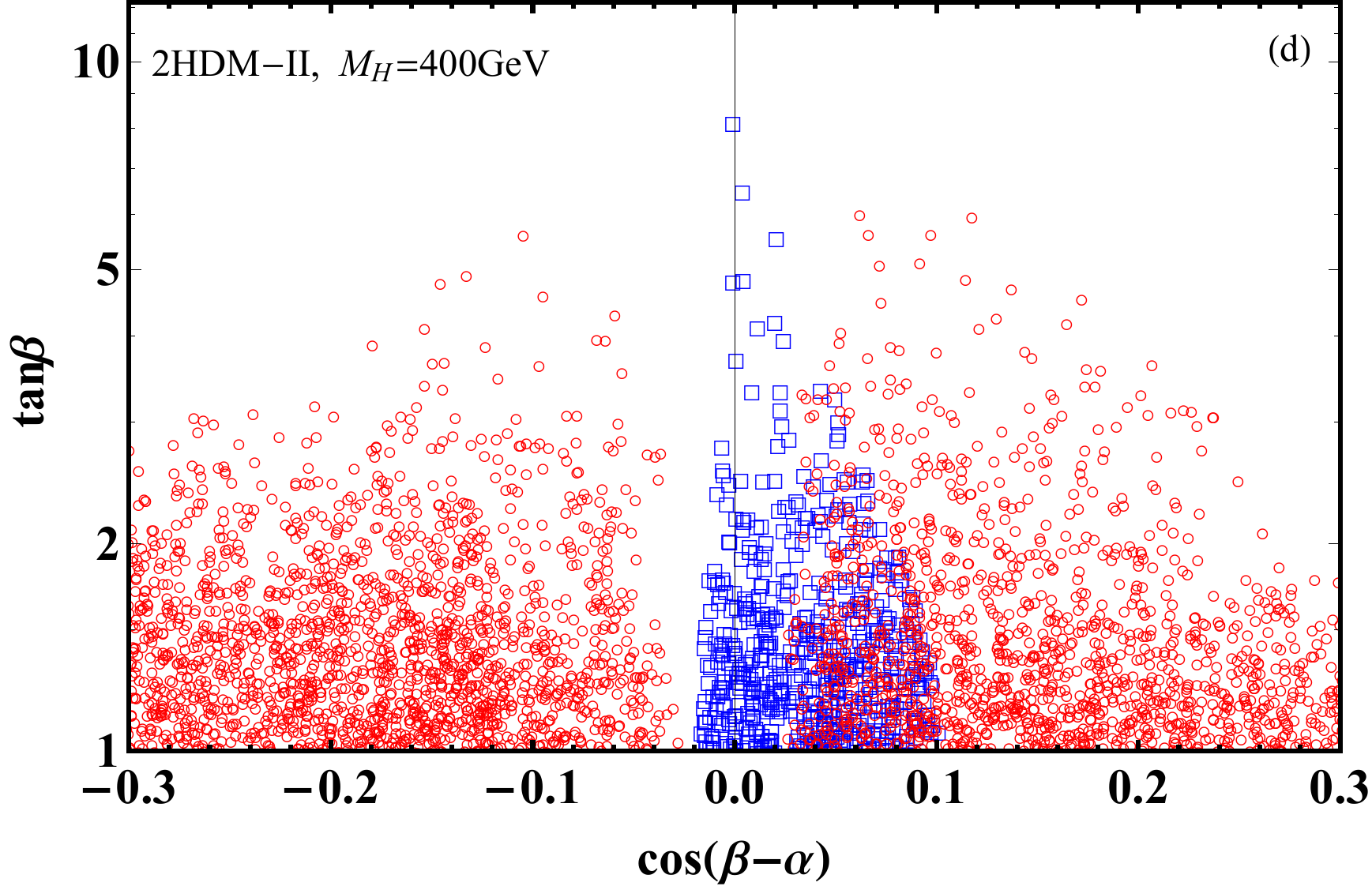}
\\[2mm]
\includegraphics[height=5.8cm,width=7.2cm]{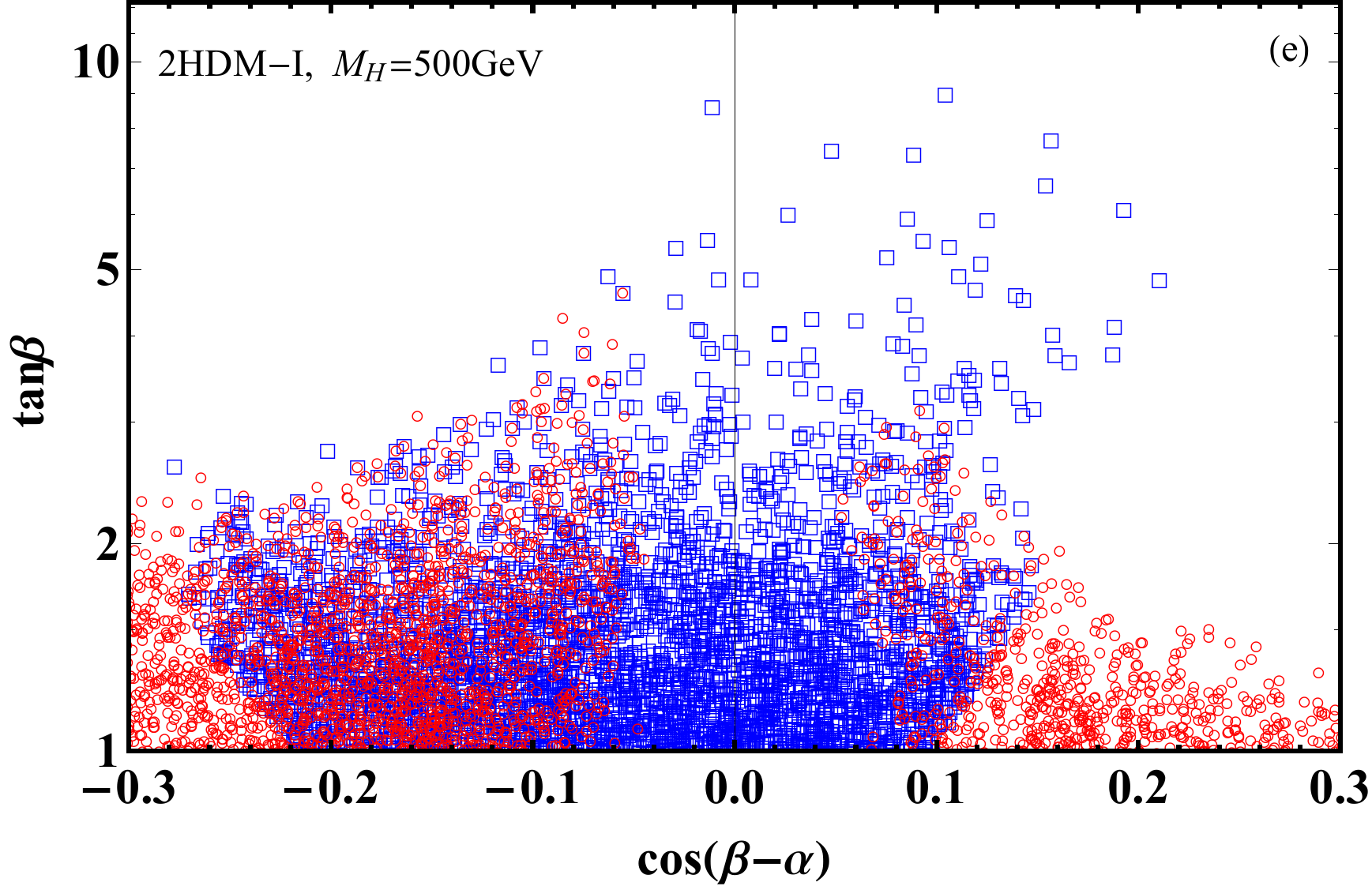}
\includegraphics[height=5.8cm,width=7.2cm]{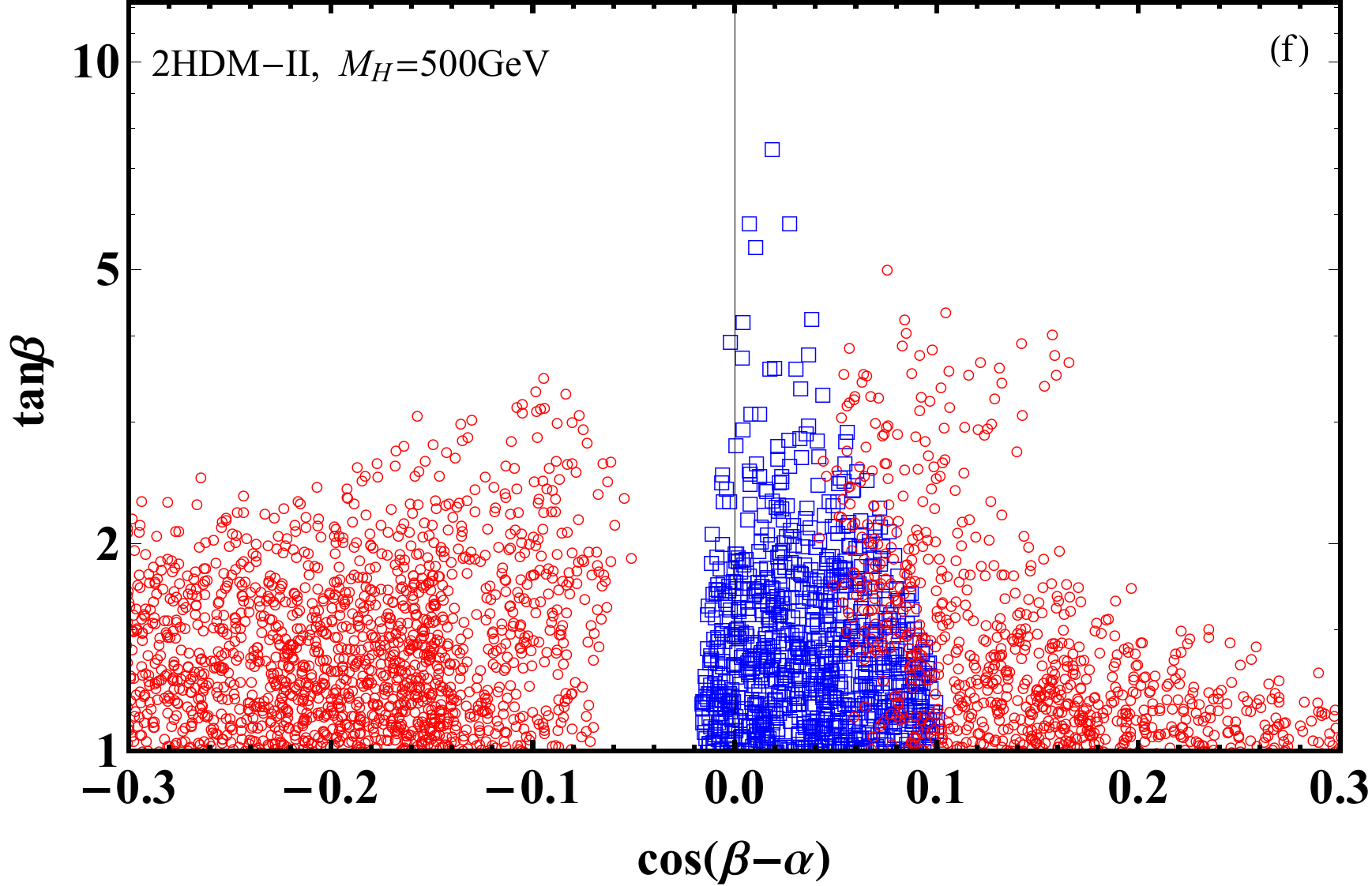}
\vspace*{-2.5mm}
\caption{Parameter space in $\cos(\alpha\!-\!\beta)-\ln(\tanb)$ plane
for the sample inputs of heavier Higgs boson mass,
$\,\MH =300$\,GeV [plots (a)-(b)],
$\,\MH =400$\,GeV [plots (c)-(d)],
$\,\MH =500$\,GeV [plots (e)-(f)],
and for 2HDM-I [plots (a),(c),(e)] and 2HDM-II [plots (b),(d),(f)].
The parameter region with red dots (circle shape) can be probed by
the LHC direct searches of the heavier Higgs boson $H$,
including the existing heavy Higgs search bounds and our study of
$\,gg\!\to\! H^0\!\!\to\! hh\!\to\! 4W$\, searches at the HL-LHC(14TeV)
with $\mathcal{L}=3000$\,fb$^{-1}$
(combined with the theoretical constraints) for all cases.
As in Fig.\,\ref{fig:7}, the blue dots (square shape)
satisfy the theoretical constraints,
the electroweak precision limits and the Higgs global fit of $h(125\text{GeV})$.
All these bounds are shown for significance $\SZZ\geqq 2$.
}
\label{fig:PSallHL}
\label{fig:8}
\label{fig:99}
\label{fig:n10}
\vspace*{-9mm}
\end{figure}

\vspace*{1.5mm}

Next, we further extend our above analysis to the HL-LHC with an integrated
luminosity $\mathcal{L}=3000\,\textrm{fb}^{-1}$.
We present this in Fig.\,\ref{fig:99},
including all three sample inputs of the heavier Higgs boson mass,
$\,\MH=300\,$GeV in plots (a)-(b), $\,\MH=400\,$GeV in plots (c)-(d),
and $\,\MH=500\,$GeV in plots (e)-(f).
Here the blue dots present the viable parameter space allowed by the
current indirect constraints, which are the same as in Fig.\,\ref{fig:88}.
For the bounds from LHC direct searches of the heavy Higgs boson $H^0$,
we identify the parameter space with red dots in the same way as in Fig.\,\ref{fig:88},
except that we include the HL-LHC probe
by the resonant di-Higgs production in the $4W$ channel with
$\mathcal{L}=3000$\,{fb}$^{-1}$.\,
In the relatively low mass range such as  $\,\MH=300\,$GeV,\,
we see that the improvements are not so visible in Figs.\,\ref{fig:99}(a)-(b),
since there are little viable
regions left after the LHC Run-2 direct searches (with $\mathcal{L}=300\,\textrm{fb}^{-1}$).
But, for higher mass range such as the case of $\,\MH=400\,$GeV,\,
Figs.\,\ref{fig:99}(c)-(d) show that
the directly probed regions (with red dots) have been significantly expanded.
For the case of $\,\MH=500\,$GeV,\,
Figs.\,\ref{fig:99}(e)-(f) demonstrate that
the projected sensitivities (regions with red dots) via
the direct heavy Higgs searches, in comparison with the current indirect bounds
(regions with blue dots). Since the current direct heavy Higgs search limits
are rather weak for $\,\MH=500\,$GeV, in Figs.\,\ref{fig:99}(e)-(f)
the regions with red dots are mainly probed by
the direct search process $\,gg\!\to\! H^0\!\to\! hh\!\to\! 4W$\,
at the HL-LHC.  We see that the projected sensitivities
in the plots (e)-(f) are somewhat weaker than the case of $\,\MH=400\,$GeV\,
in the plots (c)-(d), but they are still comparable.
This shows the power of the HL-LHC runs for probing the 2HDM parameter space
with higher masses of the heavy Higgs boson $H^0$.\,
It is clear that the HL-LHC ($\mathcal{L}=3000\,\textrm{fb}^{-1}$)
has significantly increased the sensitivity to probing
the 2HDM parameter space via the $H^0\!\to\! hh\!\to\! 4W$ channel.

\vspace*{1mm}

We may compare the current analysis of
$\,gg\to H^0\!\to hh\to WW^*WW^*\,$
with our previous study for
$\,gg\to H^0\!\to hh\to\gamma\gamma WW^*$ (including
both the pure leptonic and semi-leptonic decays of the $WW^*$ final state)\,\cite{llc}.
We find that our combined sensitivity in the $WW^*WW^*$ channel is comparable
to that of the $\gamma\gamma WW^*$ channel.
Our present BDT optimization analysis in Sec.\,\ref{sec:3}
does help to improve the sensitivity as shown in Table\,\ref{tab:8}.
Even though the $WW^*WW^*$ channel is not the most sensitive search channel for
detecting of the heavier Higgs boson $H^0$ via  resonant dihiggs production,
it is important to study all possible di-Higgs production channels
which will allow a combined analysis of the $H^0$ discovery.
In passing, we note that for other extended Higgs sectors, there could be
a new singlet scalar $S^0$ beyond the SM Higgs doublet\,\cite{bbWW}\cite{SM-S}
or beyond the two Higgs doublets\,\cite{s}.
The scalar particle $S^0$ can mix with the Higgs states $(h^0,H^0)$,\,
so it will couple to quarks $\,t\bar{t}$ and $b\bar{b}\,$,\,
and to gauge bosons $WW$ and $ZZ$\,.\,
Thus, the present analysis can be also generalized to the
resonant channels $\,gg\!\to\! S^0 \!\to\! hh \!\to\! 4W$\,
and $\,gg\!\to\! H^0 \!\to\! SS,Sh \!\to\! 4W$.\,

%%%%%%%%%%%%%%%%%%%%%%%%%%%%%%%%%%%%%%%%%%%%%%
%%============ Section 5 ===================%%
%%%%%%%%%%%%%%%%%%%%%%%%%%%%%%%%%%%%%%%%%%%%%%
\vspace*{3mm}
\section{\hspace*{-2mm}Conclusions}
\label{sec:5}

It is a pressing task for the LHC Run-2 and the upcoming HL-LHC to search for
new Higgs state(s) beyond the light Higgs boson $h^0$(125GeV), which generally exists
in all extended Higgs sectors and would point to new physics beyond the standard model (SM)
without ambiguity.  The resonant di-Higgs production is an important channel
to search for the heavy neutral Higgs state $H^0$,
which also directly probes the cubic Higgs interaction $Hhh$\,.

\vspace*{1mm}

In Section\,\ref{sec:2}, we analyzed the viable parameter space for the 2HDM
Type-I and Type-II, which satisfies the theoretical constraints and
the current experimental limits. These are described
in Figs.\,\ref{fig:1}-\ref{fig:22} and Fig.\,\ref{fig:33}.
Then, we presented in Fig.\,\ref{fig:44} the resonant di-Higgs production with
$\,\sigma(gg\!\to\!H^0)\!\times\!\text{Br}(H^0\!\!\to\! h^0h^0)\,$
as a function of the Higgs mass $\,\MH\,$ at the LHC(14TeV), under
the theoretical constraints and the current indirect and direct experimental limits.
With these, we further set up three Higgs benchmark scenarios as in Eq.\eqref{eq:benchmark}
for the subsequent LHC analyses.

\vspace*{1mm}

In Section\,\ref{sec:3}, we performed systematical Monte Carlo analysis
for the resonant di-Higgs production via 
$\,gg\!\to\! H^0\!\!\to\! h^0h^0\!\to\! WW^*WW^*$
channel at the LHC(14TeV) by using Delphes\,3 fast detector simulations.
We studied the decay channels with the same-sign di-leptons (SS2L) final state
in Section\,\ref{sec:3.2} and with the three leptons (3L) final state
in Section\,\ref{sec:3.3},
where the QCD backgrounds can be efficiently suppressed.
The top quark and $Z$ boson induced backgrounds can be suppressed by
$b$-veto and $Z$-veto, respectively. We analyzed the event distributions of
different kinematical variables in Fig.\,\ref{fig:4} for the SS2L channel
and in Fig.\,\ref{fig:6} for the 3L channel.
We further optimized the signal significance by using the BDT method
for each Higgs benchmark as presented in Fig.\,\ref{fig:5} and
Tables\,\ref{tab:33},\ref{tab:55}.
We found that the 3L channel has a higher significance than the SS2L channel.
In Section\,\ref{sec:3.4}, we derived the combined significance
for both SS2L and 3L channels, as summarized in Table\,\ref{tab:combined_Z0}
for the three Higgs benchmarks. This table shows that to discover the new Higgs
state $H^0$ at the LHC(14TeV) with a $5\sigma$ significance,
the required minimal integrated luminosity is
$\,\mathcal{L}_{\min}^{5\sigma}=(222,\,446,\,2954)$fb$^{-1}$
for $\,\MH =(300,\,400,\,500)$GeV, respectively.

\vspace*{1mm}

In Section\,\ref{sec:4},
we systematically analyzed the LHC probe of the 2HDM parameter space
by using the combined searches of both SS2L and 3L decay channels of the $4W$ final state.
In Fig.\,\ref{fig:7}, we presented the parameter space
(red dots) which can be probed by the LHC direct searches of the heavy Higgs boson $H^0$,
including the existing $H^0$ search bounds and our study of
$\,gg\!\to\! H^0\!\to\! h^0h^0\!\to\! 4W\,$ searches
at the LHC Run-2 with $\mathcal{L}=300$\,fb$^{-1}$
(combined with the theoretical constraints).
In Fig.\,\ref{fig:8}, we further demonstrated that with an integrated luminosity of
3000\,fb$^{-1}$\! at the HL-LHC, the probed parameter space of the 2HDM-I
and 2HDM-II can be significantly expanded towards the alignment region,
especially for the higher Higgs mass range
$\MH\gtrsim 400$\,GeV.
We find that considerable parts of the parameter space
are within the reach of the HL-LHC for the heavy Higgs boson mass up to
$\,\MH\! = 500\,$GeV.
Finally, we expect that extending our current study
to the future high energy circular colliders $pp(50\!-\!100\text{TeV}$)\,\cite{SPPCFCC}
will further enhance the discovery reach of the new heavy Higgs boson $H^0$
with mass well above 1\,TeV through this resonant di-Higgs production channel.

%----------------------------------------------------------------------
\addcontentsline{toc}{section}{Acknowledgments\,}
\vspace*{2mm}
\section*{Acknowledgments}
\vspace*{-2mm}

We thank Mingshui Chen for discussing the current LHC Higgs search limits,
and thank Yun Jiang for discussing the 2HDM analysis in Ref.\,\cite{Bernon:2015qea}.
We are grateful to Howard Haber for discussing the oblique corrections of the 2HDM
and Michael Ramsey-Musolf for discussing the LHC diHiggs production in the 2HDM.
JR is supported in part by the Natural Sciences and Engineering Research Council of Canada.
RQX and HJH are supported in part by the National NSF of China
(under Grants 11675086 and 11275101),
by the Shanghai Key Laboratory for Particle Physics and Cosmology
(under Grant 11DZ2260700), by the Office of Science and Technology,
Shanghai Municipal Government (under Grant 16DZ2260200), and
by the MOE Key Laboratory for Particle Physics, Astrophysics and Cosmology.
WY was supported in part by the Office of Science, Office of High Energy Physics, of the
U.S.\ Department of Energy under contract DE-AC02-05CH11231.
This work is supported in part by the CAS Center for Excellence in Particle Physics (CCEPP).
%=======================================================================

%\bibliography{SU22LHC}

\vspace*{3mm}
\addcontentsline{toc}{section}{References\,}
\begin{thebibliography}{99}


\bibitem{ATLAS2012}
G.~Aad {\it et al.}  [ATLAS Collaboration],
%``Observation of a New Particle in Search for the Standard Model Higgs Boson
%with the ATLAS Detector at the LHC'',
Phys.\ Lett.\ B\,{716} (2012) 1 [arXiv:1207.7214 [hep-ex]];
%%CITATION = ARXIV:1207.7214;%%


\bibitem{CMS2012}
S.~Chatrchyan {\it et al.}  [CMS Collaboration],
%``Observation of a New Boson at a Mass of 125\,GeV with the CMS Experiment at the LHC'',
Phys.\ Lett.\ B\,{716} (2012) 30 [arXiv:1207.7235 [hep-ex]].
%%CITATION = ARXIV:1207.7235;%%


%\bibitem{Run1-sum-h}
\bibitem{Higgs-sum}
L.\ Pontecorvo, ``Highlights and perspectives from the ATLAS experiment",
and J.\ Butler,  ``Highlights and perspectives from the CMS experiment",
plenary talks at the Fifth Annual LHC Physics conference (LHCP-2017),
May 15-20, 2017, Shanghai, China.
G.~Aad {\it et al.,} [ATLAS and CMS Collaborations],
%``Measurements of the Higgs boson production and decay rates and constraints on its couplings from
% combined ATLAS and CMS analysis of the LHC pp collision data at $ \sqrt{s}=7 $ and 8 TeV,''
JHEP {1608} (2016) 045 %doi:10.1007/JHEP08(2016)045
[arXiv:1606.02266 [hep-ex]].


\bibitem{h3}
E.g., R. Contino {\it et al.,} arXiv:1606.09408 [hep-ph];
% The measurement of the Higgs self-coupling at the LHC: theoretical status
J.\ Baglio, A.\ Djouadi, R.\ Gr\"{o}ber, M.\ M.\ M\"{u}hlleitner, J.\ Quevillon,
M. Spira,  JHEP 1304 (2013) 151 [arXiv:1212.5581];
Weiming Yao, arXiv:1308.6302 [hep-ph], in the Proceedings of Snowmass Community Summer Study
(CSS 2013), Snowmass on the Mississippi, July\,29-August\,6, 2013, MN, USA;
H.-J. He, J. Ren, and W. Yao, Phys. Rev. D 93 (2015) 015003 [arXiv:1506.03302];
F.\ Goertz, A.\ Papaefstathiou, L.\,L.\ Yang and J.\ Zurita, JHEP {1306} (2013) 016
[arXiv:1301.3492 [hep-ph]];
S.\ Di Vita, C.\ Grojean, G.\ Panico, M.\ Riembau and T.\ Vantalon,
JHEP {1709} (2017) 069 [arXiv:1704.01953 [hep-ph]];
and references therein.


\bibitem{SU2x}
%\bibitem{Wang:2013jwa}
For the minimal gauge extensions with extra SU(2) and two Higgs doublets
(including di-Higgs decay channel $H\to hh$), e.g.,
X.-F.\ Wang, C.\ Du, H.-J. He, Phys.\ Lett.\ B {723} (2013) 314 [arXiv:1304.2257];
% LHC Higgs Signatures from Topflavor Seesaw Mechanism
% \bibitem{Abe:2012fb}
T.\,Abe, N.\,Chen, H.-J. He, JHEP {1301} (2013) 082 [arXiv:1207.4103];
% LHC Higgs Signatures from Extended Electroweak Gauge Symmetry
and references therein.


\bibitem{2HDM}
For a review, {\it e.g.,}
G.\ C.\ Branco, P.\ M.\ Ferreira, L.\ Lavoura, M.\ N.\ Rebelo, M.\ Sher, and J.\ P.\ Silva,
Phys.\ Rept.\ 516 (2012) 1 [arXiv:1106.0034 [hep-ph]]; and references therein.


\bibitem{MSSM}
For a review, e.g., A.\ Djouadi, Phys.\ Rept.\ 459 (2008) 1
[arXiv:hep-ph/0503173]; and references therein.
% The Anatomy of electro-weak symmetry breaking.
% II. The Higgs bosons in the minimal supersymmetric model


\bibitem{NMSSM}
For a review, e.g., U.\ Ellwanger, C.\ Hugonie, A.\ M.\ Teixeira,
Phys.\ Rept.\ 496 (2010) 1 [arXiv:0910.1785 [hep-ph]];
and references therein.


\bibitem{Aad:2015uka}
  G.~Aad {\it et al.} [ATLAS Collaboration],
  %``Search for Higgs boson pair production in the $b\bar{b}b\bar{b}$ final state from pp collisions at $\sqrt{s} = 8$ TeVwith the ATLAS detector,''
  Eur.\ Phys.\ J.\ C {75} (2015) 412 %no.9
  % doi:10.1140/epjc/s10052-015-3628-x
  [arXiv:1506.00285 [hep-ex]].


\bibitem{Aad:2014yja}
  G.~Aad {\it et al.} [ATLAS Collaboration],
  %``Search For Higgs Boson Pair Production in the $\gamma\gamma b\bar{b}$ Final State using $pp$ Collision Data at $\sqrt{s}=8$ TeV from the ATLAS Detector,''
  Phys.\ Rev.\ Lett.\  {114} (2015) 081802  % doi:10.1103/PhysRevLett.114.081802
  [arXiv:1406.5053 [hep-ex]].


\bibitem{CMSHhh}
  CMS Collaboration,
  % ``Search for the resonant production of two Higgs bosons
  % in the final state with two photons and two bottom quarks,''
  CMS-PAS-HIG-13-032 (April 28, 2014).


\bibitem{Aad:2015xja}
  G.~Aad {\it et al.} [ATLAS Collaboration],
  %``Searches for Higgs boson pair production in the $hh\to bb\tau\tau, \gamma\gamma WW^*, \gamma\gamma bb, bbbb$ channels with the ATLAS detector,''
  Phys.\ Rev.\ D {92} (2015) 092004
  % doi:10.1103/PhysRevD.92.092004
  [arXiv:1509.04670 [hep-ex]].

%\bibitem{Aad:2015xja}
%ATLAS Collaboration, arxiv:1509.04670


\bibitem{ATLAS:2016ixk}
ATLAS Collaboration [ATLAS Note],
%``Search for pair production of Higgs bosons in the $b\bar{b}b\bar{b}$ final state
% using proton$-$proton collisions at $\sqrt{s} = 13$ TeV with the ATLAS detector,''
ATLAS-CONF-2016-049 (August\,8, 2016).


\bibitem{CMS:2016pwo}
  CMS Collaboration,
  %``Search for heavy resonances decaying to a pair of Higgs bosons in four b quark final state in proton-proton collisions at sqrt(s)=13 TeV,''
  CMS-PAS-B2G-16-008 (July\,30, 2016);
  CMS-PAS-HIG-16-002 (March\,23, 2016).


\bibitem{ATLASRun2bbgg}
  ATLAS Collaboration,
  %``Search for Higgs boson pair production in the $b\bar{b}\gamma\gamma$ final state using pp collision data at $\sqrt{s}=13$ TeV with the ATLAS detector,''
  ATLAS-CONF-2016-004 (March 14, 2016).


\bibitem{CMS:2016vpz}
  CMS Collaboration,
  %``Search for H(bb)H(gammagamma) decays at 13TeV,''
  CMS-PAS-HIG-16-032 (August 15, 2016).


\bibitem{CMS:2017orf}
  CMS Collaboration,
  %``Search for pair production of Higgs bosons in the two tau leptons and two bottom quarks final state using proton-proton collisions at $\sqrt{s} = 13~\mathrm{TeV}$,''
  CMS-PAS-HIG-17-002 (March 19, 2017).


\bibitem{CMS:2016rec}
  CMS Collaboration,
  %``Search for resonant Higgs boson pair production in the
  % $\mathrm{b}\overline{\mathrm{b}} \mathrm{l}\nu \mathrm{l}\nu$
  % final state at $\sqrt{s} = 13~\mathrm{TeV}$,''
  % CMS-PAS-HIG-16-011.
  % Search for resonant and non-resonant Higgs boson pair production
  % in the bb-->l糸l糸bb‘l糸l糸 final state at s﹟=13 TeVs=13 TeV, CMS Collaboration. 2017.
  CMS-PAS-HIG-17-006 (March\,26, 2017)
  and CMS-PAS-HIG-16-011 (March\,27, 2016).


\bibitem{ATLAS:2016qmt}
  ATLAS collaboration,
  %``Search for Higgs boson pair production in the final state of $\gamma\gamma WW^*$($\rightarrow l\nu jj$) using 13.3 fb$^{-1}$ of $pp$ collision data recorded at $\sqrt{s}= $ 13 TeV with the ATLAS detector,''
  ATLAS-CONF-2016-071 (August\,4, 2016).


\bibitem{LMPHhh}
V.~Martin-Lozano, J.~M.~Moreno and C.~B.~Park,
%``Resonant Higgs boson pair production in the $ hh\to b\overline{b}\
%  WW\to b\overline{b}{\ell}^{+}\nu {\ell}^{-}\overline{\nu} $ decay channel,''
JHEP {1508} (2015) 004 [arXiv:1501.03799 [hep-ph]].


\bibitem{llc}
L.-C.\ L\"{u}, C.\ Du, Y.\ Fang, H.-J.\ He, H.\ Zhang,
Phys.\ Letts.\ B 755 (2016) 509 [arXiv:1507.02644].
%%% DOI: 10.1016/j.physletb.2016.02.026
%%% SLACcitation = "%%CITATION = ARXIV:1507.02644;%%"


\bibitem{hh-other}
E.g.,
M.~J.~Dolan, C.~Englert, and M.~Spannowsky,
% ``New Physics in LHC Higgs boson pair production,''
Phys.\ Rev.\ D {87} (2013) 055002 [arXiv:1210.8166 [hep-ph]];
% doi:10.1103/PhysRevD.87.055002
%
J.~M.~No and M.~Ramsey-Musolf,
% ``Probing the Higgs Portal at the LHC Through Resonant di-Higgs Production,''
Phys.\ Rev.\ D {89} (2014) 095031 [arXiv:1310.6035 [hep-ph]];
% doi:10.1103/PhysRevD.89.095031
%
N.~Kumar and S.~P.~Martin,
% ``LHC search for di-Higgs decays of stoponium and other scalars in events
% with two photons and two bottom jets,''
Phys.\ Rev.\ D {90} (2014) 055007 [arXiv:1404.0996 [hep-ph]];
% doi:10.1103/PhysRevD.90.055007
%
B.~Bhattacherjee and A.~Choudhury,
% ``Role of supersymmetric heavy Higgs boson production in the self-coupling
% measurement of 125 GeV Higgs boson at the LHC,''
Phys.\ Rev.\ D {91} (2015) 073015 [arXiv:1407.6866 [hep-ph]];
% doi:10.1103/PhysRevD.91.073015
%
C.~Y.~Chen, S.~Dawson, and I.~M.~Lewis,
% ``Exploring resonant di-Higgs boson production in the Higgs singlet model,''
Phys.\ Rev.\ D {91} (2015) 035015 [arXiv:1410.5488 [hep-ph]];
% doi:10.1103/PhysRevD.91.035015
%
M.~van Beekveld, W.~Beenakker, S.~Caron, R.~Castelijn, M.~Lanfermann and A.~Struebig,
% ``Higgs, di-Higgs and tri-Higgs production via SUSY processes at the LHC with 14 TeV,''
JHEP {1505} (2015) 044 [arXiv:1501.02145 [hep-ph]];
% doi:10.1007/JHEP05(2015)044
%
V.~Mart赤n Lozano, J.~M.~Moreno and C.~B.~Park,
% ``Resonant Higgs boson pair production in the
% $ hh\to b\overline{b}\ WW\to b\overline{b}{\ell}^{+}\nu {\ell}^{-}\overline{\nu} $ decay channel,''
JHEP {1508} (2015) 004 [arXiv:1501.03799 [hep-ph]];
% doi:10.1007/JHEP08(2015)004
%
J.~Bernon, J.~F.~Gunion, H.~E.~Haber, Y.~Jiang and S.~Kraml,
%``Scrutinizing the alignment limit in two-Higgs-doublet models: m$_h$=125??GeV,''
Phys.\ Rev.\ D {92} (2015) 075004 %doi:10.1103/PhysRevD.92.075004
[arXiv:1507.00933 [hep-ph]].
%
S.~Banerjee, B.~Batell and M.~Spannowsky,
% ``Invisible decays in Higgs boson pair production,''
Phys.\ Rev.\ D {95} (2017) 035009 [arXiv:1608.08601 [hep-ph]];
% doi:10.1103/PhysRevD.95.035009
%
T.~Huang, J.~M.~No, L.~Pernie, M.~Ramsey-Musolf, A.~Safonov, M.~Spannowsky and P.~Winslow,
% ``Resonant Di-Higgs Production in the $b{\bar b}WW$ Channel:
% Probing the Electroweak Phase Transition at the LHC,''
arXiv:1701.04442 [hep-ph];
%
K.~Nakamura, K.~Nishiwaki, K.~y.~Oda, S.~C.~Park and Y.~Yamamoto,
% ``Di-higgs enhancement by neutral scalar as probe of new colored sector,''
Eur.\ Phys.\ J.\ C {77} (2017) 273 [arXiv:1701.06137 [hep-ph]];
% doi:10.1140/epjc/s10052-017-4835-4
and references therein.


\bibitem{4W-SM}
U.~Baur, T.~Plehn and D.~L.~Rainwater,
% ``Determining the Higgs boson selfcoupling at hadron colliders,''
Phys.\ Rev.\ D {67} (2003) 033003 [arXiv:hep-ph/0211224];
% doi:10.1103/PhysRevD.67.033003
%``Measuring the Higgs boson self coupling at the LHC and finite top mass matrix elements,''
Phys.\ Rev.\ Lett.\  {89} (2002) 151801 [arXiv:hep-ph/0206024].
\\
Q.~Li, Z.~Li, Q.-S.~Yan, and X.~Zhao,
% ``Probe Higgs boson pair production via the 3?2j+$\not{E}$ mode,''
Phys.\ Rev.\ D {92} (2015) 014015 (2015) [arXiv:1503.07611 [hep-ph]];
% doi:10.1103/PhysRevD.92.014015
%
% \bibitem{Zhao:2016tai}
X.~Zhao, Q.~Li, Z.~Li, and Q.-S.~Yan,
%``Discovery potential of Higgs boson pair production through
% 4$\ell$+$E\!\!/$ final states at a 100 TeV collider,''
Chin.\ Phys.\ C {41} (2017) 023105 [arXiv:1604.04329 [hep-ph]].
% doi:10.1088/1674-1137/41/2/023105


\bibitem{SM+SS4W}
Chien-Yi Chen, Jonathan Kozaczuk, Ian M.\ Lewis,
% Non-resonant Collider Signatures of a Singlet-Driven Electroweak Phase Transition
JHEP 1708 (2017) 096 [arXiv:1704.05844 [hep-ph]].


\bibitem{s}
S.\ v.\ Buddenbrock, N.\ Chakrabarty, A.\ S.\ Cornell, D.\ Kar, M.\ Kumar, T.\ Mandal,\\
Bruce Mellado, B.\ Mukhopadhyaya, R.\ G.\ Reed, X.\ Ruan,
Eur.\ Phys.\ J.\ C\,76 (2016) 580 [arXiv:1606.01674 [hep-ph]].
% Nabarun Chakrabarty, Alan S. Cornell, Deepak Kar, Mukesh Kumar,
% Tanumoy Mandal, Bruce Mellado, Biswarup Mukhopadhyaya, Robert G. Reed, Xifeng Ruan
% Phenomenological signatures of additional scalar bosons at the LHC


\bibitem{lambda}
J.\ F.\ Gunion and H.\ E.\ Haber,
%``The CP conserving two Higgs doublet model: the approach to the decoupling limit",
Phys.\ Rev.\ D\,67 (2003) 075019 [arXiv:hep-ph/0207010].


\bibitem{Bernon:2015qea}
  J.~Bernon, J.~F.~Gunion, H.~E.~Haber, Y.~Jiang and S.~Kraml,
  %``Scrutinizing the alignment limit in two-Higgs-doublet models: m$_h$=125GeV,''
  Phys.\ Rev.\ D {92} (2015) 075004
  %doi:10.1103/PhysRevD.92.075004
  [arXiv:1507.00933 [hep-ph]];
  see also: Yun Jiang, Ph.D thesis, ``Higgs Boson Physics beyond the Standard Model",
  UC Davis, 2015.


\bibitem{Misiak:2017bgg}
M.\ Misiak and M.\ Steinhauser,
Eur.\ Phys.\ J.\ C\,77 (2017) 201 [arXiv:1702.04571].
% "Weak radiative decays of the B meson and bounds on MH㊣MH㊣
% in the Two-Higgs-Doublet Model"


\bibitem{unitarity1}
C.\ H.\ Llewellyn Simth, Phys.\ Lett.\ 46B (1973) 233;
D.\ A.\ Dicus and V.\ S.\ Mathur, Phys.\ Rev.\ D\,7 (1973) 3111;
J.\ M.\ Cornwall, D.\ N.\ Levin, and G.\ Tiktopoulos,
Phys.\ Rev.\ D\,10 (1974) 1145;
M.\ Veltman, Acta Phys.\ Pol.\ B\,8 (1977) 475;
B.\ W.\ Lee, C.\ Quigg, and H.\ B.\ Thacker, Phys.\ Rev.\
D\,16 (1977) 1519;
M.\ S.\ Chanowitz and M.\ K.\ Gaillard,
Nucl.\ Phys.\ B\,261 (1985) 379; \\
%\bibitem{unitarity2}
%\bibitem{Dicus:2004rg}
D.\ A.\ Dicus and H.-J. He,
%Scales of Fermion Mass Generation and Electroweak Symmetry Breaking
Phys.\ Rev.\ D 71 (2005) 093009 [arXiv:hep-ph/0409131]; \\
%\bibitem{Dicus:2005ku}
Phys.\ Rev.\ Lett.\ 94 (2005) 221802 [arXiv:hep-ph/0502178];
and references therein.


\bibitem{ET}
For a comprehensive review,
H.\ J.\ He, Y.\ P.\ Kuang and C.\ P.\ Yuan,
arXiv:hep-ph/9704276 and DESY-97-056,
in the proceedings of the workshop on ``Physics at the TeV Energy Scale",
vol.72 (1996) p.119. See also,
H.\ J.\ He and W.\ B.\ Kilgore,
Phys.\ Rev.\ D 55 (1997) 1515; %[arXiv:hep-ph/9609326];
% DOI: 10.1103/PhysRevD.55.1515 
% SLACcitation   = "%%CITATION = HEP-PH/9609326;%%"
H.\ J.\ He, Y.\ P.\ Kuang, C.\ P.\ Yuan,
Phys.\ Rev.\ D 51 (1995) 6463; %[arXiv:hep-ph/9410400];
% DOI: 10.1103/PhysRevD.51.6463 
% SLACcitation   = "%%CITATION = HEP-PH/9410400;%%"
Phys.\ Rev.\ D 55 (1997) 3038; %[arXiv:hep-ph/9611316]
% DOI: 10.1103/PhysRevD.55.3038 
% SLACcitation   = "%%CITATION = HEP-PH/9611316;%%"
H.\ J.\ He, Y.\ P.\ Kuang, X.\ Li,
Phys.\ Lett.\ B 329 (1994) 278; 
% DOI: 10.1016/0370-2693(94)90772-2 
% SLACcitation   = "%%CITATION = HEP-PH/9403283;%%"
Phys.\ Rev.\ D 49 (1994) 4842;
% DOI: 10.1103/PhysRevD.49.4842
% SLACcitation   = "%%CITATION = PHRVA,D49,4842;%%"
Phys.\ Rev.\ Lett.\ 69 (1992) 2619; 
% DOI: 10.1103/PhysRevLett.69.2619
% SLACcitation   = "%%CITATION = PRLTA,69,2619;%%"
and references therein.


\bibitem{STU}
%Estimation of oblique electroweak corrections
M.\ E.\ Peskin and T.\ Takeuchi, Phys.\ Rev.\ D\,46 (1992) 381.


\bibitem{STUHaber}	
H.\ E.\ Haber, %Introductory low-energy supersymmetry,
arXiv:hep-ph/9306207,
in the proceedings of {\it Recent Directions in Particle Theory,} p.589, %p.589-686,
edited by J.\ Harvey and J.\ Polchinski,
Boulder, Colorado, USA, 1992; \\
H.\ E.\ Haber and D.\ O'Neil,
Phys.\ Rev.\ D\,83 (2011) 055017 [arXiv:1011.6188 [hep-ph]].
% Basis-independent methods for the two-Higgs-doublet model III:
% The CP-conserving limit, custodial symmetry, and the oblique parameters S, T, U


\bibitem{stu1}
%\bibitem{He:2001tp}
H.-J. He, N. Polonsky and S.\ Su,
Phys.\ Rev.\ D 64 (2001) 053004 [arXiv:hep-ph/0102144].


\bibitem{stu2}
M.\ Baak {\it et al.,} [Gfitter Group Collaboration],
Eur.\ Phys.\ J.\ C\,74 (2014) 3046 [arXiv:1407.3792 [hep-ph]].


\bibitem{2HDMC}
D.\ Eriksson, J.\ Rathsman, and O.\ Stal,
``2HDMC: Two-Higgs-Doublet Model Calculator Physics and Manual",
Comput.\ Phys.\ Commun.\ 181 (2010) 189 [arXiv:arXiv:0902.0851].


\bibitem{Run1-sum-h}
  G.~Aad {\it et al.} [ATLAS and CMS Collaborations],
  %``Measurements of the Higgs boson production and decay rates and constraints on its couplings from a combined ATLAS and CMS analysis of the LHC pp collision data at $ \sqrt{s}=7 $ and 8 TeV,''
  JHEP {1608} (2016) 045 %doi:10.1007/JHEP08(2016)045
  [arXiv:1606.02266 [hep-ex]].



\bibitem{Atlas2-gaga-ZZ}
ATLAS Collaboration [ATLAS Note], ATLAS-CONF-2017-047 (July\,6, 2017).
% Measurement of fiducial, differential and production cross sections
% in the H↙污污 decay channel with 13.3 fb?1?1 of 13 TeV
% proton-proton collision data with the ATLAS detector
%ATLAS Collaboration [ATLAS Note], ATLAS-CONF-2016-067
%and ATLAS-CONF-2016-081 (August\,8, 2016).
%
%\bibitem{Atlas2-gaga-ZZ}
% Study of the Higgs boson properties and search for high-mass scalar resonances
% in the H↙ZZ?↙4?H↙ZZ?↙4? decay channel at s﹟s = 13 TeV with the ATLAS detector
%ATLAS Collaboration [ATLAS Note], ATLAS-CONF-2016-079
%
% \bibitem{hZZaaATLAS13}
% The ATLAS collaboration [ATLAS Collaboration],
% ``Combined measurements of the Higgs boson production and decay rates
% in $H\to ZZ^*\to 4\ell$ and $H\to\gamma\gamma$ final states
% using $pp$ collision data at $\sqrt{s}=$ 13 TeV
% in the ATLAS experiment,'' Aug 8, 2016.
%and ATLAS-CONF-2016-081 (August\,8, 2016).
%ATLAS-CONF-2017-047


\bibitem{tthATLAS}
ATLAS Collaboration [ATLAS Note], ATLAS-CONF-2016-068 (August\,3, 2016).
% Combination of the searches for Higgs boson production in association with top quarks
% in the $\gamma\gamma$, multilepton, and $b\bar{b}$ decay channels
% at $\sqrt{s}=13$TeV with the ATLAS Detector


\bibitem{Atlas2-WW}
% Measurements of the Higgs boson production cross section via Vector Boson Fusion
% and associated WHWH production in the WW*↙l糸l糸 decay mode with the ATLAS detector
% at s﹟s = 13 TeV
ATLAS Collaboration [ATLAS Note], ATLAS-CONF-2016-112 (November\,10 2016).


\bibitem{tthATLASbb}
ATLAS Collaboration [ATLAS Note], ATLAS-CONF-2016-080
(August\,4, 2016).
% Search for the Standard Model Higgs boson produced in association with top quarks
% and decaying into $b\bar{b}$ in $pp$ collisions at $\sqrt{s} = 13$TeV with the ATLAS detector


\bibitem{Atlas2-bb}
% Search for the Standard Model Higgs boson produced in association
% with a vector boson and decaying to a bb pair in pp collisions
% at 13 TeV using the ATLAS detector
ATLAS Collaboration [ATLAS Note], ATLAS-CONF-2016-091 (August\,6, 2016).


\bibitem{CMS2-ZZ}
% Measurements of properties of the Higgs boson decaying into four leptons
% in pp collisions at sqrt{s} = 13 TeV
CMS Collaboration [CMS Note], CMS-PAS-HIG-16-041 (March\,21, 2017).


\bibitem{CMS2-gaga}
CMS Collaboration [CMS Note], CMS-PAS-HIG-16-040 (May\,15, 2017).
% Measurements of properties of the Higgs boson
% in the diphoton decay channel with the full 2016 data set
%
% \bibitem{haaCMS13}
% CMS Collaboration [CMS Collaboration],
% ``Updated measurements of Higgs boson production in the diphoton decay channel
% at $\sqrt{s}=13~\textrm{TeV}$ in pp collisions at CMS.,''
% CMS-PAS-HIG-16-020.


\bibitem{CMS2x-gaga}
CMS Collaboration [CMS Note], CMS-PAS-HIG-16-020 (August\,4, 2016).
% Updated measurements of Higgs boson production in the diphoton decay channel
% at $\sqrt{s}=13$TeV in pp collisions at CMS.


\bibitem{CMS2-WW}
CMS Collaboration [CMS Note], CMS-PAS-HIG-16-023 (August\,4, 2016).
% Search for high mass Higgs to WW with fully leptonic decays using 2015 data


\bibitem{CMS2-tautau}
% Observation of the SM scalar boson decaying to a pair of 而而 leptons
% with the CMS experiment at the LHC
CMS Collaboration [CMS Note], CMS-PAS-HIG-16-043 (May\,15, 2017).


\bibitem{tthbb}
CMS Collaboration [CMS Note], CMS-PAS-HIG-16-038 (November\,9, 2016).
% Search for ttH production in the H --> bb decay channel
% with 2016 pp collision data at $\sqrt{s} = 13$TeV


\bibitem{CMS2-bb}
CMS Collaboration [CMS Note], CMS-PAS-HIG-17-010 (May\,29, 2017).
% Inclusive search for the standard model Higgs boson produced in pp collisions
% at s﹟=s= 13 TeV using H↙bb‘H↙bb‘ decays


\bibitem{Haber2015}
H.\ E.\ Haber and O.\ Stal,
Eur.\ Phys.\ J.\ C\,75 (2015) 491 [arXiv:1507.04281 [hep-ph]].
% New LHC Benchmarks for the CP-conserving Two-Higgs-Doublet Model


% \bibitem{htt-talk2017}
% For a recent review, G.\ Petrucciani [on behalf of ATLAS and CMS Collaboration],
% ``Evidence for the $t\bar{t}h$ Production with 13\,TeV Data?",
%% https://indico.in2p3.fr/event/13763/session/0/contribution/79/material/slides/0.pdf
% invited presentation at the 50th Recontres de Moriond,
% ``Electroweak Interactions and Unified Theories",
% March 18-25, 2017, La Thuile, Valle d'Aosta, Italy.


\bibitem{ATLASHVV}
  G.~Aad {\it et al.,} [ATLAS Collaboration],
  %``Search for an additional, heavy Higgs boson in the $H\rightarrow ZZ$ decay channel at $\sqrt{s} = 8\;\text{ TeV }$ in $pp$ collision data with the ATLAS detector,''
  Eur.\ Phys.\ J.\ C\,76 (2016) 45
  % doi:10.1140/epjc/s10052-015-3820-z
  [arXiv:1507.05930 [hep-ex]].


\bibitem{CMSHVV}
  V.~Khachatryan {\it et al.} [CMS Collaboration],
  %``Search for a Higgs boson in the mass range from 145 to 1000 GeV decaying to a pair of W or Z bosons,''
  JHEP {1510} (2015) 144 % doi:10.1007/JHEP10(2015)144
  [arXiv:1504.00936 [hep-ex]].


\bibitem{HWWlvlvATLAS13}
  ATLAS collaboration [ATLAS Note],
  %``Search for a high-mass Higgs boson decaying to a pair of $W$ bosons in $pp$ collisions at $\sqrt{s}$=13 TeV with the ATLAS detector,''
  ATLAS-CONF-2016-074 (August\,5, 2016).


\bibitem{HZZ4lATLAS13}
  ATLAS collaboration [ATLAS Note],
  %``Study of the Higgs boson properties and search for high-mass scalar resonances in the $H \rightarrow ZZ^* \rightarrow 4\ell$ decay channel at $\sqrt{s}$ = 13 TeV with the ATLAS detector,''
  ATLAS-CONF-2016-079 (August\,4, 2016).


\bibitem{HZZ2l2vATLAS13}
  ATLAS collaboration [ATLAS Note],
  %``Search for new phenomena in the $Z(\rightarrow\ell\ell) + E_{\mathrm{T}}^{\mathrm{miss}}$ final state at $\sqrt{s}$ = 13 TeV with the ATLAS detector,''
  ATLAS-CONF-2016-056 (August\,3, 2016).


\bibitem{tataATLAS}
ATLAS Collaboration [ATLAS Note], ATLAS-CONF-2016-085 (August\,4, 2016).
% Search for the Minimal Supersymmetric Standard Model Higgs bosons H/A
% in the 而而 final state in up to 13.3 fb?1 of p p collision data
% at ﹟s = 13 TeV with the ATLAS detector


\bibitem{tataCMS}
CMS Collaboration [CMS Note],
% Search for a neutral MSSM Higgs boson decaying into 而而 with 12.9 fb?1 of data at ﹟s = 13 TeV
CMS-PAS-HIG-16-037 (November\,9, 2016).


\bibitem{Hbb98}
%\bibitem{Balazs:1998nt}
E.g., C.\ Balazs, J.\ L.\ Diaz-Cruz, H.\ J.\ He, T.\ Tait, C.\ P.\ Yuan,
% Probing Higgs Bosons with Large Bottom Yukawa Coupling at Hadron Colliders
Phys.\ Rev.\ D\,59 (1999) 055016 [arXiv:hep-ph/9807349];
% DOI: 10.1103/PhysRevD.59.055016 
% SLACcitation = "%%CITATION = HEP-PH/9807349;%%"
% \bibitem{DiazCruz:1998qc}
J.\ L.\ Diaz-Cruz, H.\ J.\ He, T.\ Tait and C.\ P.\ Yuan,
% Higgs bosons with large bottom Yukawa coupling at Tevatron and LHC
Phys.\ Rev.\ Lett.\ 80 (1998) 4641 [arXiv:hep-ph/9802294];
% DOI: 10.1103/PhysRevLett.80.4641 
% SLACcitation = "%%CITATION = HEP-PH/9802294;%%"
and references therein.


%\bibitem{HigMssm-Rev}
%For a review,
%A.\ Djouadi, Phys.\ Rept.\ 459 (2008) 1 [arXiv:hep-ph/0503173];
%The Anatomy of electro-weak symmetry breaking. II. The Higgs bosons in the minimal supersymmetric model
%and references therein.


\bibitem{Hdecay}
E.g., A. Djouadi, J. Kalinowski, and M. Spira,
Comput.\ Phys.\ Commun.\ 108 (1998) 56 [arXiv:hep-ph/9704448];
% HDECAY: A Program for Higgs boson decays in the standard model and its supersymmetric extension
A.\ Djouadi, Phys.\ Rept.\ 459 (2008) 1 [arXiv:hep-ph/0503173]; \\
and references therein.
%The Anatomy of electro-weak symmetry breaking. II. The Higgs bosons
%in the minimal supersymmetric model


\bibitem{Alloul:2013bka}
  A.~Alloul, N.~D.~Christensen, C.~Degrande, C.~Duhr and B.~Fuks,
  %``FeynRules  2.0 - A complete toolbox for tree-level phenomenology,''
  Comput.\ Phys.\ Commun.\  {185} (2014) 2250 % doi:10.1016/j.cpc.2014.04.012
  [arXiv:1310.1921 [hep-ph]].


\bibitem{Alwall:2014hca}
  J.~Alwall {\it et al.},
  %``The automated computation of tree-level and next-to-leading order differential cross sections, and their matching to parton shower simulations,''
  JHEP {1407} (2014) 079  % doi:10.1007/JHEP07(2014)079
  [arXiv:1405.0301 [hep-ph]].


\bibitem{Sjostrand:2006za}
  T.~Sjostrand, S.~Mrenna and P.~Z.~Skands,
  %``PYTHIA 6.4 Physics and Manual,''
  JHEP {0605} (2006) 026  %doi:10.1088/1126-6708/2006/05/026
  [hep-ph/0603175].


\bibitem{deFavereau:2013fsa}
  J.~de Favereau {\it et al.} [Delphes-3 Collaboration],
  %``DELPHES 3, A modular framework for fast simulation of a generic collider experiment,''
  JHEP {1402} (2014) 057    %doi:10.1007/JHEP02(2014)057
  [arXiv:1307.6346 [hep-ex]].


\bibitem{Djouadi-SMh2005}
For a review,
A.\ Djouadi, Phys.\ Rept.\ 457 (2008) 1-216 [arXiv:hep-ph/0503172].
% "The Anatomy of electro-weak symmetry breaking I: The Higgs boson in the standard model"


\bibitem{SusHI}
R.\ V.\ Harlander, S.\ Liebler, and H.\ Mantler,
Comput.\ Phys.\ Commun.\ 184 (2013) 1605 [arXiv:1212.3249 [hep-ph]].
% SusHi: A program for the calculation of Higgs production
% in gluon fusion and bottom-quark annihilation in the Standard Model and the MSSM


\bibitem{Hhh-2HDM}
%\bibitem{Hespel:2014sla}
E.g., B.~Hespel, D.~Lopez-Val and E.~Vryonidou,
%``Higgs pair production via gluon fusion in the Two-Higgs-Doublet Model,''
JHEP {1409} (2014) 124 %doi:10.1007/JHEP09(2014)124
[arXiv:1407.0281 [hep-ph]].


\bibitem{Agashe:2014kda}
  K.~A.~Olive {\it et al.} [Particle Data Group],
  %``Review of Particle Physics,''
  Chin.\ Phys.\ C\,{38} (2014) 090001.
  %doi:10.1088/1674-1137/38/9/090001


\bibitem{ATL-PHYS-PUB-2015-022}
  ATLAS Collaboration, %[ATLAS Collaboration],
  %``Expected performance of the ATLAS b-tagging algorithms in Run-2''
  % http://cds.cern.ch/record/2037697/files/ATL-PHYS-PUB-2015-022.pdf?version=1
  ATL-PHYS-PUB-2015-022 (July\,24, 2015).


\bibitem{criteria}
ATLAS Collaboration, ATLAS-CONF-2015-006 (March\,17, 2015).
% Search for the associated production of the Higgs boson with a top quark pair
% in multi-lepton final states with the ATLAS detector


\bibitem{ATL-PHYS-PUB-2013-004}
  ATLAS Collaboration, %[ATLAS Collaboration],
% http://cds.cern.ch/record/1527529/files/ATL-PHYS-PUB-2013-004.pdf
% Performance assumptions for an upgraded ATLAS detector at a High-Luminosity LHC
ATL-PHYS-PUB-2013-004(v2), (April\,3, 2014).


\bibitem{ATL-PHYS-PUB-2016-026}
  ATLAS Collaboration, %[ATLAS Collaboration],
  %``Expected performance for an upgraded ATLAS detector at High-Luminosity LHC''
  % http://cds.cern.ch/record/2223839/files/ATL-PHYS-PUB-2016-026.pdf
  ATL-PHYS-PUB-2016-026 (October\,11, 2016).


\bibitem{PhysRevLett.107.152003}
  G.~Ferrera, M.~Grazzini and F.~Tramontano,
  %``Associated WH production at hadron colliders: a fully exclusive QCD calculation at NNLO,''
  Phys.\ Rev.\ Lett.\  {107} (2011) 152003   %doi:10.1103/PhysRevLett.107.152003
  [arXiv:1107.1164 [hep-ph]].


\bibitem{Campbell2012}
  J.~M.~Campbell and R.~K.~Ellis,
  %``$t \bar{t} W^{+-}$ production and decay at NLO,''
  JHEP {1207} (2012) 052 %doi:10.1007/JHEP07(2012)052
  [arXiv:1204.5678 [hep-ph]].


\bibitem{1310.1132}
  J.~Adelman, A.~Loginov, P.~Tipton and J.~Vasquez,
  %``Study of $t\bar{t}H$ ($H \to \mu^+ \bar{\mu}^-$) in the three lepton channel at $\sqrt{s}=14$ TeV;
  %A Snowmass white paper,''
  Snowmass white paper, SNOW13-00197 [arXiv:1310.1132 [hep-ex]].


\bibitem{PhysRevD.60.113006}
  J.~M.~Campbell and R.~K.~Ellis,
  %``An Update on vector boson pair production at hadron colliders,''
  Phys.\ Rev.\ D\,{60} (1999) 113006  %doi:10.1103/PhysRevD.60.113006
  [arXiv:hep-ph/9905386].


\bibitem{https://twiki.cern.ch/twiki/bin/view/LHCPhysics/TtbarNNLO}
M. Czakon and A. Mitov, Comput.\ Phys.\ Commun.\ 185 (2014) 2930
[arXiv:1112.5675 [hep-ph]];
% Top$++$: A Program for the Calculation of the Top-Pair Cross-Section at Hadron Colliders,
and ``ATLAS-CMS recommended predictions for top-quark-pair cross sections
using the Top$^{++}$v2.0 program",
https://twiki.cern.ch/twiki/bin/view/LHCPhysics/TtbarNNLO


\bibitem{Hocker:2007ht}
  A.~Hocker {\it et al.},
  %``TMVA - Toolkit for Multivariate Data Analysis,''
  PoS ACAT (2007) 040 and CERN-OPEN-2007-007
  [arXiv:physics/0703039].


\bibitem{Z0}
G.\ Cowan, K.\ Cranmer, E.\ Gross, and O.\ Vitells,
Eur.\ Phys.\ J.\ C\,71 (2011) 1554 [arXiv:1007.1727 [physics.data-an]].
% Asymptotic formulae for likelihood-based tests of new physics


\bibitem{bbWW}
T.~Huang, J.~M.~No, L.~Pernie, M.~Ramsey-Musolf, A.~Safonov, M.~Spannowsky and P.~Winslow,
% ``Resonant Di-Higgs Production in the $b{\bar b}WW$ Channel:
% Probing the Electroweak Phase Transition at the LHC,''
Phys.\ Rev.\ D\,96 (2017) 035007 [arXiv:1701.04442 [hep-ph]].


\bibitem{0804.0350}
  T.~Binoth, G.~Ossola, C.~G.~Papadopoulos and R.~Pittau,
  %``NLO QCD corrections to tri-boson production,''
  JHEP {0806} (2008) 082
  % doi:10.1088/1126-6708/2008/06/082
  [arXiv:0804.0350 [hep-ph]].


\bibitem{SM-S}
E.g., Ian M. Lewis and Matthew Sullivan,
Phys.\ Rev.\ D96 (2017) 035037 [arXiv:1701.08774 [hep-ph]].
% Benchmarks for the Singlet Extended Standard Model at the LHC
% Chien-Yi Chen, Jonathan Kozaczuk, and Ian M.\ Lewis,
% Non-resonant Collider Signatures of a Singlet-Driven Electroweak Phase Transition
%arXiv:1704.05844 [hep-ph];
For a recent review including the singlet extension, e.g., 
I.\ P.\ Ivanov, Prog.\ Part.\ Nucl.\ Phys.\ 95 (2017) 160 [arXiv:1702.03776 [hep-ph]];
and references therein.


\bibitem{SPPCFCC}
FCC Collaboration,
M.\ Bicer {\it et al.,} JHEP 1401 (2014) 164 [arXiv:1308.6176 [hep-ex]],
http://tlep.web.cern.ch;
and CEPC/SPPC Collaboration, http://cepc.ihep.ac.cn


\end{thebibliography}
%\bibliographystyle{h-physrev5}
%\bibliographystyle{JHEP}

%\newpage

\end{document}